\documentclass{aastex61}
\usepackage{graphicx}
\usepackage{amsmath,amssymb,latexsym,wrapfig}
\usepackage{epstopdf}
\usepackage[flushleft]{threeparttable}
\usepackage{color}
\usepackage{float}

\newcommand{\Hb}{\ensuremath{{\rm H}\beta}}
\newcommand{\Ha}{\ensuremath{{\rm H}\alpha}}

\newcommand{\beq}{\begin{equation}}
\newcommand{\eeq}{\end{equation}}

\newcommand{\be}{\begin{equation}}
\newcommand{\ee}{\end{equation}}

\accepted{\today}
\submitjournal{ApJ}

\shorttitle{Anomalously Low Mass Black Hole in IC 750}
\shortauthors{Zaw et al.}

\begin{document}


\title{An Accreting, Anomalously Low Mass Black Hole at the Center of Low Mass Galaxy IC 750}

\correspondingauthor{Ingyin Zaw}
\email{iz6@nyu.edu}

\author[0000-0002-5208-1426]{Ingyin Zaw}
\affiliation{New York University Abu Dhabi,
    P.O. Box 129188, Abu Dhabi, UAE}
    
\author{Michael J. Rosenthal}
\affiliation{New York University Abu Dhabi,
    P.O. Box 129188, Abu Dhabi, UAE}    
    
\author[0000-0002-6425-6879]{Ivan Yu. Katkov}
\affiliation{New York University Abu Dhabi,
    P.O. Box 129188, Abu Dhabi, UAE}
    \affiliation{Sternberg Astronomical Institute, Moscow M.V. Lomonosov State University, 
    Universitetskij pr., 13, Moscow, 119234, Russia}

\author[0000-0003-4679-1058]{Joseph D. Gelfand}
\affiliation{New York University Abu Dhabi,
    P.O. Box 129188, Abu Dhabi, UAE}
    
\author[0000-0001-8821-0309]{Yan-Ping Chen}
\affiliation{New York University Abu Dhabi,
    P.O. Box 129188, Abu Dhabi, UAE}        

\author[0000-0003-4912-5974]{Lincoln J. Greenhill}
\affiliation{Harvard-Smithsonian Center for Astrophysics,
    60 Garden St., Cambridge, MA 02138, USA}

\author{Walter Brisken}
\affiliation{National Radio Astronomy Observatory, Soccoro, NM 87801, USA}

\author[0000-0002-4187-4981]{Hind Al Noori}
\affiliation{New York University Abu Dhabi,
    P.O. Box 129188, Abu Dhabi, UAE}
\affiliation{Department of Physics, University of California, Santa Barbara,
Santa Barbara, CA 93106, USA}
    






\begin{abstract}
We present a multi-wavelength study of the active galactic nucleus in the nearby ($D=14.1$ Mpc) low mass galaxy IC 750, which has circumnuclear 22 GHz water maser emission. The masers trace a nearly edge-on, warped disk $\sim$0.2 pc in diameter, coincident with the compact nuclear X-ray source which lies at the base of the $\sim$kpc-scale extended X-ray emission. The position-velocity structure of the maser emission indicates the central black hole (BH) has a mass less than $1.4 \times 10^5~M_\odot$. Keplerian rotation curves fitted to these data yield enclosed masses between $4.1 \times 10^4~M_\odot$ and $1.4 \times 10^5~M_\odot$, with a mode of $7.2 \times 10^4~M_\odot$. Fitting the optical spectrum, we measure a nuclear stellar velocity dispersion $\sigma_* = 110.7^{+12.1}_{-13.4}$~{\rm km~s}$^{-1}.$ From near-infrared photometry, we fit a bulge mass of  $(7.3 \pm 2.7) \times 10^8~M_\odot$ and a stellar mass of $1.4 \times 10^{10}~M_\odot$. The mass upper limit of the intermediate mass black hole in IC 750 falls roughly two orders of magnitude below the $M_{\rm BH}-\sigma_*$ relation and roughly one order of magnitude below the $M_{\rm BH}-M_{\rm Bulge}$ and $M_{\rm BH}-M_*$ relations -- larger than the relations' intrinsic scatters of (0.58 $\pm$ 0.09) dex, 0.69 dex, and (0.65 $\pm$ 0.09) dex, respectively. These offsets could be due to larger scatter at the low mass end of these relations. Alternatively, black hole growth is intrinsically inefficient in galaxies with low bulge and/or stellar masses, which causes the black holes to be under-massive relative to their hosts, as predicted by some galaxy evolution simulations.

\end{abstract}




\section{Introduction} 
\label{sec:intro}

Galaxies contain two types of black holes (BHs); stellar mass black holes have masses of $\sim$10 $M_\odot$ and are found scattered throughout the galaxy and massive black holes (mBHs), most with $M_{\rm BH} \gtrsim 10^5 M_\odot$, are usually found in the galactic nucleus. A Milky Way-sized galaxy typically has millions of stellar mass black holes and $\sim$1 mBH. Evidence of mBHs in dwarf ($M_* \lesssim 3 \times 10^9 M_\odot$) and low mass ($M_* \lesssim 10^{10} M_\odot$) galaxies is scarce. Currently, the largest sample \citep{Reines13} consists of 136 dwarf galaxies which show optical spectroscopic signatures of an active galactic nucleus (AGN) from a parent sample of $\sim$25,000 emission line galaxies in SDSS, a detection rate of $\sim$0.5\%. This rarity could be due to a low occupation fraction of black holes in small haloes, as seen in simulations \citep[e.g.,][]{Bellovary11}. In addition, low accretion rates and/or a low Eddington limit leading to a low bolometric luminosity make them difficult to distinguish from other sources, such as star formation and ultra-luminous X-ray sources (ULXs). 

Although rare, mBHs in low mass galaxies constitute an important population for understanding galaxy evolution. They occupy the low mass end of the black holes at galactic centers, shedding light on the existence and demographics of intermediate mass black holes (IMBHs), $10^2 M_\odot \lesssim M_{\rm BH} \lesssim 10^5 M_\odot$. In addition, their positions on the BH-galaxy relations such as $M_{\rm BH}-\sigma_*$ provide clues to black hole seed formation in the early universe \citep[e.g.,][]{Volonteri10} and test models of the accretion and feedback physics which regulate the co-evolution of mBHs and their host galaxies \citep[e.g.,][]{Alexander14, BHsonFIRE, Bower17, Dekel19}. In order to do so, precise and accurate measurements of $M_{\rm BH}$ are necessary.

At present, most of the mass measurements of low mass mBHs are for mBHs in broad line AGNs, and come from single epoch optical spectroscopy, based on the width of the broad H$\alpha$ line \citep[e.g.,][]{Reines13, Reines15, Schutte19, Xiao11}, with a small fraction from reverberation mapping \citep[e.g.,][]{Bentz18, Woo19}. Reverberation mapping, the more precise and accurate method of the two, relies on broad line region dynamics dominated by the central black hole. A dimensionless correctional factor, $f$, known as the virial factor, is needed to characterize the unknown affects due to the geometry, kinematics, and inclination of the broad line region. This factor cannot be determined for most individual BHs but is derived for the whole sample by calibrating to the $M_{\rm BH}-\sigma_*$ relation \citep[e.g.,][]{Park12, Peterson14, Woo10, Woo13, Woo15}. This can lead to uncertainties of up to a factor of $\sim$2 depending on the sample used \citep[e.g.][]{Peterson14, Greene19} but in several galaxies where dynamical mass measurements and reverberation mapped masses are available, the measurements agree \citep[e.g.,][]{Pancoast14,Zoghbi19}. The single epoch spectroscopic masses are, in turn, calibrated to reverberation mapping, relying on the empirical correlations between the radius of the broad line region and the continuum luminosity at 5100\AA, between the optical continuum luminosity and Balmer emission line luminosity, and the width of the broad H$\alpha$ line and the width of the braod H$\beta$ line \citep[e.g.,][]{Greene05, Peterson14, Reines13, Reines15}. Consequently, the single epoch spectroscopic mass measurements have errors that are at least 0.5 dex. Both reverberation mapped masses and single epoch spectroscopic masses might have some residual dependence on the $M_{\rm BH}-\sigma_*$ relation. Stellar dynamical masses of mBHs are available for a handful of low mass galaxies \citep[e.g.,][]{Nguyen19}. 

Keplerian fits of water megamasers in circumnuclear disks provide the most accurate and precise masses beyond the local group \citep[e.g.,][]{Humphreys13, Kuo11}. Water maser emission at 22.235 GHz ($\lambda$ = 1.35 cm) is the only known tracer of warm (T = $\sim400-1000$ K), dense ($\rho$ = $\sim10^7 - 10^{11}$ cm$^{-3}$) gas in the central parsec of AGNs ($10^4 \lesssim R_g \lesssim 10^6$) that is resolvable in both position and velocity. In some systems where very long baseline interferometry (VLBI) has resolved angular structure and rotation curves of water maser emission \citep[e.g.,][]{Miyoshi95}, emission has been shown to trace highly inclined accretion disks. The best working model for ``disk masers'' is that (nearly) edge-on orientation creates long amplification gain paths, and the disk structure is outlined by maser emission, producing a distinct radio line emission of Doppler components close to the host galaxy systemic velocity (systemic masers), which are moving across the line of sight, and features symmetrically offset by the rotation speed(s) of the disk (high velocity masers), which are moving towards the observer (blue-shifted) or away from the observer (redshifted). The systemic masers have line-of-sight velocities typically within a few km s$^{-1}$ of the velocity of the galaxy while the redshifted and blue-shifted masers have velocities which are the velocity of the galaxy plus or minus the rotational velocity at the given radius, respectively. Consequently, the velocities of the high velocity masers generally obey a Keplerian fall-off with radius. Since the masers are within the sphere of influence of the BH, the mass can be derived from first principles, making the method accurate, precise, and independent of BH-galaxy scaling relations. Furthermore, masers are almost exclusively in narrow line (Type 2) AGNs, a population where mass measurements are not possible using reverberation mapping and single epoch spectroscopy. The main challenge in the study of circumnuclear water maser emission in AGNs is that it is rare. Only $\sim$3-5\% of AGNs host water masers and, of those, only $\sim$1/3 are in disk systems \citep[e.g.,][]{Zhu11}. 

\citet{Chen17} report that IC 750 is a dwarf galaxy, $M_* \sim 1.3 \times 10^9~M_\odot$, which hosts a Type 2 AGN identified from its narrow optical lines. This highly inclined ($i \sim 66^\circ$) spiral galaxy also hosts water maser emission \citep{Darling14} where the radio spectrum is indicative of a disk maser, the only such system currently known. Therefore, IC 750 provides a rare opportunity to measure an accurate mBH mass and understand the subparsec scale geometry of accretion in a dwarf AGN. We present the first VLBI map of the maser emission in IC 750, using the Very Long Baseline Array (VLBA). In addition, we have analyzed archival, multi-wavelength data to derive the AGN and galaxy properties of IC 750 and examine the black hole-galaxy relations at the low mass end. As discussed in Section~\ref{sec:distance}, we adopt a distance of $(14.1\pm1.1)$ Mpc to IC 750.

This paper is organized as follows. The multi-wavelength observations and data reduction are described in Section~\ref{sec:data}. The results are detailed in Section~\ref{sec:results}. Specifically, the subparsec disk geometry is presented in Section~\ref{sec:maserdisk}, the measurement of $M_{\rm BH}$ is described in Section~\ref{sec:BHmass}, and the AGN and galaxy properties are reported in Section~\ref{sec:hostgal}. The location of IC 750 on the black hole-galaxy relations and the implications for models of galaxy evolution are discussed in Section~\ref{sec:discussion} and we conclude in Section~\ref{sec:conclusions}.

\section{Observations and Data Reduction}
\label{sec:data}
We have reduced and analyzed publicly available X-ray, optical, infrared data, and radio data, in addition to our VLBA observation of IC 750. The X-ray data are from {\it Chandra}, {\it XMM-Newton}, and {\it NuSTAR}. The optical data are from the Sloan Digitial Sky Survey (SDSS). The infrared data are from the {\it Hubble Space Telescope} ({\it HST}) and {\it Spitzer}. The radio data are from NSF's Karl G. Jansky Very Large Array (VLA) and the Robert C. Byrd Green Bank Telescope (GBT). The details of the multi-wavelength observations used in this paper are summarized in Table~\ref{tab:obs}. 

\begin{table*}[tbh]
 \caption{Properties of the Radio, X-ray, Optical, and Infrared Observations of IC 750}
    \begin{tabular}{ccccccccc}
      \hline
      Telescope & Date & Obs. ID & $\alpha_{\rm
        J2000}$ & $\delta_{\rm J2000}$ & $\theta_{\rm dist}$ & Obs. Time & Resolution \\
      \hline 
      {\bf VLBA} & {\bf 2018 Mar. 19} & {\bf BZ074A} & {\bf 11:58:52.252} & {\bf +42:43:20.245} & $\bf 0.\!\!^{\prime \prime}0126$ & {\bf 12 hours} & $\bf 0.90~{\rm mas} \times 0.58~{\rm mas}^1$\\
      VLA & 2012 Mar. 03 & 12A-283 & 11:58:52.225 & 42:43:20.65 & $0.\!\!^{\prime \prime}537$ & 2 hours & 1\arcsec\\
      GBT &  2012 Feb. 28 & AGBT12A\_046 & 11:58:52.18 & +42:43:21.3 & $0.\!\!^{\prime \prime}0107$ & 1 hour & 34\arcsec \\
      GBT & 2015 Apr. 23 & AGBT14B\_342 & 11:58:52.21 & +42:43:21.0 & $0.\!\!^{\prime \prime}0050$  & 10 mins & 34\arcsec \\
      \hline
      {\it XMM-Newton} & 2004 Nov. 28 & 0744040301 & 11:58:52.60 &
      +42:34:13.2 & $9.\!^{\prime}1$ & 23.9~ks & 4.5\arcsec (MOS), 15\arcsec (pn) \\
      {\it XMM-Newton} & 2004 Nov. 30 & 0744040401 & 11:58:52.60 &
      +42:34:13.2 & $9.\!^{\prime}1$ & 22.9~ks & 4.5\arcsec (MOS), 15\arcsec (pn) \\
      {\it NuSTAR} & 2012 Oct. 28 & 60061217002 & 11:58:52.75 &
      +42:34:12.4 & $9.\!^{\prime}1$ & 13.9~ks & 1\arcmin \\
      {\it NuSTAR} & 2013 Feb. 04 & 60061217004 & 11:58:52.75 &
      +42:34:12.4 & $9.\!^{\prime}1$ & 100.4~ks & 1\arcmin \\
      {\it NuSTAR} & 2013 May 23 & 60061217006 & 11:58:52.75 &
      +42:34:12.4 & $9.\!^{\prime}1$ & 44.0~ks & 1\arcmin \\
      {\it Chandra} & 2014 Oct. 5 & 17006 & 11:58:52.20 & +42:43:20.2
      & $0.\!\!^{\prime \prime}5$ & 29.7~ks & 1\arcsec \\
      {\it NuSTAR} & 2014 Nov. 28 & 60001148002 & 11:58:20.60 &
      +42:34:13.2 & $9.\!^{\prime}1$ & 53.4~ks & 1\arcmin \\
      {\it NuSTAR} & 2014 Nov. 30 & 60001148004 & 11:58:20.60 &
      +42:34:13.2 & $9.\!^{\prime}1$ & 49.8~ks & 1\arcmin \\
      \hline
      SDSS & 2004 Apr. 25 & -- & 11:58:52.2 & +42:43:20.7 & 0 & 2220 s & 3\arcsec \\
      \hline
      {\it HST} & 1998 Jun. 17 & 7919 & 11:58:51.64 & 42:43:28.4 & $0.\!\!^{\prime \prime}177$ & 192 s & $0.\!\!^{\prime \prime}22$ \\
      {\it Spitzer} & 2005 Dec. 24 & IG\_AO2/20140 & 11:58:42.23 & +42:43:45.2 & $1.\!^{\prime}9$ & 536 s & $1.\!\!^{\prime \prime}1$\\
      \hline
    \end{tabular}
  \begin{tablenotes}
           \small
            \item $\theta_{\rm dist}$ is the angular distance between the target $\alpha_{\rm J2000}$, $\delta_{\rm J2000}$ of a particular observation and the SDSS optical center of IC~750 ($\alpha_{\rm J2000}$ = 11:58:52.2, $\delta_{\rm J2000}$ = +42:43:20.7). 
            \item The exposure time is given in units customarily used for the wavelength of the observation.
            \item $^1$This is the size of the beam, which has a position angle of $-1^\circ$. Positional errors are centroiding errors given by $\sigma_x = \theta_{min}/(2{\rm SNR})$ and $\sigma_y = \theta_{maj}/(2{\rm SNR})$. We require the signal-to-noise, SNR, to be $\geq$ 5, so the positional errors are a tenth of the size of the beam or smaller.
            \item At the distance of IC 750, 14.1 Mpc, 1\arcsec\ = 68 pc.
   \end{tablenotes}         
  \label{tab:obs}
\end{table*}

\subsection{Radio Data}
The radio data used in this work include both a new VLBI dataset taken with the VLBA and publicly available interferometric data taken with the VLA and single-dish data taken with the GBT, as listed in Table~\ref{tab:obs}.

\subsubsection{VLBA Data}
\label{sec:VLBA}
The 12-hour VLBA dataset, BZ074A, was taken on 2018 March 19. The dataset has three $\sim$1-hour blocks of scans, at the beginning, middle, and end, targeting strong calibrators distributed over a wide range of zenith angles, that are intended to be used for estimation of atmospheric delay to benefit astrometry. The remainder of the time was used to observe the target, IC 750, and calibrators, J1146+3958 and J1150+4332. J1146+3958 was used for bandpass and coarse delay calibrations and J1150+4332 was used for phase calibration to get absolute astrometry. J1150+4332 has positional errors of 0.240 mas in right ascension and 0.251 mas in declination. Most of the observing time was split into blocks of 20-minute observations of IC 750, alternating with 6-minute observations of J1146+3958, to be used for self-calibration. In addition, there are five 5-minute blocks which alternate between J1150+4332 and IC 750 every minute, concentrated in the middle of the observation when IC 750 is highest in the sky.

The maser emission was centered within two overlapping 16 MHz, dual polarization, intermediate frequency (IF) bands, covering the velocity ranges $650-865$ km s$^{-1}$ (IF2, center frequency 22178 MHz) and $510-725$ km s$^{-1}$ (IF3, center frequency 22189 MHz) with 31.25 kHz ($\sim$0.42 km s$^{-1}$) channels. The source position was set to be $\alpha$ = 11:58:52.25180 and $\delta$ = +42:43:20.24462, from a fit to the image of the continuum emission from the publicly available VLA observation in the C configuration, with a beam size of $\sim$1'', as described in Section~\ref{sec:VLA}. The dataset contains two additional 16 MHz bands, one below the maser emission and one above, with center frequencies of 22050 MHz and 22332 MHz, to attempt to detect the continuum emission. 

The data were edited, calibrated, and imaged using the Astronomical Image Processing System (AIPS, {\tt www.aips.nrao.edu}). Standard procedures were used to correct for ionospheric delays, errors in Earth Orientation Parameters (EOP), digital sampling effects, instrumental phases and delays, amplitude based on gains and system temperatures, and effects of the parallactic angle of the source on fringe phase. IC 750 data were aligned in frequency to account for the Earth's rotation and motion. The geodetic data were used to solve for the multiband delays, zenith atmospheric delays, and clock errors, and applied to the IC 750, J1146+3958, and J1150+4332 data. The bandpass solutions were determined using a scan of J1146+3958 found to have good fringes for all antennas. A global fringe fit was performed on J1146+3958 to derive coarse phase, delay, and rate solutions and applied to IC 750 and J1150+4332. Phase solutions were found for J1150+4332, which has positional uncertainties of 0.17 mas in right ascension and declination (ICRF3, Charlot et al. (2020), submitted to Astron. Astrophys). and applied to IC 750 in the phase-calibration blocks. The data from the two polarizations were averaged. The strongest maser feature, found at 780.7 km s$^{-1}$ and extending one channel on each side, was imaged in the phase-calibrated blocks and was detected with a signal-to-noise ratio (SNR) of $\sim$34. The estimated astrometric position of this maser feature was found to be $\alpha$ = 11:58:52.25184 and $\delta$ = +42:43:20.24230. The $u-v$ data for IC 750 were corrected to account for the true position of the strongest maser feature relative to the nominal position specified for the observation. The strongest maser feature was then used to solve for the phase and amplitude corrections and applied to the rest of the frequencies for the IC 750 data in IF2. The calibrations were also applied to IF3. After applying all the calibrations, IC 750 was imaged using the task IMAGR with 0.05 mas $\times$ 0.05 mas pixels, using natural weighting (ROBUST=5). The resulting image cubes have 31.25 kHz ($\sim$0.4 km s$^{-1}$) wide frequency channels, each with a 512 pixel $\times$ 512 pixel image. Both IF2 and IF3 have beams of size $\theta_{maj} \times \theta_{min} = 0.90~{\rm mas} \times 0.58~{\rm mas}$ and position angle $-1^\circ$, i.e., the major axis lies roughly North-South and the minor axis lies roughly East-West. The calibrations were also applied to the continuum IFs, IF1 and IF4. The channels in each of these IFs were averaged into a single image, using 0.05 mas $\times$ 0.05 mas pixels and natural weighting, after removing 5\% of the lowest and 10\% of the highest channels to avoid edge effects.

The root mean square (RMS) noise value in each spectral channel of the image cubes was computed from a large, line-free region using IMSTAT. The channels in IF2 have an average RMS of 4.2 mJy/beam and those in IF3 have an average RMS of 7.8 mJy/beam. The positions and the peak and integrated fluxes of the maser spots in each spectral channel are obtained by fitting two dimensional elliptical Gaussians to the images, using the task JMFIT. In the channels where there were two (blended) maser spots, double Gaussians were used to fit for both simultaneously. To avoid double counting in the frequency channels where IF2 and IF3 overlap, only the maser spots from IF2, which has lower noise, were used. For the maser spots in these channels, the fluxes from IF2 and IF3 were consistent to $\sim$10\%. The SNR of each maser spot was calculated by dividing the peak flux by the image RMS in the corresponding spectral channel. Only the maser spots detected with a SNR $\geq$ 5 were kept. The (relative) positional errors of the maser spots are centroiding errors and given by $\sigma_x = \theta_{min}/(2{\rm SNR})$ and $\sigma_y = \theta_{maj}/(2{\rm SNR})$. Due to our SNR requirement, all the positional errors are a tenth of the size of the beam or smaller, corresponding to $\lesssim$0.004 pc in right ascension and $\lesssim$0.006 pc in declination, at the distance of IC 750.

\subsection{VLA Data}
\label{sec:VLA}
In order to have a position of the water maser emission accurate enough for VLBA observations, we reduced the K band archival VLA data, from project 12A-283, taken on 2012 March 03 in the C configuration, which has a resolution of 1\arcsec, corresponding to $\sim$68 pc at the distance of IC 750. In addition to IC 750, three calibrators were observed: J1229+0203 for bandpass calibration, J1331+3030 for flux calibration, and J1146+3958 for phase calibration. The data were edited and calibrated using the Common Astronomy Software Applications (CASA, {\tt www.casa.nrao.edu}). 

The data were edited using both automated algorithms (shadow and quack) and by visual inspection. Antenna positions were corrected as needed. Antenna gain curves were generated using the opacities based on the weather for the observation. The flux model for the flux calibrator, J1331+3030, was set for K band. Delay and bandpass calibrations were preformed using strong calibrator J1229+0203. The phase solutions for each antenna were obtained for each integration using J1146+3958. Using the phase calibrations, amplitude solutions were obtained, also using J1146+3958. IC 750 was split into a separate file after applying all the calibrations. After inspection of the resultant spectrum, the line free channels were used to model the continuum and the continuum was subtracted from the data. Doppler corrections were applied to the spectral line data. The spectral line data and continuum data were imaged separately. All the maser emission was combined into one image. Both the continuum and maser emission appear as unresolved point sources. The position and fluxes of the continuum and maser emission were determined using the task IMFIT.

\subsubsection{GBT Data}
We have reduced two epochs of publicly available GBT data for IC 750, AGBT12A\_046 taken on 2012 February 28 and AGBT14B\_342 taken on 2015 April 23, using GBTIDL ({\tt http://gbtidl.nrao.edu/}). The data were taken in nodding mode with dual polarization. All the data in each observation were averaged together. A baseline was fit to the regions of the spectrum without spectral line emission, using a fifth degree polynomial, and subtracted from the data. The residual spectrum was boxcar smoothed to channels of $\sim$0.3 km s$^{-1}$, then Hanning smoothed. At 22 GHz, the GBT has a FWHM beam width of 34\arcsec, which corresponds to $\sim$2.3 kpc at the distance of IC 750.

\subsection{X-Ray Data}
IC~750 is within the field of view of several recent {\it XMM-Newton, Chandra, and NuSTAR}
observations, whose properties are listed in Table \ref{tab:obs}. Of these, {\it Chandra} has the best angular resolution, 
namely a radius of $\sim$1\arcsec\ which contains 90\% of the energy at 1.5 keV, corresponding to $\sim$68 pc at the distance of IC 750.
{\it XMM-Newton} has FWHM resolutions of $\sim$4.5\arcsec\ for the {\tt MOS} detectors and $\sim$15\arcsec\ for the {\tt pn} detector, 
corresponding to $\sim$310 pc and $\sim$1.0 kpc, respectively. {\it NuSTAR} has a half-power diameter of $\sim$1\arcmin, which corresponds to $\sim$4.1 kpc.

Both {\it XMM-Newton} \citep{xmm} observations were analyzed using
{\sc HEASoft} v6.25 -- a combined release of the FTOOLS \citep{ftools} 
and Xanadu packages -- and the {\it XMM-Newton} Science Analysis ({\sc XMMSAS}) v17.0.0 using the Current Calibration Files as
of 2019 March 7. In both {\it XMM-Newton} observations, IC~750 was
outside the active CCDs on the {\tt MOS1} detector, but emission from
the galaxy was detected in both the {\tt MOS2} the {\tt pn}
detectors.  However, too few photons were detected by
the {\tt MOS2} detector, so the spectral analysis discussed below only
used data recorded by the {\tt pn} detector.  For both observations,
the {\tt pn} data were first reprocessed using the {\tt epproc} task,
and then additionally filtered using {\tt evselect} to only include
events with ``(PATTERN$<=$12) \&\& (PI in [200:15000]) \&\&
\#XMMEA\_EP.''  Good time intervals (GTIs), free from abnormally high
or low count rates, were then created using the {\tt tabgtigen} task,
and applied to the filtered event files with {\tt evselect}.  After
such filtering, the ``good'' exposure times of the {\it XMM-Newton}
observations 0744040301 and 0744040401 were 14.2~ks and 16.2~ks,
respectively. The spectrum of IC~750 was calculated using events from
a circular region with a radius of 36$^{\prime \prime}$ centered on
this galaxy, extracted using {\tt evselect} and applying the
additional filters ``(FLAG==0) \&\& (PATTERN$<=$4)'' to obtain the
higest quality spectrum.  A similar procedure was used to generate the
background spectrum, using events in a source-free rectangular region
on the same chip as IC~750 but outside its optical extent. The response
matrix file (RMF) and ancillary response file (ARF) of both the source
and background spectra were generated using {\tt rmfgen} and {\tt
  arfgen}, respectively, with the area of the each spectrum calculated
using the {\tt backscale} task.  Finally, the source spectrum was
binned to ensure a minimum of 25 counts per channel using the FTOOL
{\tt grppha}, and then modeled using {\sc Xspec} v12.10.1
\citep{xspec}.

The {\it Chandra} observation listed in Table \ref{tab:obs} was
analyzed using version 4.11 Chandra Interactive Analysis of
Observations ({\sc ciao}) software package \citep{ciao} and version
4.8.2 of the Calibration Database (CALDB).  This data was first
reprocessed with the {\tt chandra\_repro} script, and then the spectra
and response files (RMF and ARF) of the source and background regions
were generated using the {\tt specextract} script, which grouped the
spectrum into bins with a minimum of 15~photons.  

The {\it NuSTAR} observations listed in Table \ref{tab:obs} were
first reprocessed with the {\tt nupipeline} script provided in the
{\it NuSTAR} subpackage of FTOOLS.  The images produced by the
resultant event files were then examined for emission coincident with
IC~750, but none was found.

\subsection{SDSS Optical Spectrum}
We use the publicly available Sloan Digital Sky Survey \citep[SDSS]{SDSS} spectrum for IC 750, taken on 2004 April 25, with an exposure time of 2220 s and the 3\arcsec\ fiber ($\sim$205 pc) centered on $\alpha$ = 11:58:52.20 and $\delta$ = +42:43:20.71. SDSS spectra have a wavelength range of 3800-9200 \AA\ and a spectral resolution of $\frac{\lambda}{\Delta\lambda}\sim$1800-2000. In addition the spectra have absolute flux calibration and have been corrected for telluric contamination, if present, before public release. 

\subsection{Infrared Data}
\subsubsection{{\it HST} Infrared Image}
The optical images of IC~750 are strongly affected by dust absorption features. To mitigate the effects of the dust obscuration we use the publicly available {\it HST} images of the IC~750 obtained with the NICMOS3 camera and the F160W filter (1.6$\mu$m) which roughly match the near-infrared H-band. In addition, {\it HST} has a better angular resolution, with the FWHM a point spread function (PSF) of $0.\!\!^{\prime \prime}22$, corresponding to $\sim$15 pc at the distance of IC 750. The data were taken on 1998 June 17 (Proposal ID 7919) with an exposure time of 192 s.

Comparing the pipeline processed F160W image extracted from the MAST archive with SDSS optical images, we found errors in the World Coordinate System (WCS) information. To fix the image WCS, we found 6 sources identified in GAIA DR2 within the field of view of the IC 750 observation. Two of these sources have obvious point-like counterparts in the NICMOS image, another two could be identified with smoothed knots on the spirals. Using these four sources, we estimated a 1.3\arcsec\ linear offset correction (without rotation) roughly towards the north.

\subsubsection{{\it Spitzer} Infrared Image}
In order to better understand the fainter, outer parts of the galaxy, we use the publicly available infrared image of IC 750 from the {\it Spitzer} Survey of Stellar Structure in Galaxies (S$^4$G) \citep{Sheth10, Munoz13, Querejeta15}. {\it Spitzer} has a PSF with a Gaussian core of $2.\!\!^{\prime \prime}1$, corresponding to $\sim$145 pc at the distance of IC 750. We downloaded a fully reduced image of IC 750 in 3.6~$\mu$m obtained with the Infrared Array Camera \citep[IRAC;][]{Fazio04} of the {\it Spitzer} telescope \citep{Spitzer}, from the NASA/IPCA Infrared Science Archive website ({\tt http://irsa.ipac.caltech.edu/}). The data were taken on 2005 December 24 with an exposure time of 536 s.

\section{Results}
\label{sec:results}
In this section, we report the measurements of the BH mass, sub-parsec scale circumnuclear disk structure, and AGN luminosity, as well as galaxy bulge mass, stellar mass, and stellar velocity dispersion, using the multi-wavelength data described above. 

\subsection{Distance to IC 750}
\label{sec:distance}
A key parameter that underpins our results is the distance to IC 750. The BH mass and disk size are linearly proportional to distance, whereas the luminosity, bulge mass, and stellar mass are proportional to the distance squared. There are three different distance measures for IC 750, namely Tully-Fisher measurements, the distance based on recessional velocity, and the distance based on the velocity measurements of the group IC 750 belongs to. IC 750 is interacting with IC 749, 3$\farcm{41}$ away, and even the largest of the radio telescopes used in the Tully-Fisher measurements does not have the angular resolution to separate the two \citep[e.g.,][]{Theureau07}. As a result, the Tully-Fisher measurements for IC 750 are biased due to confusion with its neighbor, and we do not use them. The systemic velocity measurement of 700.6 $\pm$ 0.9 km s$^{-1}$ \citep{Verheijen01}, however, is from HI measurements made with the Westerbork telescope which have a beam of 60\arcsec $\times$ 60\arcsec\ for the lowest smoothed angular resolution, and can separate IC 750 from IC 749. The redshift-derived distance in the reference frame of the 3 K cosmic microwave background, with cosmological parameters of $H_0$ = 67.8 km s$^{-1}$, $\Omega_{\rm matter}$ = 0.308, and $\Omega_{\rm vacuum}$ = 0.692, given in NED, is 13.8 $\pm$ 1.0 Mpc. We also determine the distance from the dynamics of the NGC 4111 group, to which IC 750 belongs. We take the average velocity of the group, 880 km s$^{-1}$ in the Local Group reference frame \citep{Makarov11} and correct it to the Galactic Standard of Rest frame. We then input this velocity and the sky position of IC 750 into the Cosmicflow3 Distance-velocity Calculator \citep[{\tt http://edd.ifa.hawaii.edu/NAMcalculator}]{Shaya17}, which yields a distance of 14.1 Mpc. This distance, which accounts for any peculiar velocity IC 750 may have, agrees with the redshift derived distance. For the uncertainty in this measurement, we search for measurements of the size of the group reported in literature. The sizes reported differ depending on the method and which galaxies were included, e.g. $R=1.1$ Mpc \citep{Huchra82}, $R=0.3$ Mpc \citep{Tully87}, $R=1.0$ Mpc \citep{Crook07}, and $R=0.5$ Mpc \citep{Makarov11}. We take the largest, 1.1 Mpc, as the distance error and adopt $D=(14.1\pm1.1)$ Mpc as the distance to IC 750.

\subsection{Maser Disk}
\label{sec:maserdisk}

\begin{figure}[tbh]
\begin{center}
\includegraphics[width=6.5in]{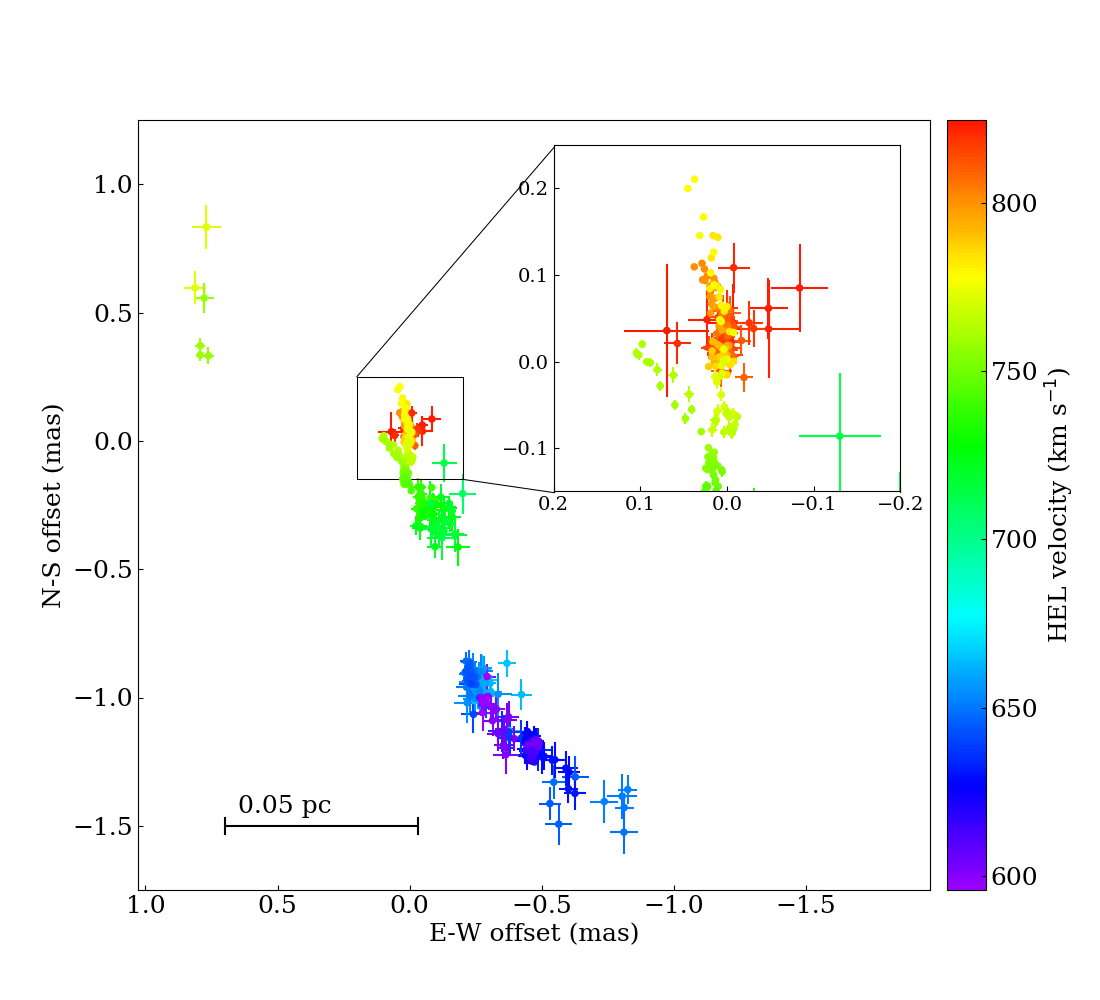}
\end{center}
\caption{\footnotesize \em{Map of the maser emission in IC 750. The colors indicate the line-of-sight heliocentric velocities of the maser spots. The maser spots trace a warped, nearly edge-on disk $\sim$0.2 pc in diameter. The inset shows a zoom in of the redshifted maser spots, in the velocity range $\sim$725-825 km s$^{-1}$.}}
\label{fig:mapzoom}
\end{figure}

The map of the maser emission in IC 750, from the VLBA data, is shown in Figure~\ref{fig:mapzoom}. The $x$-axis and $y$-axis show the East-West offset and North-South offset, respectively, from the maser feature used for self-calibration, located at ($\alpha$ = 11:58:52.25184, $\delta$ = +42:43:20.24230). The colors indicate the line-of-sight heliocentric velocity of the maser spots. The maser emission lies in a roughly linear structure, extending $\sim$0.2 pc from northeast to southwest, and spans a line-of-sight velocity range of 596 km s$^{-1}$ to 825 km s$^{-1}$. The northeast set of spots are redshifted relative to the systemic velocity of 700.6 $\pm$ 0.9 km s$^{-1}$ \citep{Verheijen01}. Those in the central parts are around the systemic velocity and those the southwest are blue-shifted relative to the systemic velocity. This is evidence for a nearly edge-on circumnuclear disk rotating around the BH. The positional spread of the blue-shifted maser emission transverse to the radial direction is consistent with positional errors indicating that the disk is geometrically thin. The outer parts of the disk have a position angle of $\sim 51^\circ$ while the inner part has a position angle of $\sim 14^\circ$. This indicates that the disk has a positional angle warp. Warped disks are seen in a number of maser systems including NGC 4258 \citep[e.g.,][]{Herrnstein99} and Circinus \citep{Greenhill03}.

The inset shows a zoom in of the redshifted maser spots, in the velocity range $\sim$725-825 km s$^{-1}$. They lie in two distinct lines that diverge, with the lower velocity (green) spots veering towards the northeast and the higher velocities spots (yellow, orange, and red) extending almost directly northward. A corresponding fork is not seen in the blue-shifted emission. As described in Section~\ref{sec:xrays}, the X-ray emission shows a $\sim$kpc-scale extended emission. Therefore, there is likely also a subparsec scale disk wind. The masers in IC 750 likely originate in both the disk and the wind. Maser emission from both a disk and a wind or jet has been seen in other maser systems such as Circinus \citep{Greenhill03}, NGC 1068 \citep{Gallimore04}, and NGC 3079 \citep{Kondratko05}.

\begin{figure}[tbh]
\begin{center}
\includegraphics[width=5.0in]{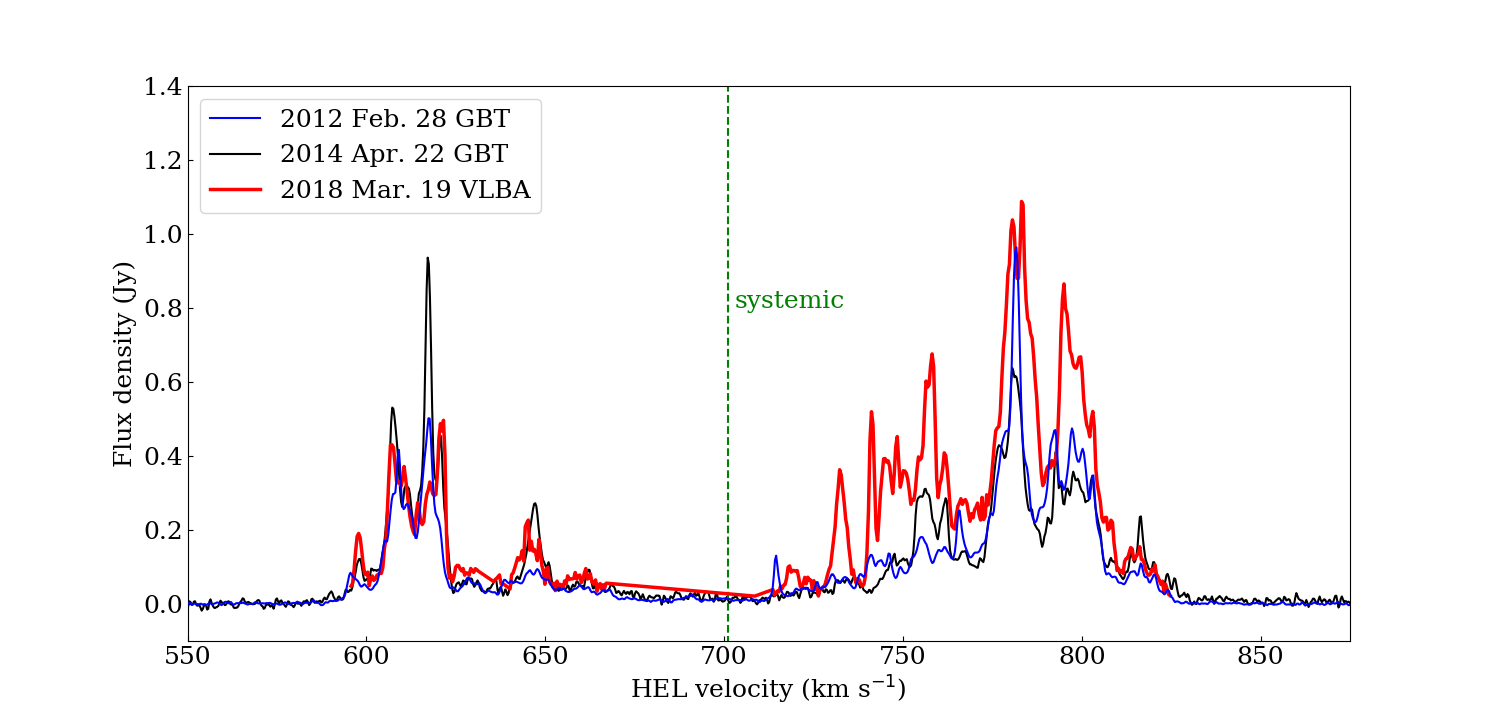}
\end{center}
\caption{\footnotesize \em{Spectrum of the maser emission in IC 750. The maximum flux and overall extent in velocity are consistent between the VLBA and GBT spectra. There is no maser feature at the systemic velocity.}}
\label{fig:radiospectra}
\end{figure}

Figure~\ref{fig:radiospectra} shows the spectra of the maser emission in IC 750. The single dish spectra taken with the GBT in 2012 and 2015 are overlaid with the spectrum constructed from the VLBA dataset by adding the flux in all the significant (SNR $\geq$ 5) maser emission features in each spectral channel. The flux calibrations for the VLBA and GBT should both be accurate to order 10\%. We see variability in the emission from different epochs, as is expected for maser emission, but the maximum flux and overall extent in velocity are consistent between the VLBA and GBT spectra. There is no maser emission detected between 667 km s$^{-1}$ and 708 km s$^{-1}$, around IC 750's systemic velocity of 700.6 $\pm$ 0.9 km s$^{-1}$ \citep{Verheijen01}, corresponding to the gap in the map at the center of the disk. 

\subsubsection{Radio Continuum Emission}
\label{sec:radiocontinuum}
A continuum source at 22 GHz was detected in the VLA archival data taken in the C configuration. The source is unresolved in the image which has a beam size of $\sim$1\arcsec. It has a peak flux of $4.39\pm0.69$ mJy beam$^{-1}$ and an integrated flux of $4.34\pm1.1$ mJy, confirming that it is compact at this scale. We attempted to image the continuum in the VLBA data in order to help pin down the location of the central BH, either from the location of a compact nuclear source or by intersecting the direction of an extended jet with the maser disk. The continuum was not detected despite sufficient image sensitivity (RMS $\sim$ 0.2 mJy beam$^{-1}$). The major axis of the VLBA beam is $\sim$1 mas, suggesting that the continuum emission subtends at least a few mas.

\subsection{Mass of the Nuclear Black Hole}
\label{sec:BHmass}

In order to determine the mass of the BH, we fit Keplerian rotation curves to the high velocity maser emission, allowing for the uncertainty in the BH position and velocity. In addition, we find the upper limit of the BH mass by fitting a Keplerian envelope that lies above all the maser emission in a position-velocity (PV) diagram.

\subsubsection{BH Mass Fits}
\label{sec:massfits}
As described in Section~\ref{sec:maserdisk}, some of the maser emission in IC 750 traces a nearly edge-on geometrically thin disk. In order to fit for the BH mass, we assume that the disk in IC 750 is edge-on. Geometrically thin maser disks have measured inclination angles within $\lesssim~5^\circ$ \citep[e.g.][]{Humphreys13, Kuo11} of edge-on. Deviation of the line-of-sight velocity from rotational velocity for the high velocity masers on the midline will, therefore, be $\lesssim$0.4\%. In systems where the systemic and high velocity masers are detected and clearly separated, such as in NGC 4258 \citep[e.g.][]{Herrnstein99, Humphreys13, Kuo11}, the BH position and velocity are roughly that of the systemic masers and the high velocity masers are used in mass fit. In IC 750, there is no maser feature at the systemic velocity, corresponding to a gap in both the map and the spectrum. Consequently, we do not know the exact position and velocity of the BH. Furthermore, due to the complex structure of the emission, with different velocity maser spots at the same projected locations, it is also necessary to disentangle the high velocity disk emission from the rest of the maser emission.

\begin{figure}[htb]
\begin{center}
\includegraphics[width=3.5in]{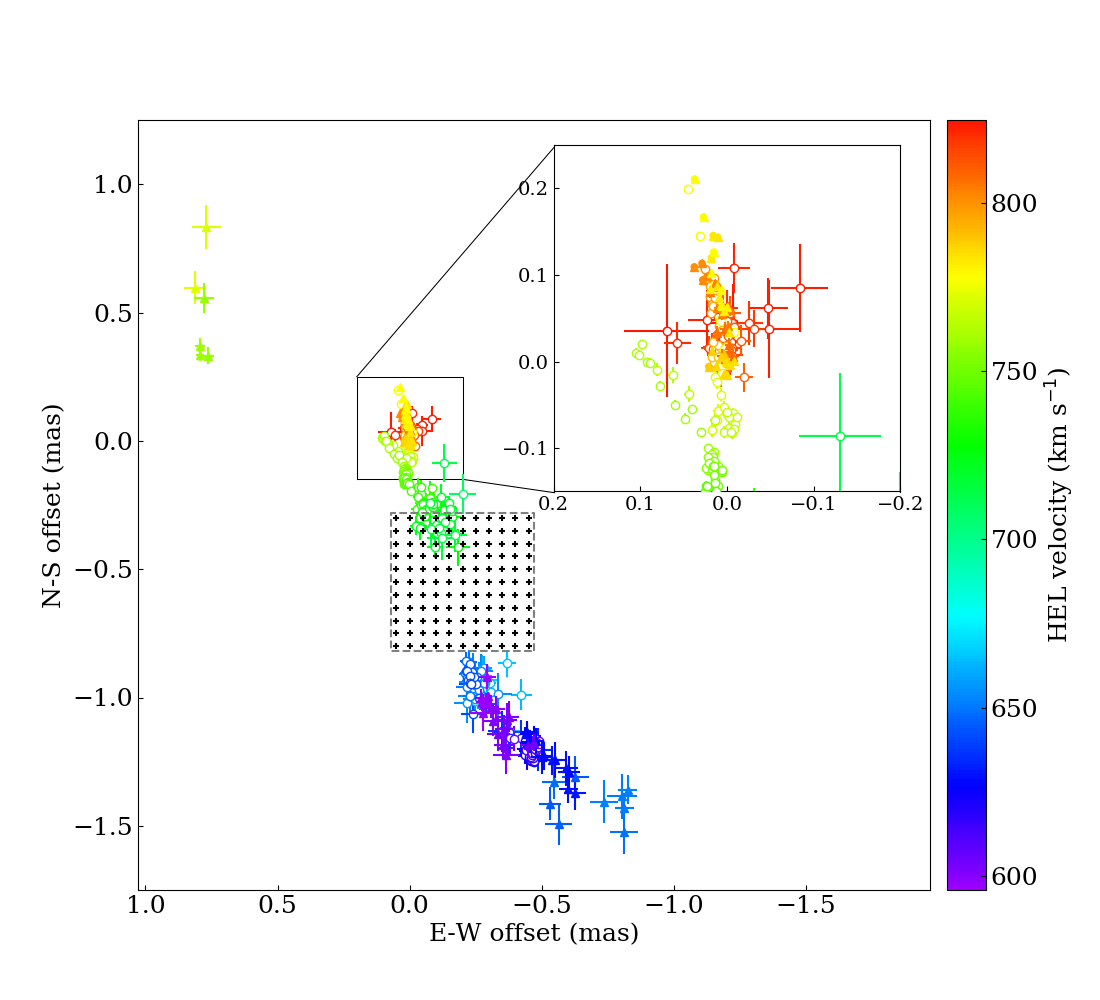}
\includegraphics[width=3.5in]{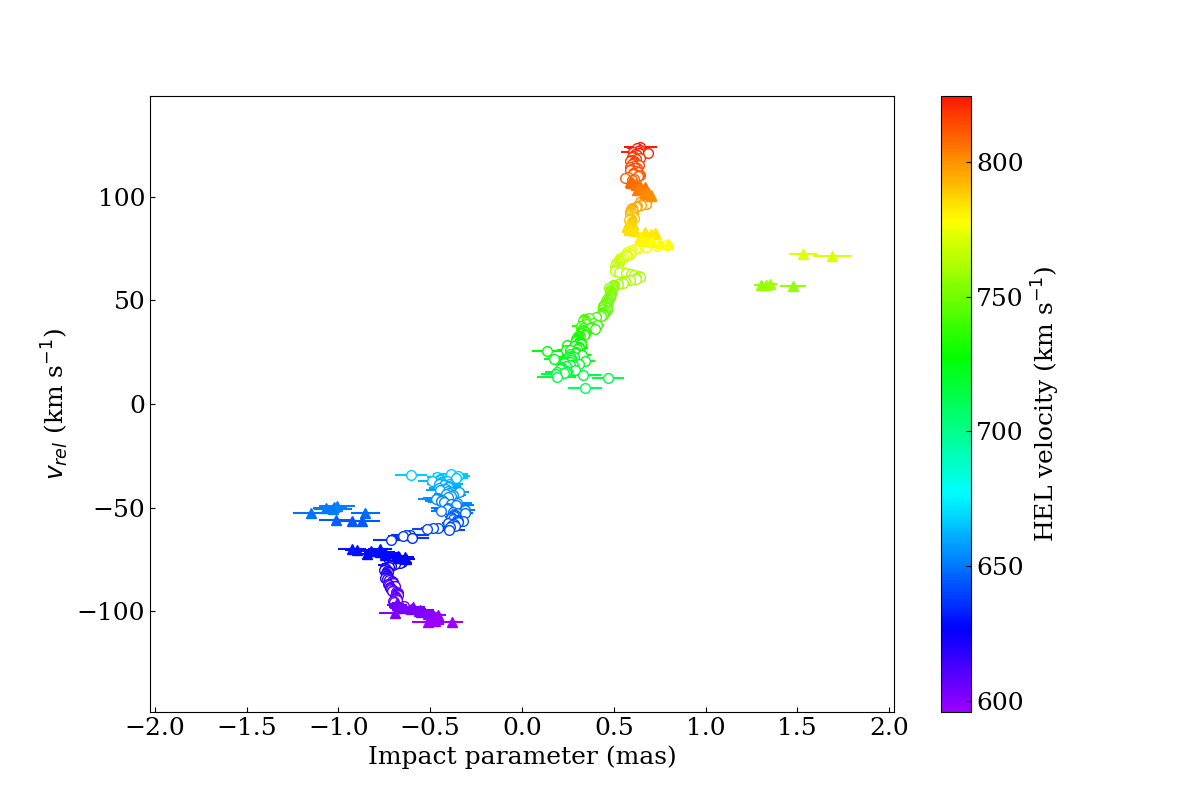}
\end{center}
\caption{\footnotesize \em{In both plots, the open circles indicate maser spots not used in the fits for BH mass and the solid triangles indicate maser spots used in the fits, based on their position and velocity structure. The colors of the solid triangles and open circles indicate the line-of-sight velocities of the maser spots. Left: Map of the maser emission. The inset shows a zoom in of the redshifted maser spots, in the velocity range $\sim$725-825 km s$^{-1}$. The box with the dashed grey border shows the extent of the allowed BH positions in the mass fits. The black points mark the grid of putative BH positions used in our BH mass fits. Right: An example PV diagram assuming that the BH is at center position of the grid and has a velocity of 701 km s$^{-1}$. The impact parameter is the (great-circle) angular distance from the assumed BH location and the relative velocity, $v_{rel}$, is the difference between the assumed BH velocity and maser line-of-sight velocity.}} 
\label{fig:kepreg}
\end{figure}

{\it Possible BH locations in position and velocity:} The BH should be centered within the disk that orbits it. Therefore, we assume that it is in the gap between the redshifted and blue-shifted maser emission, indicated by the $0.5~{\rm mas} \times 0.5~{\rm mas}$ dashed box in the map shown in the left panel of Figure~\ref{fig:kepreg}. We set up a grid of possible BH positions in this region in steps of 0.05 mas, the pixel size in our images. The highest velocity red and blue emission are roughly symmetric about the systemic velocity, 700.6 $\pm$ 0.9 km s$^{-1}$ \citep{Verheijen01}, so we also assume that the BH velocity is close to the systemic velocity. Specifically, we assume 697 km s$^{-1}$ $\leq~v_{\rm BH}~\leq$ 705 km s$^{-1}$ and vary the velocity in steps of 1 km s$^{-1}$.

{\it Maser spots used in the fits:} As discussed in Section~\ref{sec:maserdisk}, the maser emission is complex and likely originates in both a wind and a disk. The PV diagram of the maser emission is shown in the right panel of Figure~\ref{fig:kepreg}. The velocity of each maser spot relative to the putative BH velocity, $v_{sys}$ = 701 km s$^{-1}$, is plotted versus the impact parameter, the angular distance from the maser spot to the center of the grid of possible BH positions. It is not clear from the map and PV diagram which maser spots are in the high velocity regions of the disk. We cannot determine if individual spots are part of the high velocity region of the disk but we can analyze the position-velocity structure of series of maser spots. This is possible because, as described in Section~\ref{sec:VLBA}, the maser data is composed of image cubes, i.e. a separate image is made for each frequency channel in steps of 31.25 kHz, roughly corresponding to 0.4 km s$^{-1}$. While there are hundreds of maser spots in total, shown together on the map and PV diagram, each frequency channel (which has maser emission) has only one, or in a few cases two, maser spots. We examine the maser spots in consecutive frequency/velocity channels.

In the high velocity regions of the disk, the relative velocity of the maser spots should decrease with distance from the BH. Series of maser spots where the relative velocity increases with distance, or show no correlations with distance from the BH, are inconsistent with Keplerian rotation in the high velocity regions of the disk\footnote{Increasing velocity with distance is consistent with systemic emission where masers are emitted in a wedge. Those along the direct line of sight will have zero relative velocity and those off the direct line of sight will have line of sight velocity which increases with distance \citep[e.g.][]{Humphreys13, Gallimore04}. Maser spots which do not show any relation between distance and velocity are consistent with being in a wind such as in Circinus \citep{Greenhill03}.}. Therefore, we exclude them from the fits. We retain only the series of maser spots which have decreasing relative velocity with distance\footnote{We have used the central position of the grid and 701 km s$^{-1}$ for determining the position and velocity structure of the series of maser spots. We note that using different points in the grid shown in the left panel of Figure~\ref{fig:kepreg} and other velocities in the range of putative BH velocities will shift the relative positions and velocities of the maser spots. However, the overall structure of each series of maser spots, i.e. whether the relative positions increase, decrease, or show no relation, with increasing relative velocity, remain the same for different putative BH positions and velocities and we include or exclude the series rather than individual maser spots.}. We do not further require agreement with Keplerian rotation, $v \propto r^{-0.5}$, for these series of maser spots. This is because masers can be emitted in a wedge \citep[e.g.,][]{Humphreys13, Reid13} and those off the midline of the disk will have line-of-sight velocities less than rotational velocity. There can also be fragmentation within the disk, or additional velocity contribution from a wind which is likely to be present based on the extended emission seen in X-rays, described in Section~\ref{sec:xrays}. We cannot further isolate the masers which have contributions only from the gravity of the BH, so we keep all the series of maser spots which have decreasing velocity with distance, i.e. the ones for which Keplerian rotation cannot be ruled out as the dominant contribution. Other maser systems where only a fraction of the maser spots are in Keplerian rotation about the BH include Circinus \citep{Greenhill03} and NGC 1068 \citep{Greenhill96, Greenhill97}. The maser spots excluded from the fits are shown as open circles and those used in the fits are shown as solid triangles in Figure~\ref{fig:kepreg}. The map of each individual series of maser spots is shown in Appendix~\ref{ap:regions} to better illustrate its position-velocity structure.

{\it BH Mass Fitting:} For each combination of putative BH position and velocity, we perform a least square fit of the positions and velocities of the maser spots to a Keplerian curve, $r_{rel}(v_{rel}) = GM_{\rm BH}/v_{rel}^2$, where $r_{rel}$ is the position of the maser spot relative to the assumed BH position and $v_{rel}$ the velocity relative to the assumed BH velocity. $M_{\rm BH}$ is the free parameter in the fits. We designate the position as the dependent variable because the uncertainties in the positions are larger than the uncertainties in the velocities. 

We first fit for the mass from the redshifted maser emission ($M_{red}$) and from the blue-shifted maser emission ($M_{blue}$) separately. For the true position and velocity of the BH, the two masses should be equal, i.e. $M_{red}/M_{blue} = 1$, if the mass of the disk is small compared to the mass of the BH. However, since we are varying the putative position and velocity, presumably around the true BH position and velocity, we allow the mass ratio to be between 0.95 and 1.05. Furthermore, we allow for the possibility of a disk that could be massive relative to the BH mass. For maser systems with geometrically thin disks, the highest reported disk masses relative to BH masses are in NGC 3393 \citep{Kondratko08} and NGC 1068 \citep{Hure02, Lodato03}, where the disks are roughly as massive as the BH. The redshifted maser emission extends out to a radius of $\sim$1.7 mas while the blue-shifted emission only extends to a radius of $\sim$1.2 mas. Consequently, if the disk is massive, the redshifted maser emission will enclose more mass than the blue-shifted maser emission. Therefore, we allow $M_{red}$ to be up to 25\% larger than $M_{blue}$. The detailed reasoning for possible differences in $M_{red}$ and $M_{blue}$ is given in Appendix~\ref{ap:diskmass}. {\it We accept the fit and consider a BH position and velocity combination as viable if $0.95 \leq M_{red}/M_{blue} \leq 1.25$.} For each acceptable combination of BH position and velocity, we refit for a single value for the BH mass using the redshifted and blue-shifted maser spots together. The resultant mass is the sum of the BH mass and the average (weighted by the positional uncertainties) disk mass enclosed by the maser emission\footnote{The uncertainties in our data do not allow us to accurately fit separate BH and disk masses.}. The average (statistical) error on the mass in each individual fit is $\sim$1.5\%.

\begin{figure}[htb]
\begin{center}
\includegraphics[width=3.5in]{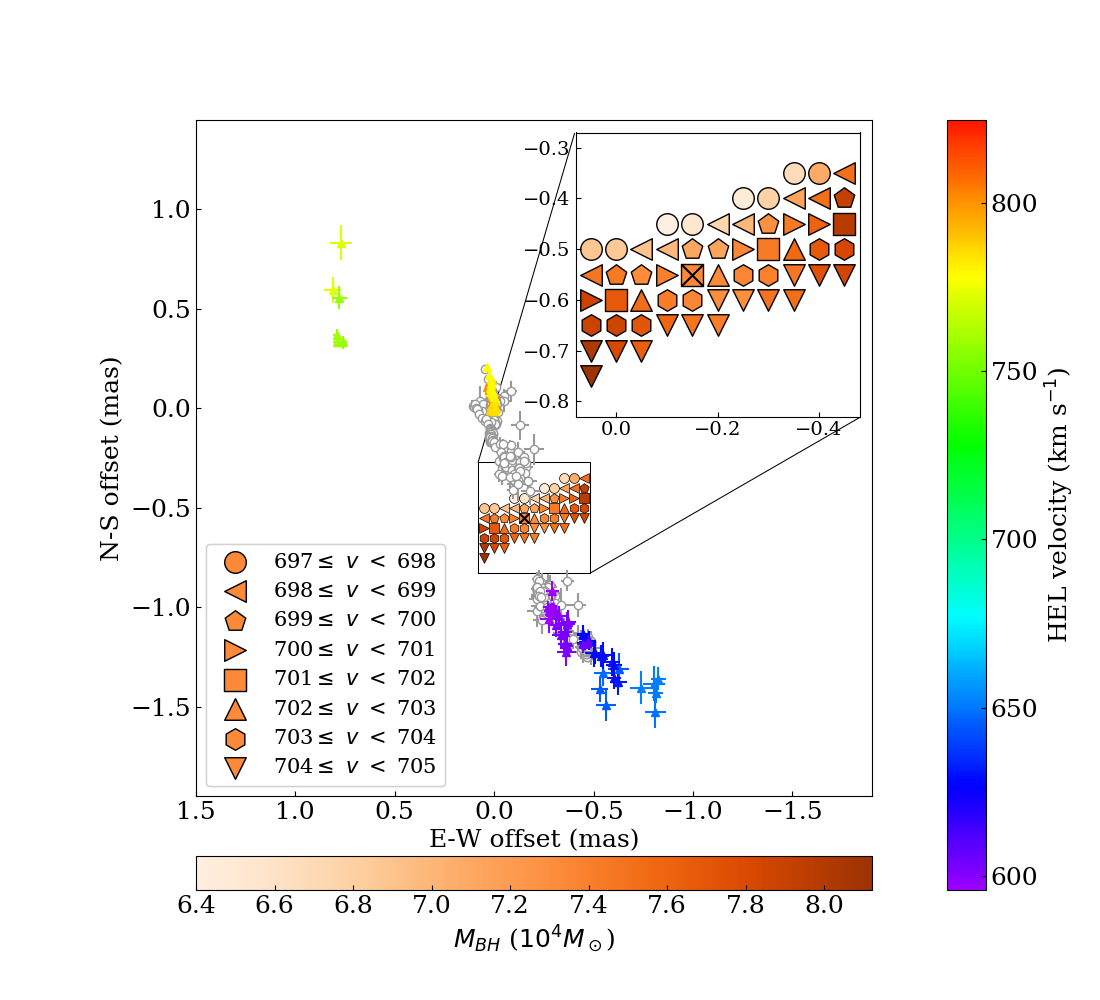}
\includegraphics[width=3.5in]{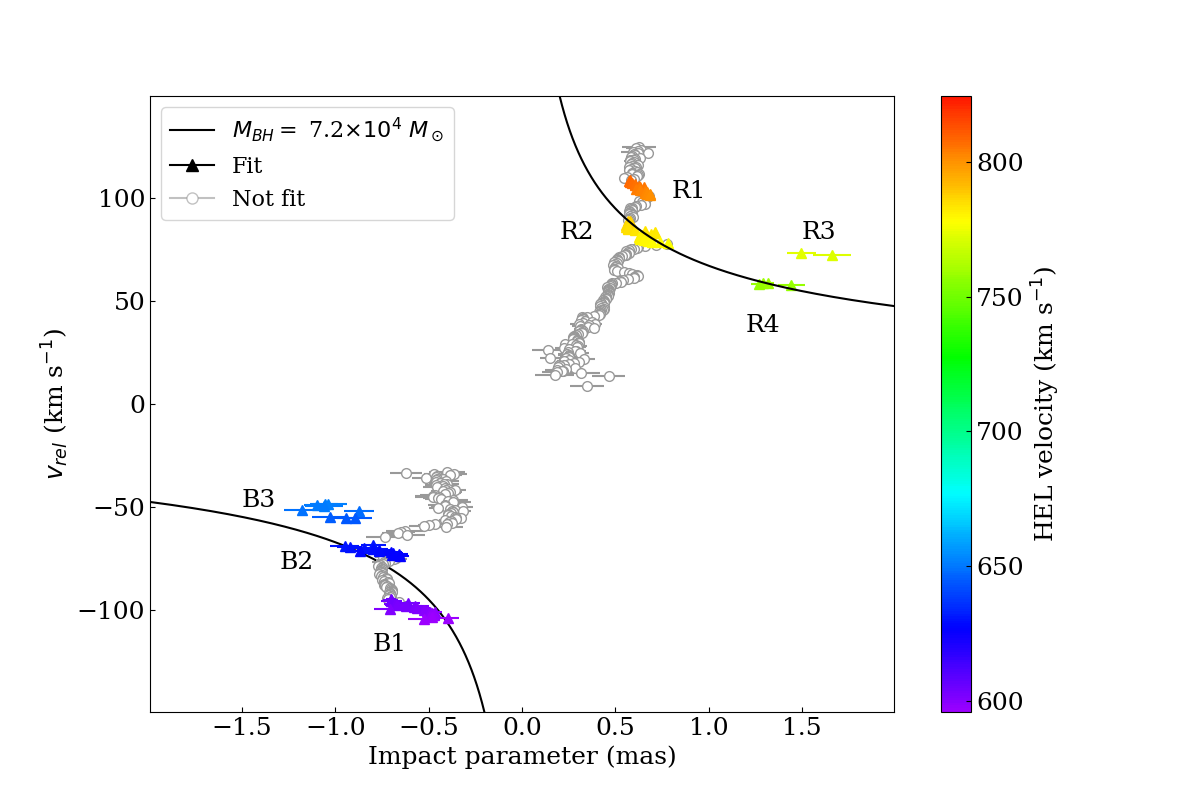}
\end{center}
\caption{\footnotesize \em{Map and PV diagram for the Keplerian fits. In both plots, the grey circles indicate maser spots not used in the fit and the solid triangles indicate maser spots used in the fit. The colors of the solid triangles indicate the line-of-sight velocities of the maser spots. Left: The map of the viable positions and velocities for the central black hole. The 0.5 mas $\times$ 0.5 mas box outlined in black shows the region over which we vary the putative position of the BH, in steps of 0.05 mas. We also vary the putative velocities between 697 km s$^{-1}$ and 705 km s$^{-1}$, in steps of 1 km s$^{-1}$. For each putative BH position and velocity, the redshifted and blue-shifted emission are fit separately to give $M_{red}$ and $M_{blue}$, respectively. A fit is considered to be acceptable if $0.95 \leq M_{red}/M_{blue} \leq 1.25$ and the corresponding combination of putative BH position and velocity is considered to be viable. Then a single mass for the position and velocity combination is (re)fit using both the redshifted and blue-shifted emission. The symbols indicate the viable velocity at the given location. Each position has one to five viable velocities. For the positions with more than one viable velocity, we average the velocities and the resulting masses. The orange color bar indicates the averaged viable mass at each position. Right: The PV diagram for the BH location at an East-West offset of -0.15 mas and a North-South offset of -0.55 mas, marked with a black cross on the map, and a velocity of 700 km s$^{-1}$. The impact parameter is the (great-circle) angular distance from the assumed BH location and the relative velocity, $v_{rel}$, is the difference between the assumed BH velocity and maser line-of-sight velocity. The line is the Keplerian fit of the maser spots. This position and velocity gives a fitted mass of $(7.2 \pm 0.1) \times 10^4~M_\odot$.}}
\vspace{-0.3cm}
\label{fig:bestfitmass}
\end{figure}

\begin{figure}[htb]
\begin{center}
\includegraphics[width=3.5in]{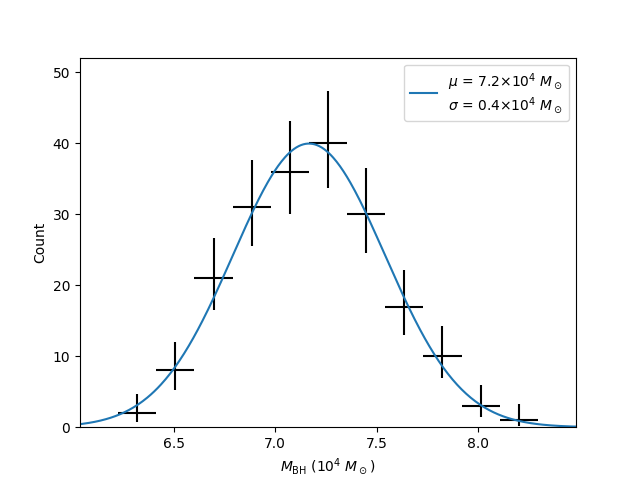}
\includegraphics[width=3.5in]{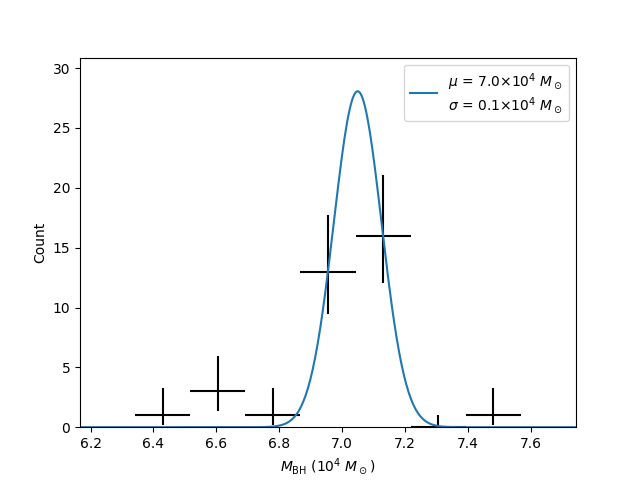}
\end{center}
\caption{\footnotesize \em{Left: Distribution of masses from the fits when varying the putative position and velocity of the BH using all the series of maser spots which have decreasing relative velocity with distance from the BH. Right: Distribution of masses from the fits when different series of maser spots are used in the fit while keeping the center position and velocity fixed at (-0.15 mas, -0.55 mas; 700 km s$^{-1}$).}}
\vspace{-0.3cm}
\label{fig:masshists}
\end{figure}

The viable positions and velocities of the BH are shown in the left panel of Figure~\ref{fig:bestfitmass}, along with the masses for the combinations. The viable positions extend roughly perpendicular to the central region of the maser disk. They are closer to the redshifted emission than the blue-shifted emission, consistent with expectation since the nearest redshifted emission is closer in velocity to the systemic velocity of the galaxy than the nearest blue-shifted emission. Each position has one to five viable velocities, with the more central positions allowing more possible velocities. For the positions with more than one viable velocity, the color of the marker for the position indicates the average of the masses from the viable velocities. In the direction perpendicular to the inner disk, masses are lower towards the center than at the ends. This is as expected because BH positions at the ends are more distant from all the emission than those closer to the center. In the direction extending along the inner disk, the masses are lower when the positions are closer to the redshifted emission than when they are closer to the blue-shifted emission. The viable velocities decrease as the center positions move towards the redshifted emission and increase as they move towards the blue-shifted emission. If there was a maser at the same (projected) position as the black hole, the two should have roughly the same velocity. Therefore, positions closer to the redshifted emission should have higher velocities and those towards the blue should have lower velocities. The fact that the fits show the opposite trend suggests that the most central position and velocity are the most physically well motivated position and velocity for the BH. The right panel of Figure~\ref{fig:bestfitmass} shows the PV diagram for a central position at an East-West offset of -0.15 mas and a North-South offset of -0.55 mas, marked with black cross on the map, and a central velocity of 700 km s$^{-1}$, which gives a mass of $(7.2 \pm 0.1) \times 10^4~M_\odot$. For this position, viable velocities range between 699 km s$^{-1}$ and 703 km s$^{-1}$ and the average mass is $7.0 \times 10^4~M_\odot$. The distribution of the masses from all the acceptable fits, for all the viable position and velocity combinations, is shown in the left panel of Figure~\ref{fig:masshists}. Each acceptable fit is included separately, i.e., not averaged for a given position.

There are three series of blue-shfited maser spots and four series of redshifted maser spots which have decreasing velocity with increasing distance, labeled as B1 through B3 and R1 through R4 in the right panel of Figure~\ref{fig:bestfitmass}. The fitted line, shown in the right panel of Figure~\ref{fig:bestfitmass} does not go through all the series used in the fit. More specifically, the fitted Keplerian curve misses R1 and R3, which fit to a higher mass, and B3, which fits to a lower mass. As discussed when describing the procedure for determining which maser spots are included in the mass fits, there can be additional contributions which can make the position-velocity structure of maser spots deviate from Keplerian rotation about the BH; these include masers that are not on the midline of the disk, which will decrease the rotational velocity, disk mass, which will increase the estimate for the enclosed mass but flatten the rotation curve, winds, which can increase or decrease the velocity of the maser spots, and additional structure in the disk such as warps in inclination angle and eccentricity. Inclination angle warps have been seen in other systems, such as NGC 4258 \citep{Humphreys13} but are rare. Eccentricity has not yet been observed in masers disks. A model can be constructed which includes all the disk parameters such as positional angle and inclination angle warps, and eccentricity \citep[e.g.,][]{Humphreys13, Reid13}. However, to do so requires repeated mappings of the maser emission to better sample the disk and to differentiate the emission originating from a disk and wind, e.g. from differences in variability of masers in a disk vs. wind. 

Since full model fitting is not possible with only one epoch of maser data, we assume that at least one of the series of redshifted maser spots and one of the series of blue-shifted maser spots have dynamics dominated by the gravity of the BH, and take different combinations of the different series of maser spots. At least one series of blue-shifted maser spots and one series of redshifted maser spots are required to be in each fit, in order to apply the $M_{red}-M_{blue}$ agreement. There are seven different combinations of the blue-shifted series of maser spots (B1; B2; B3; B1+B2; B1+B3; B2+B3; B1+B2+B3) and fifteen different combinations of the redshifted series of maser spots (R1; R2; R3; R4; R1+R2; R1+R3; R1+R4; R2+R3; R2+R4; R3+R4; R1+R2+R3; R1+R2+R4; R1+R3+R4; R2+R3+R4; R1+R2+R3+R4), leading to a total of 105 different combinations of redshifted and blue-shifted series of maser spots. We repeat the mass fits for each of these possible combinations, first by fixing the BH position and velocity at (-0.15 mas, -0.55 mas; 700 km s$^{-1}$), the most central position and velocity determined from varying the BH putative position and velocity, and then allowing the position and velocity to vary over the full range, which has the effect of realigning the series of maser spots relative to each other on the PV diagram. 

{\it BH Mass:} The procedure of varying the putative position and velocity constitutes repeated measurements of the BH mass sampling the uncertainties in the BH position and velocity. If the true BH position and velocity are within the tested range of BH positions and velocities, and the errors are random, we expect the distribution of masses from the fits to be roughly Gaussian. As seen in the left panel of Figure~\ref{fig:masshists}, the distribution of masses is well fit by a Gaussian, with a mean of $\mu = 7.2 \times 10^4~M_\odot$ and a standard deviation of $\sigma = 0.4 \times 10^4~M_\odot$. We can take the mean as the measured mass and the standard deviation as the error due to the uncertainty on the BH position and velocity. 

The distribution of masses from the acceptable fits when varying the series of maser spots used in fits, while keeping the BH position and velocity fixed at (-0.15 mas, -0.55 mas; 700 km s$^{-1}$) is shown in the right panel of Figure~\ref{fig:masshists}. For this position 35 possible combinations of redshifted and blue-shifted series of maser spots meet the $M_{red}-M_{blue}$ mass agreement requirement. For example, B3+R1+R3 does not yield an acceptable fit. The distribution has a central Gaussian with a mean of $\mu = 7.0 \times 10^4~M_\odot$ and a standard deviation of $\sigma = 0.1 \times 10^4~M_\odot$. However, the distribution also shows higher and lower peaks from combinations that favor higher and lower masses. Consequently, the width of the Gaussian does not fully capture the error due to the variations of the series of maser spots used in the fits. Additionally, this error cannot be added in quadrature with the error from the uncertainty in the position and velocity of the BH.

\begin{figure}[htb]
\begin{center}
\includegraphics[width=3.5in]{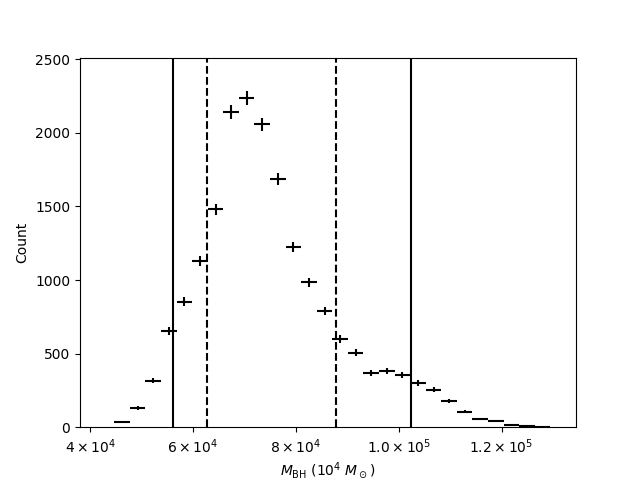}
\end{center}
\caption{\footnotesize \em{Distribution of masses from fits when varying the position and velocity of the BH vary while also varying the series of maser spots used in the fit. The distribution still peaks at $7.2 \times 10^4~M_\odot$ but it is not Gaussian. The two dashed and solid vertical lines indicate the ranges within which 68\% and 90\% of the masses fall, respectively. The 68\% range extends between $(6.3-8.8)~\times~10^4~M_\odot$, the 90\% range extends between $(5.6-10.2)~\times~10^4~M_\odot$, and all the fitted masses are between $4.5 \times 10^4~M_\odot$ and $1.3 \times 10^5~M_\odot$.}}
\vspace{-0.3cm}
\label{fig:varyeverything25}
\end{figure}

To account for uncertainties in the BH position and velocity as well as which series of maser spots have dynamics dominated by the gravity of the BH, we vary the position and velocity of the BH while also varying the series of maser spots used in the fit, i.e. we take each possible combination of BH position, BH velocity, and combination of redshifted and blue-shifted series of maser spots and repeat the mass fits, requiring that $0.95 \leq M_{red}/M_{blue} \leq 1.25$. All the possible combinations of series of maser spots, except for B3+R3, give acceptable fits for at least five BH position and velocity combinations, with the most viable combinations of series of maser spots allowing for 302 BH position and velocity combinations. This indicates that the assumption that Keplerian rotation dominates for at least some of the series of maser spots is valid. Since additional velocity contributions can raise or lower the maser velocities, and consequently raise or lower the fitted mass, they will add noise to the mass fits and increase the range of the fitted masses but they are unlikely to cause an overall systematic underestimation or overestimation. Therefore, the true enclosed mass should fall within the range of fitted masses. The distribution of masses from the acceptable fits is shown in Figure~\ref{fig:varyeverything25}. The distribution still peaks at $7.2 \times 10^4~M_\odot$ but it is not Gaussian, as expected since we are effectively convolving the Gaussian distribution, from varying the BH position and velocity, with a non-Gaussian distribution, from varying the series of maser spots included in the fit. Therefore, we do not quote a $1\sigma$ error for the mass. The two dashed and solid vertical lines indicate the ranges within which 68\% and 90\% of the masses fall, respectively. The 68\% range extends between $(6.3-8.8)~\times~10^4~M_\odot$, the 90\% range extends between $(5.6-10.2)~\times~10^4~M_\odot$, and all the fitted masses are between $4.5 \times 10^4~M_\odot$ and $1.3 \times 10^5~M_\odot$. This includes the uncertainties from the least square fitting, $\sim0.1 \times 10^4~M_\odot$, the uncertainty of the BH position and velocity, and uncertainty due to the series of maser spots used in the fits.

The maser BH mass measurement is proportional to the distance to the galaxy. Consequently, the error in distance to IC 750 will contribute linearly to the mass error. As discussed in Section~\ref{sec:distance}, the Hubble distance using the CMB reference frame, 13.8 $\pm$ 1.0 Mpc, and the distance based on the dynamics of the group to which IC 750 belongs, 14.1 $\pm$ 1.1 Mpc, agree. We apply the larger of the two distance errors, 8\%, to the range of fitted masses. Therefore, the enclosed mass within a diameter of $\sim$0.2 pc, likely dominated by the central BH, is between $4.1 \times 10^4~M_\odot$ and $1.4 \times 10^5~M_\odot$, with a mode of $7.2 \times 10^4~M_\odot$.

\subsubsection{Mass Upper Limit}
\label{sec:upperlimit}
\begin{figure}[htb]
\begin{center}
\includegraphics[width=3.5in]{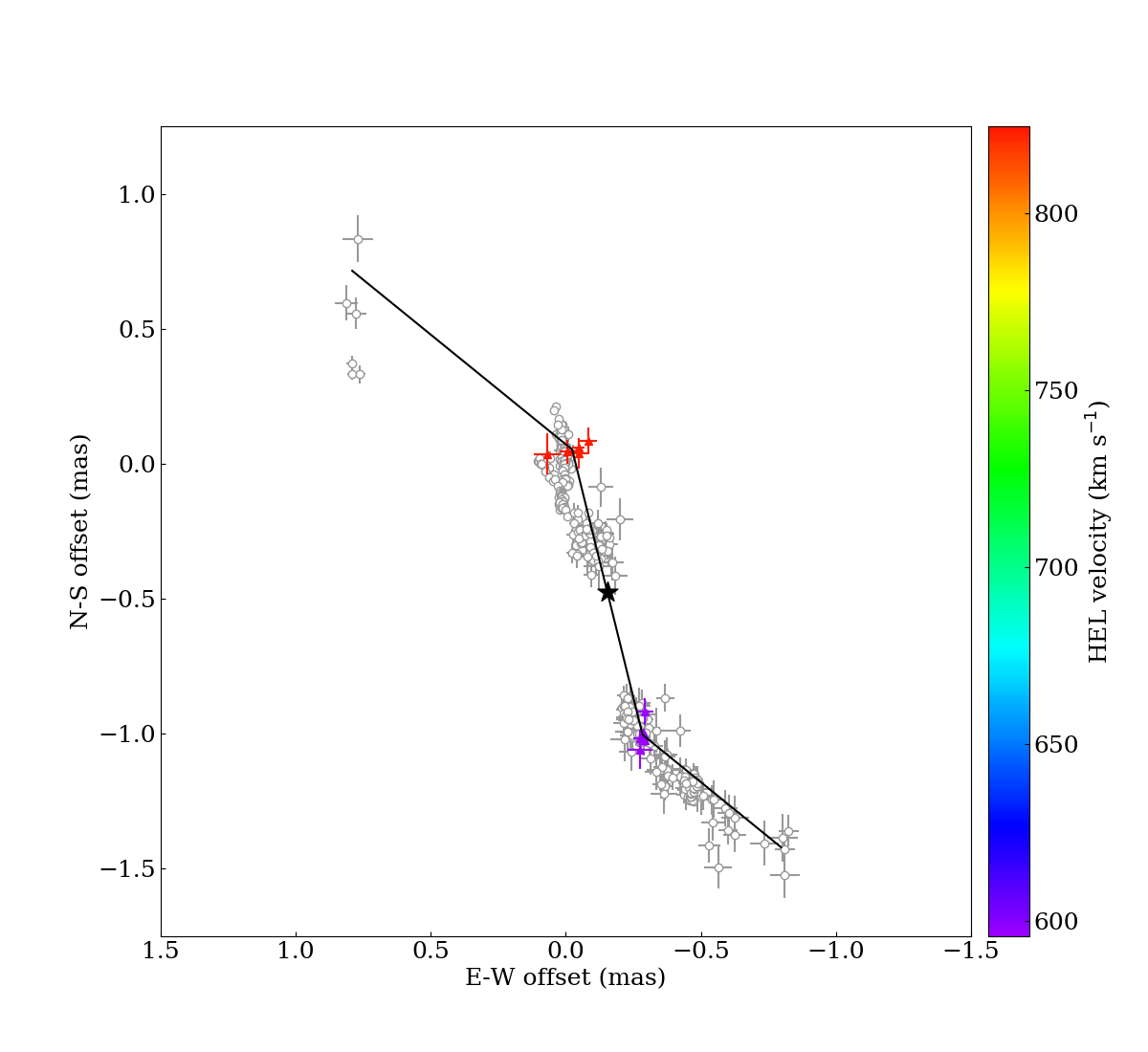}
\includegraphics[width=3.5in]{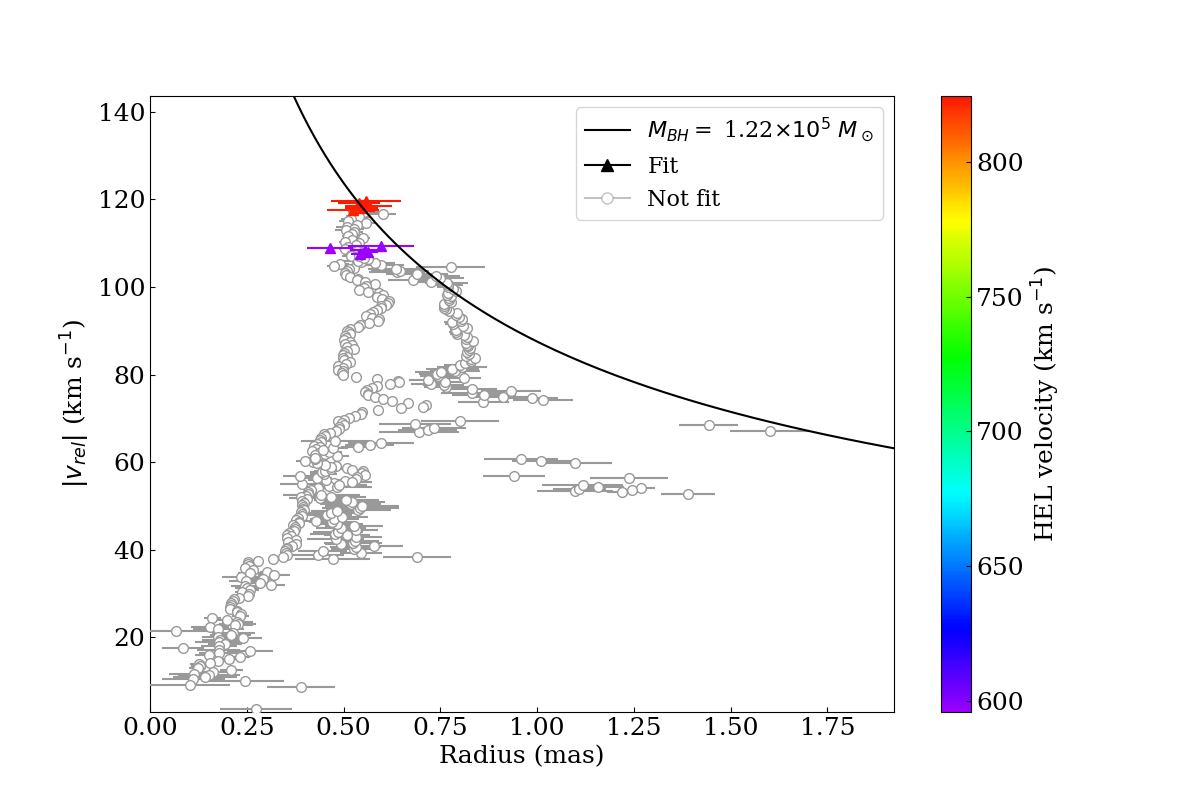}
\end{center}
\caption{\footnotesize \em{Map and PV diagram for the Keplerian envelope. In both plots, the grey circles indicate maser spots not used in the fit and the solid triangles indicate maser spots used in the fit. The colors of the solid triangles indicate the line-of-sight velocities of the maser spots. Left: Map of the maser emission showing the rough position angles for the inner and outer disks. The black star marks the location of the center between the highest velocity redshifted and blue-shifted emission. Right: The PV diagram for the position shown as the black star and a velocity of 705 km s$^{-1}$. The radius is the absolute value of the impact parameter and the relative velocity, $v_{rel}$, is the difference between the assumed BH velocity and maser line-of-sight velocity. The Keplerian envelope, shown as the black line, lies above virtually all the maser emission, corresponds to a mass of $1.22 \times 10^5~M_\odot$.}}
\label{fig:envelope}
\end{figure}

We also determine the upper limit for the mass of the black hole by fitting to only the very highest velocity features. These masers should be at the inner edge of the disk as well as the midline of the emission. To be as conservative as possible, we do not exclude any maser spots. In order to locate the center, we take the highest velocity redshifted emission, between 824.71 km s$^{-1}$ and 822.59 km s$^{-1}$, and the highest velocity blue-shifted emission, which has a Gaussian profile with the center at 597.87$\pm$0.02 km s$^{-1}$ and a width of 1.27 km s$^{-1}$, as the inner-most points on the disk, and fit a rotation curve with only those points. We draw a line between them, as shown in the left panel of Figure~\ref{fig:envelope}, and use the midpoint, shown as the black star, as the location of the BH. We find that for a velocity of 705 km s$^{-1}$, this BH position yields the PV diagram shown in the right panel of Figure~\ref{fig:envelope}. This PV diagram shows the radius, i.e. absolute value of the impact parameter, and the absolute value of the relative velocity to better assess the symmetry between the redshifted and blue-shifted emission. The Keplerian curve, shown as the black line, corresponds to a mass of $1.22 \times 10^5~M_\odot$ and lies above virtually all the maser emission within positional uncertainties, and is defined as the Keplerian envelope. As the mass of the BH increases, the Keplerian curve moves up and to the right, as $M_{\rm BH} = (r_{rel} \cdot v_{rel}^2)/G$. Higher masses will lead to higher velocities at a given radius or, conversely, a larger radius at a given velocity. Therefore, if any of the maser spots are from clouds of water vapor in Keplerian rotation around the BH, the BH mass cannot be higher than $1.22 \times 10^5~M_\odot$\footnote{We imaged a region ten times larger than the region of the detected emission and the velocity coverage of the GBT spectrum is also an order of magnitude wider than the detected maser emission, so we are confident that there is no additional maser emission, above our sensitivity limit, which lies above the Keplerian curve corresponding to $M_{\rm BH} = 1.22 \times 10^5~M_\odot$.}. Applying the uncertainty on the distance to IC 750, we get $1.3 \times 10^5~M_\odot$. This mass also includes contributions from the disk at the radii spanned by the maser emission.

\begin{figure}[htb]
\begin{center}
\includegraphics[width=3.5in]{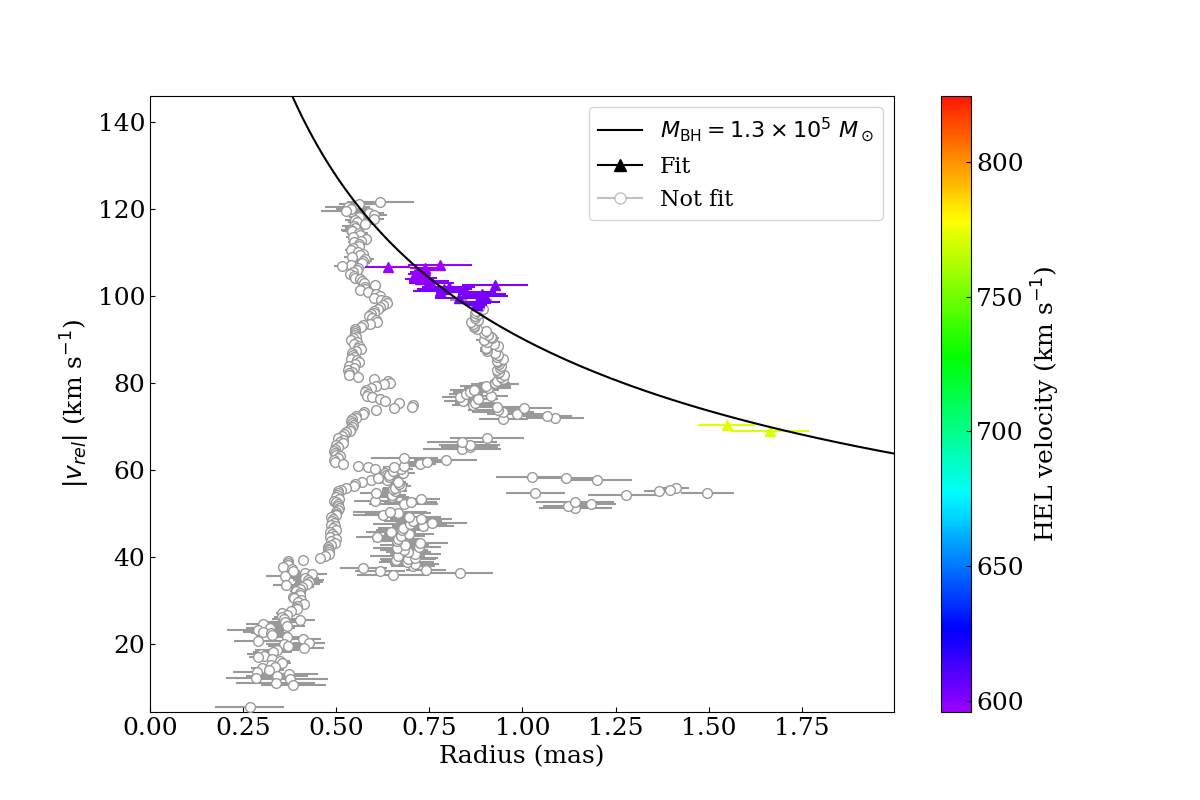}
\end{center}
\caption{\footnotesize \em{PV diagram for the fit with the maximum mass when the position and velocity of the BH are allowed to vary at the same time as varying which maser regions are included in the mass fit. The fit is for a BH position and velocity of -0.45 mas, -0.3 mas; 703 km s$^{-1}$). The radius is the absolute value of the impact parameter and the relative velocity, $v_{rel}$, is the difference between the assumed BH velocity and maser line-of-sight velocity. The grey circles indicate maser spots not used in the fit and the solid triangles indicate maser spots used in the fit. The colors of the solid triangles indicate the line-of-sight velocities of the maser spots. The fitted mass is $1.3 \times 10^5~M_\odot$.}}
\label{fig:varyall25}
\end{figure}

We also check the upper limit of the mass by simultaneously allowing the position and velocity of the BH to vary while also varying the series of maser spots used in the fit, again requiring $0.95 \leq M_{red}/M_{blue} \leq 1.25$. The highest mass from these fits, for a BH position and velocity of (-0.45 mas, -0.3 mas; 703 km s$^{-1}$) is $1.3 \times 10^5~M_\odot$, consistent within fitting error to that from the Keplerian envelope described above. The fitted Keplerian curve for these parameters, shown in Figure~\ref{fig:varyall25}, also lies above virtually all the maser emission. Applying the error on this mass, we get $1.4 \times 10^5~M_\odot$, and take this as our upper limit on the BH mass.

The appendices describe additional variations in the fits, namely, different constraints on the difference between $M_{red}$ and $M_{blue}$, extending the range of possible velocities for the BH, and using additional maser spots in the fits. These do not change the results described above.

\subsection{AGN and Galaxy Properties}
\label{sec:hostgal}
We analyze the available, archival multi-wavelength data for IC 750 in order to derive the AGN and galaxy properties of IC 750. 

\subsubsection{Bolometric Luminosity and Stellar Velocity dispersion}
\label{sec:sigma}

\begin{figure}[tbh]
\begin{center}
\includegraphics[width=5.0in]{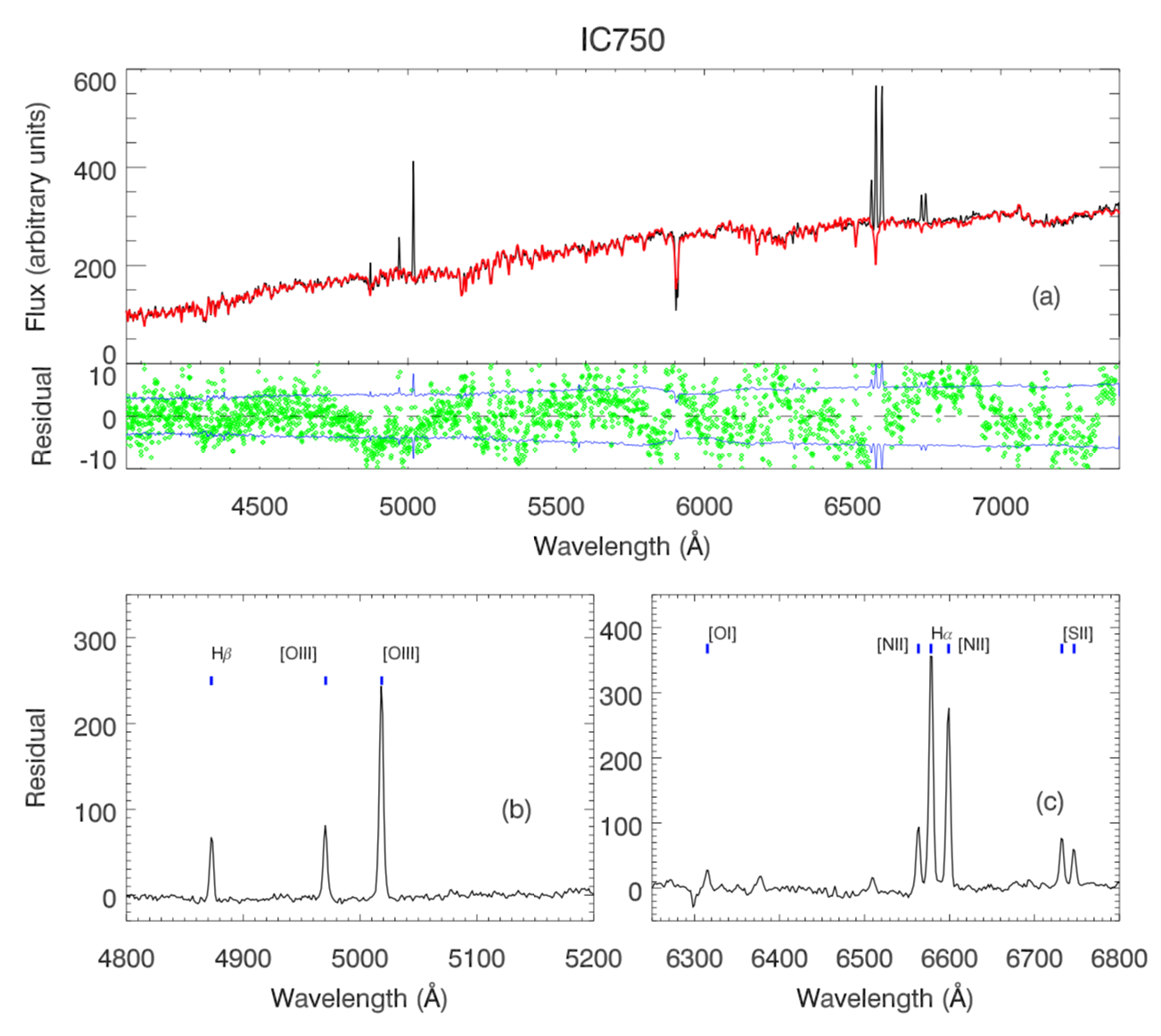}
\end{center}
\caption{\footnotesize \em{Full-spectrum-fitting of the SDSS spectrum for IC 750. Panel (a): DR8 data spectrum is shown as a black line, the error spectrum is shown as a blue line, and the best stellar population model fit of  host-galaxy spectrum in red. The green dots show the residual in the line-free regions. The bottom panel shows the residuals, i.e. emission lines, in the fit. Panel (b) and (c): The residual spectrum zoomed in on the two wavelength ranges with the emission lines used for AGN identification.}}
\label{fig:opticalspectrum}
\end{figure}

In order to understand the optical emission lines in IC 750, we first subtract the stellar absorption and continuum emission from the host galaxy. The procedure is described in detail by \citet{ZCF} and we briefly summarize it here. After masking out the emission lines, a full spectrum fitting was performed on the SDSS spectrum using pPXF \citep{ppxf}. The fit consists of a linear combination of input model spectra, which are shifted to the redshift of the observed spectrum and broadened to account for the stellar velocity dispersion. After subtracting the host galaxy contribution to the spectrum, the residual emission lines are fit using Gaussians. \citet{CZF} showed that the fluxes and line ratios from this procedure are dependent on the stellar population models used. To measure the line fluxes, we use the MILES \citep{MILES} stellar population templates which are based on the current largest and best calibrated empirical stellar library. The spectral fit for IC 750 using MILES is shown in Figure~\ref{fig:opticalspectrum}. For the measurement of velocity dispersion $\sigma_*$, however, we use the PEGASE \citep{pegase} stellar templates since they have better spectral resolution. 

The line ratios we obtain fall above the \citet{Kewley01} AGN identification line in the BPT diagram \citep{BPT}, confirming that IC 750 is a Type 2 AGN. The Balmer decrement, \Ha/\Hb, is $\sim$10.3, much higher than the 3.0 for unobscured AGN. This is consistent with the high obscuration expected in maser host AGN as well as the dust lane seen in the optical images. The observed flux for the [OIII]$\lambda$5007 line is $1.04 \times 10^{-14}$ erg cm$^{-2}$ s$^{-1}$. After correcting for the Balmer decrement \citep[e.g,][]{Bassani99}, this yields a luminosity of $L_{{\rm [OIII]},cor} = 7.4 \times 10^{38}$ erg s$^{-1}$. The [OIII]$\lambda$5007 luminosity is known to be correlated with the bolometric luminosity of the AGN \citep{Heckman04}. Applying the bolometric correction factor of 600 for extinction corrected [OIII]$\lambda$5007 luminosity \citep{Heckman14}, we find that the bolometric luminosity of the AGN in IC 750 is $4.5 \times 10^{41}$ erg s$^{-1}$. Comparing the mass of the black hole in IC 750, from Section~\ref{sec:BHmass}, to the bolometric luminosity derived from [OIII]$\lambda$5007 luminosity, we find that the black hole is accreting at $\sim$5\% of its Eddington limit. This is similar to the AGNs in other maser systems.

To measure the velocity dispersion, we use the PEGASE 3.1 stellar population templates, which have a resolution of R$\sim$10,000. We convolve the PEGASE template into the same resolution of the SDSS spectrum by applying the LSF as a function of wavelength. Then the fit is performed on the common wavelength range between SDSS and PEGASE, 3900-6800 Angstrom, using the full-spectrum-fitting package ULySS \citep{Koleva09}. Bad pixels masked in the SDSS data, emission lines, and the NaD line regions are not used in the full spectrum fitting. We use a pixel size of 70 km s$^{-1}$ in the fits. A multiplicative fifth order polynomial is used to compensate for the possible extinction residuals of the data. The derived velocity dispersion is an averaged value from the full spectrum fitting result. We also performed Monte Carlo simulation on the fitting to determine the errors of the velocity dispersion. Three hundred simulations were performed where random Gaussian noise, based on the SDSS error spectrum, was added to the data before fitting the resultant spectra.

The fit to the SDSS gives a stellar velocity dispersion $\sigma_*=110.7$ km s$^{-1}$. The Monte Carlo simulations yield a statistical error of 3.6 km s$^{-1}$. The size of the SDSS fiber, $1.\!\!^{\prime \prime}5$ in radius which corresponds to $\sim$100 pc at the distance of IC 750, is roughly the size of the bulge of IC 750, with an effective radius of $2\farcs{30} \pm 0\farcs{39}$ ($\sim160 \pm 30$ pc) fit from infrared data as described in Section~\ref{sec:IRnuc}. IC 750 is a spiral galaxy with a high inclination angle, with reported values of 66$^\circ$ \citep{Verheijen01} and $71^\circ \pm 3^\circ$ \citep{Mao10}. Therefore, rotational broadening can contribute to the measured velocity dispersion. Simulations show that the line-of-sight stellar velocity dispersion in spiral galaxies can be over-estimated by up to $\sim$10\% for an edge-on galaxy, due to the motion of the stars in the disk \citep{Bellovary14}. The study by \citet{Bellovary14} from which we applied the error for the inclination angle of IC 750 assumes spatially resolved spectra. We note that for IC 750, the bulge dominates the SDSS aperture, with $L_{\rm Bulge}/L_{\rm disk} \sim 1.86$ based the photometric decomposition described in Section~\ref{sec:IRnuc}. Conversely, obscuration in the galaxy can lead to an underestimate of the stellar velocity dispersion by roughly the same amount \citep{Stickley12}. IC 750 is highly obscured with a Balmer decrement of 10.3. We take the maximal fractional errors for inclination and obscuration which gives a systematic error of 11.1 km s$^{-1}$. 

The SDSS measurement is for a 3\arcsec~diameter fiber and is not as spatially resolved as long slit measurements with small slit widths. \citet{Heraudeau99}, using a long slit with a width of $2.\!\!^{\prime \prime}2$, measure a stellar velocity dispersion in the central pixel, with width $1.\!\!^{\prime \prime}2$, of $118 \pm 25$ km s$^{-1}$. This is consistent with our result and the larger error in \citet{Heraudeau99} is due to a poorer spectral resolution of $\sim$80 km s$^{-1}$ and a lower quality spectrum. Using the parameters from our photometric decomposition of IC 750, we find that $L_{\rm Bulge}/L_{\rm disk} \sim 2.43$ for their spatial resolution of $2.\!\!^{\prime \prime}2 \times 1.\!\!^{\prime \prime}2$.

\citet{Bennert15} carried out a systematic comparison between different definitions of the velocity dispersion parameter taken from the literature and their fiducial measurement ($\sigma_{\rm spat,reff}$). The paper finds that single aperture measurements can overestimate the stellar velocity dispersion, with the biggest discrepancies for galaxies seen close to face-on. For galaxies where $v/\sin(i) < \sigma$, the ratio is $1.06\pm0.01$. IC 750 is a slowly rotating galaxy. Both the stellar rotational velocity \citep{Heraudeau99} and gas rotational velocity \citep{Catinella05} are $\leq 60$ km s$^{-1}$ within $R < 1.\!\!^{\prime \prime}5$ of the SDSS fiber. The lower reported value for the inclination of IC 750 is 66$^\circ$ \citep{Verheijen01} which gives $\sin(i) \sim 0.91$. Therefore, $v/\sin(i) \sim 66$ km s$^{-1}$, which is less than the velocity dispersion. Conversely, the reported ratio of the stellar velocity dispersion within the SDSS fiber to the stellar velocity dispersion within the effective radius of the bulge is $\sigma_{\rm spat,SDSS}/\sigma_{\rm spat,reff}=0.97\pm0.01$. We therefore, take the systematic error on the stellar velocity dispersion due to the spatial resolution of the SDSS fiber to be -6\% = -6.6 km s$^{-1}$ and +3\% = 3.3 km s$^{-1}$. Adding these errors in quadrature with the statistical error on the measurement, $\pm3.6$ km s$^{-1}$, and the systematic errors due to the inclination angle and dust obscuration, $\pm11.1$ km s$^{-1}$. The resulting measurement, $\sigma_* = 110.7^{+12.1}_{-13.4}$ km s$^{-1}$, is consistent with the previously published measurements \citep[e.g.,][]{Heraudeau99, Prugniel01, Zasov04}, and well above the $\sim$70 km s$^{-1}$ SDSS spectral resolution limit ({\tt http://www.sdss.org/dr7/algorithms/veldisp.html}).

The correction factors applied above for inclination and spatial resolution are average values. We also construct a toy model specifically for IC 750 using the constraints on the velocity gradients and stellar velocity dispersion at the center, from \citet{Catinella05} and \citet{Heraudeau99}, assuming a bulge and disk component based on our photometric decomposition, accounting for the inclination angle and the SDSS aperture. The model, described in detail in Appendix~\ref{ap:sigma}, yields a true bulge stellar velocity dispersion of $\gtrsim$ 95 km s$^{-1}$ for the higher gas velocity gradient and the upper limit on the central disk stellar velocity dispersion, consistent with our measurement above. This gives us extra confidence that our measurement does not overestimate the true stellar velocity dispersion of the bulge or underestimate its error by a large amount.

\subsubsection{Nuclear Point Source and Extended Emission in X-rays}
\label{sec:xrays}

\begin{figure}[htb]
\begin{center}
\includegraphics[width=5.5in]{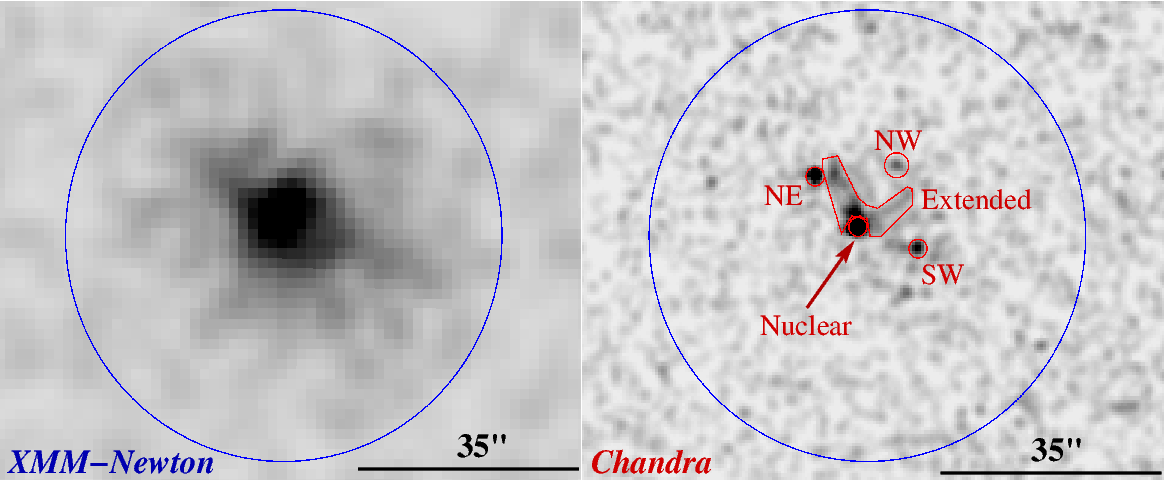}
\end{center}
\caption{\footnotesize \em{Left: The {\it XMM-Newton} image with the extraction region. Right: The {\it Chandra} image with extraction regions for the sources. The large circle shows the {\it XMM-Newton} source region.}}
\label{fig:xrayregions}
\end{figure}

\begin{figure}[htb]
\begin{center}
\includegraphics[width=3.0in]{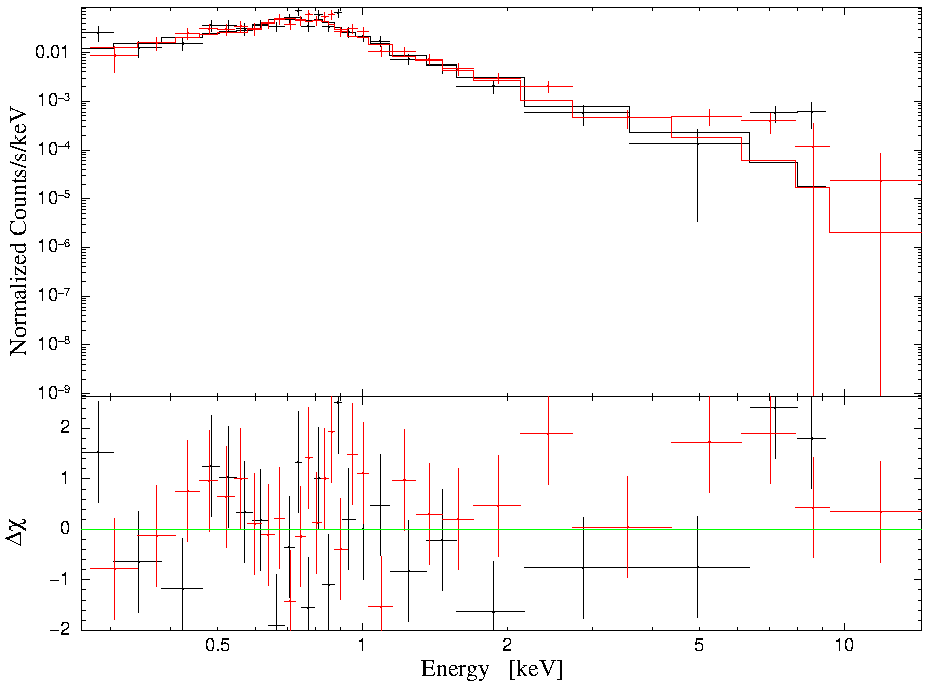}
\includegraphics[width=3.0in]{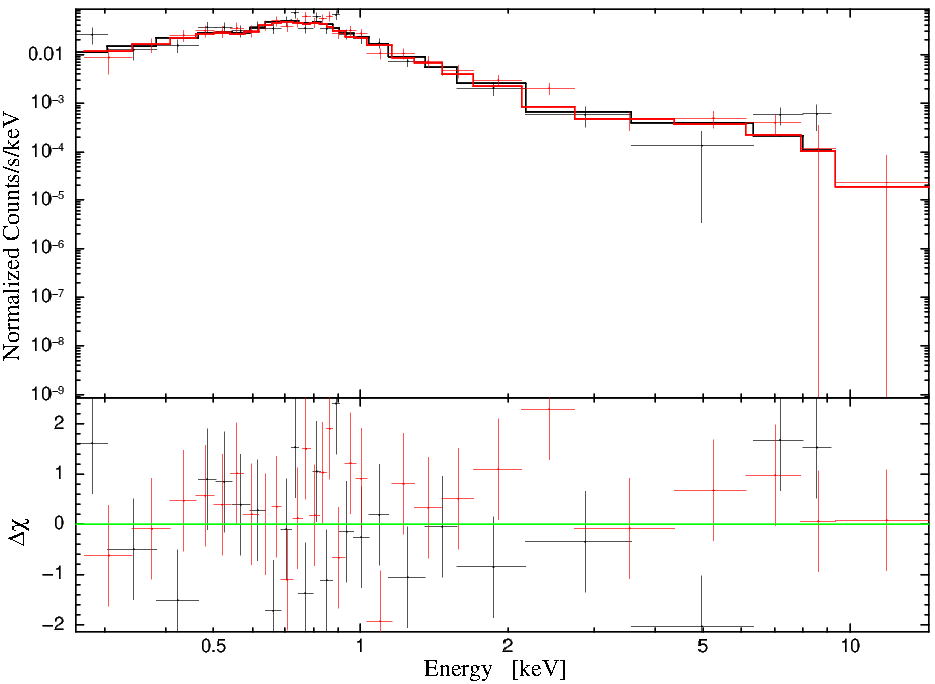}
\end{center}
\caption{\footnotesize \em{The {\it XMM-Newton} spectrum for IC 750 (top) and Residuals to fits (bottom). Left: The spectrum fit with an absorbed power law and MEKAL component. Right: The spectrum fit with an absorbed power law, MEKAL component, and a Gaussian at the energy of the Fe K$\alpha$ line. The solid lines show the fits to the data points. The black data and model are for the 2004 Nov. 28 observation and the red are for the 2004 Nov. 30 observations. The model parameters are constrained to be the same for both datasets.}}
\label{fig:xmmspecfits}
\end{figure}

The X-ray emission in IC 750 is complex. The {\it Chandra} image, Figure~\ref{fig:xrayregions} (right panel), reveals four compact sources, one nuclear and three off-nuclear, and a soft extended source. However, in the {\it XMM-Newton} image, Figure 2 (left panel), the sources cannot be resolved individually. We begin by analyzing the combined X-ray emission in IC 750 from the {\it XMM-Newton} observation. We extract a combined spectrum of 1050 counts, after background subtraction, as shown in Figure~\ref{fig:xmmspecfits}. The solid lines show the fits to the data points. The black data and model are for the 2004 Nov. 28 observation and the red are for the 2004 Nov. 30 observations. The model parameters are constrained to be the same for both datasets. The data are well fit by an absorbed ($N_H=1.26\times10^{21}~{\rm cm}^{-2}$), thermal (MEKAL, $kT = 0.32~{\rm keV}$) and non-thermal (power-law, $\Gamma=2.3$) emission model as seen in the left panel of Figure~\ref{fig:xmmspecfits}. There is an excess around 6.4 keV and adding a Gaussian for a putative Fe K$\alpha$ line improves the fit, as seen in Figure~\ref{fig:xmmspecfits}, right panel. However, since the emission is from five sources with different photon indices, the excess could be an artifact due to the addition of these components and the equivalent width (EW) of the line is highly uncertain. 

We then analyze the individual sources seen in the {\it Chandra} image. We fit the spectra of the nuclear compact source and the extended emission in XSPEC and estimate their 0.3-10 keV and 2-10 keV fluxes. The off-nuclear compact sources do not have enough counts to fit a spectrum, so we have used WebPIMSS ({\tt https://heasarc.gsfc.nasa.gov/cgi-bin/Tools/w3pimms/w3pimms.pl}), with a fixed photon index of 2.0, to estimate the fluxes. The sources, counts, photon indices, fluxes, and luminosities are shown in Table~\ref{tab:xraysources}.

\begin{table}[htb]
\begin{center}
\caption{Table of the Properties of the Sources in the {\it Chandra} Image}
\label{tab:xraysources}
\footnotesize
\begin{tabular}{lcccccc}
\hline
Source & Counts & $\Gamma$ & $F_{0.3-10 keV, obs}$  & $F_{2-10 keV, obs}$ & $L_{0.3-10 keV, obs}$ & $L_{2-10 keV, obs}$ \\
& (Bkg Subtracted) & & (erg cm$^{-2}$ s$^{-1}$) & (erg cm$^{-2}$ s$^{-1}$) & (erg s$^{-1}$) & (erg s$^{-1}$) \\
\hline
Nuclear & 76.7 & 2.0 & $2.7 \times 10^{-14}$ & $1.7 \times 10^{-14}$ & $5.4 \times 10^{38}$ & $3.4 \times 10^{38}$ \\
Extended & 109.9 & 2.2 & $3.0 \times 10^{-14}$ & $9.9 \times 10^{-15}$ & $6.2 \times 10^{38}$ & $2.0 \times 10^{38}$ \\
NE & 26.9 & 2.0 (fixed) & $1.5 \times 10^{-14}$ & $9.1 \times 10^{-15}$ & $3.1 \times 10^{38}$ & $1.9 \times 10^{38}$ \\
SW & 13.9 & 2.0 (fixed) & $7.8 \times 10^{-15}$ & $4.7 \times 10^{-15}$ & $1.6 \times 10^{38}$ & $9.6 \times 10^{37}$ \\
NW & 10.6 & 2.0 (fixed) & $5.9 \times 10^{-15}$ & $3.6 \times 10^{-15} $& $1.3 \times 10^{38}$ & $7.3 \times 10^{37}$ \\
\hline
\end{tabular}
\end{center}
\begin{tablenotes}
\item The off-nuclear compact sources (NE, SW, and NW) did not have enough counts for spectral fitting, so we fix their photon index, $\Gamma$ to be 2.0 and use WebPIMSS to estimate the fluxes.
\end{tablenotes}
\end{table}

\begin{figure}[htb]
\begin{center}
\includegraphics[width=3.0in]{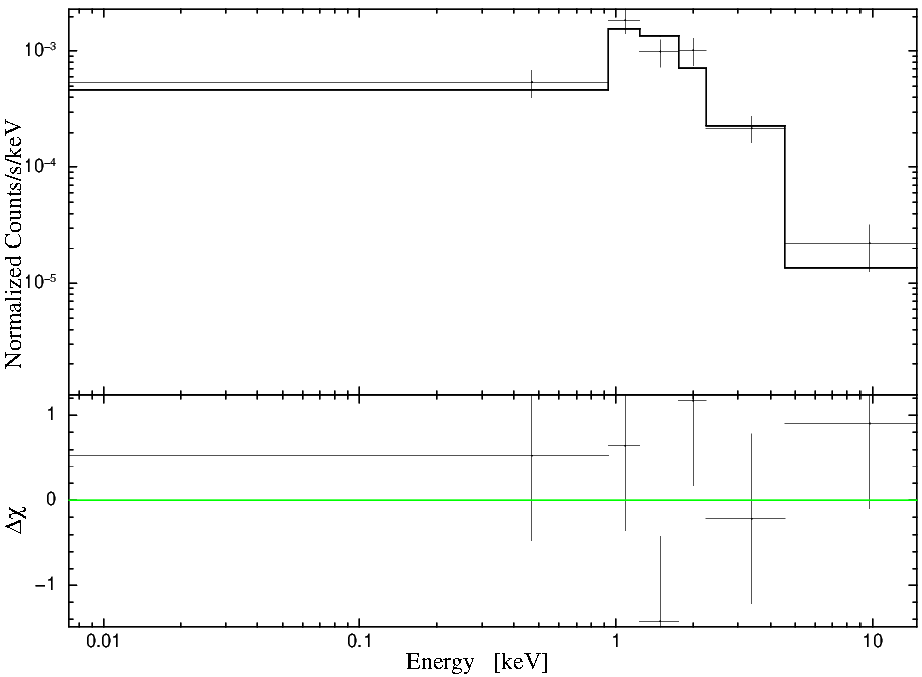}
\includegraphics[width=3.0in]{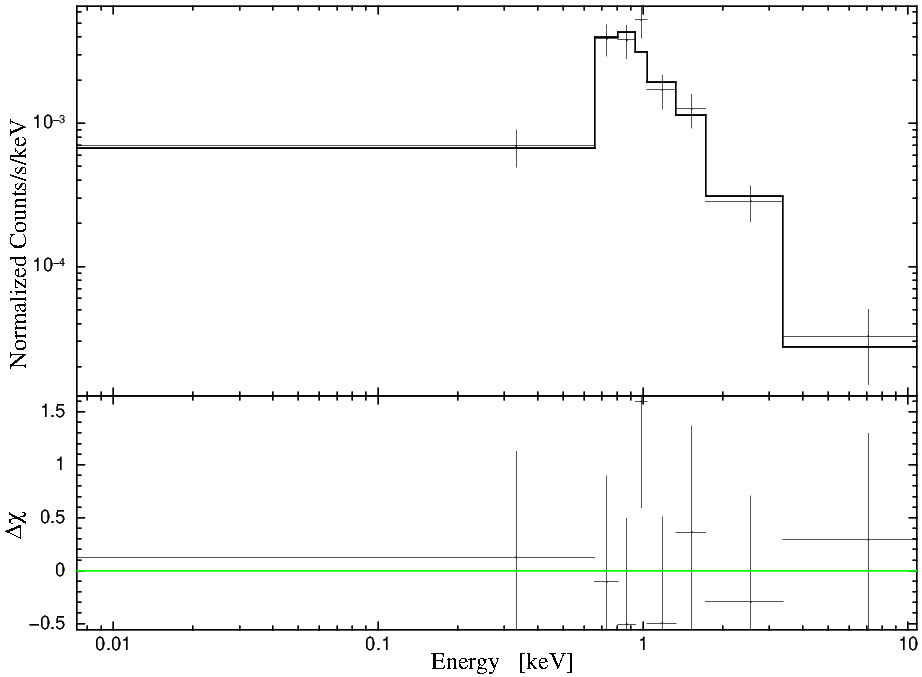}
\end{center}
\caption{\footnotesize \em{Left: The {\it Chandra} spectrum for the nuclear compact source. Right: The {\it Chandra} spectrum for the extended source.}}
\label{fig:chandraspecfits}
\end{figure}

The fits to the spectra for the nuclear source and extended source are shown in Figure~\ref{fig:chandraspecfits}. The observed 2-10 keV luminosity and photon index from our fit of the spectrum for the nuclear source are consistent with that published by \citet{Chen17}. However, we do not find evidence for the high column density, $N_H = 1.2^{+1.4}_{-1.0} \times 10^{23}$ cm$^{-2}$ claimed by these authors. We get an acceptable fit ($\chi^2_{red} = 1.2$) with an absorbed power law, as seen in Figure~\ref{fig:chandraspecfits} (left panel), with a column density fixed at $1.26 \times 10^{21}$ cm$^{-2}$ as derived from the {\it XMM-Newton} spectrum. When the value of $N_H$ was allowed to float, the value returned by the fit was consistent with zero. 

Although we could not definitively conclude that there is high obscuration in IC 750 from the X-ray data, there is some suggestive evidence. The ratio of the observed 2-10 keV luminosity, $3.4 \times 10^{38}$ erg s$^{-1}$, to the corrected [OIII]$\lambda$5007 luminosity, $7.4 \times 10^{38}$ erg s$^{-1}$, is 0.46. Absorbed 2-10 keV to intrinsic [OIII] ratios below 0.1 generally indicate Compton-thick AGNs but heavily obscured Compton thin AGNs and a significant fraction of Compton-thick AGNs are also found to have ratios between 0.1 and 1.0 \citep[e.g.,][]{Goulding11}. Furthermore, $\sim$85\% of disk water maser systems are hosted by Compton thick AGN \citep[e.g.,][]{Castangia13, Greenhill08}, since masers require a nearly edge-on disk (within $\sim 10^\circ$) to have a sufficient amplification path. As discussed in Section~\ref{sec:sigma}, the Balmer decrement is high, 10.3 compared to an unobscured value of 3.0. All these indicate that IC 750 may host a highly obscured nucleus but deeper X-ray observations are needed to confirm it.

For the spectrum of the extended emission in the {\it Chandra} image, an absorbed power law alone does not yield a good fit but adding a MEKAL component ($kT = 0.27~{\rm keV}$) does, as shown in Figure~\ref{fig:chandraspecfits} (right panel). This suggests that the emission results from a combination of thermal and non-thermal processes, possibly a large scale ($\sim$1 kpc) wind with shocks. The most luminous feature is the part extending NE-SW which contains roughly two thirds of the flux. The off-nuclear sources appear to be X-ray binaries but the brightest source could potentially be a ULX. 

IC 750 is a source in the {\it NuSTAR} serendipitous survey catalog \citep{lansbury2017nustar} with large uncertainties for the flux but we could not find a significantly detected source in the publicly available data. It is at the edge of the field of view (FoV) in all but one of the archival {\it NuSTAR} datasets. Our (re-)analysis found no significant detection at the position of IC 750 in the observation in which it is within the FoV. 

\subsubsection{Bulge Radius and Mass}
\label{sec:IRnuc}

\begin{figure*}[htb]
\includegraphics[width=\textwidth]{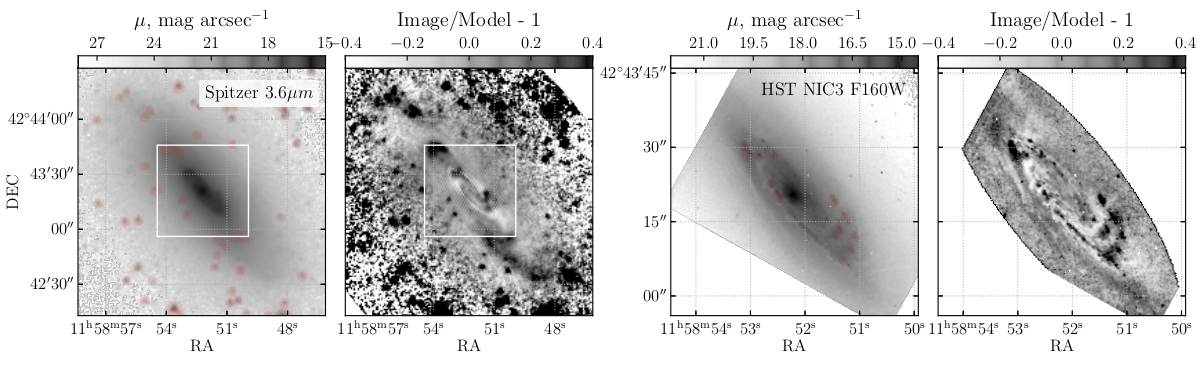}
\includegraphics[width=\textwidth]{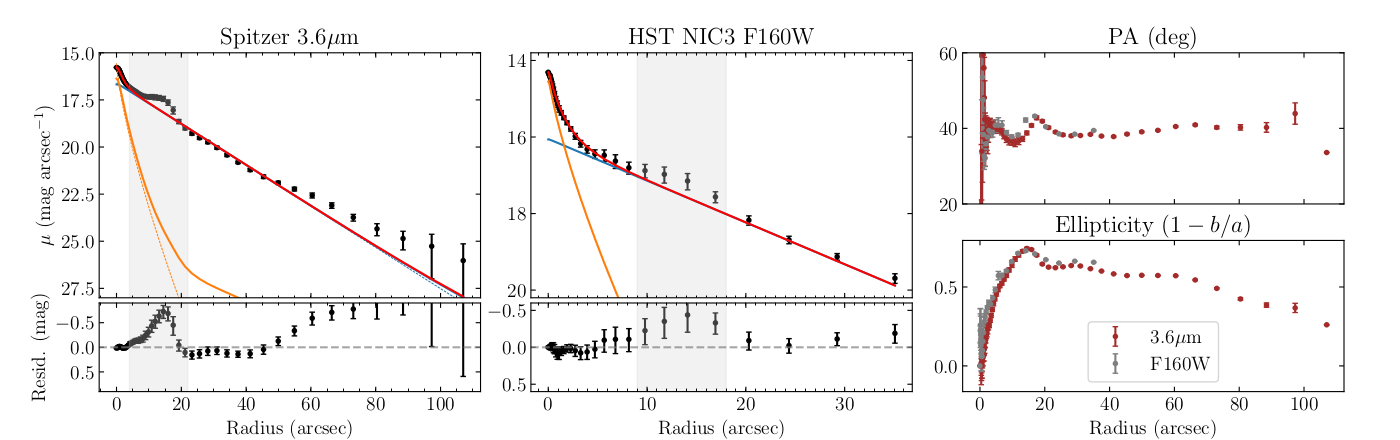}
\caption{\footnotesize \em{Results of the isophotal decomposition of the infrared images of IC 750. The top row shows the images and residual after the isophotal model subtraction for the data from {\it Spitzer} with the IRAC1 (3.6 $\mu$m) filter on the left and {\it HST} with the NICMOS F160W filter on the right. The white boxes on the {\it Spitzer} image and residual correspond to the extent of {\it HST} image. The regions masked in the isophotal analysis are marked in transparent red. The bottom row shows the radial profiles of the isophotal parameters including average surface brightness, positional angle of the major axis, and ellipticity. The grey regions show the radii excluded from the fit due to the presence of spiral arms. Both the {\it Spitzer} and {\it HST} brightness profiles are simultaneously fit by the model (red line) composed of a S\`{e}rsic component (orange line) and an exponential disk (blue line). In the fit, we require that the radial disk scale, the effective radius, and the index of the S\`{e}rsic component to be the same for both datasets. The S\`{e}rsic and disk components of the model were convolved with the corresponding PSFs of the instruments. The unconvolved S\`{e}rsic and disk components are shown as the dashed orange and blue lines, respectively.}}
\label{fig:bulgefit}
\end{figure*}

In order to derive the bulge parameters, we simultaneously fit the infrared images from {\it Spitzer} (IRAC1, 3.6 $\mu$m) and {\it HST} (NICMOS F160W). The {\it HST} image has a much higher spatial resolution and, therefore, better constrains the bulge while the larger and more sensitive {\it Spitzer} data better constrain the disk radial scale. First, we performed an isophotal analysis applying the IRAF/ELLIPSE task \citep{Jedrzejewski87} and calculated radial profiles of the surface brightness, ellipticity, and position angles for the two images, after masking out stars. The regions with prominent spiral arms were also masked out. Then we simultaneously decompose both surface brightness profiles into S\`{e}rsic and exponential disk components by using the non-linear minimization \textsc{lmfit} package \citep{Newville16}. The effective radius and the S\`{e}rsic index of the bulge, and the radial scale of the disk are constrained to be identical for both datasets. In order to account for PSF effects, we convolve the model profiles of each minimization iteration with the PSFs for {\it HST}, from the Tiny Tim PSF modelling tool \citep{Krist11}, and {\it Spitzer} \citep{Comeron18}. The derived bulge parameters from the simultaneous fit are consistent with fitting for the {\it HST} data alone. The fits and residuals are shown in Figure~\ref{fig:bulgefit}.

The effective radius of the S\`{e}rsic component is found to be $2\farcs{30} \pm 0\farcs{39}$, which corresponds to $\sim160 \pm 30$ pc. This component has a S\`{e}rsic index $n=1.34 \pm 0.11$, strongly suggesting that the component does not represent a classical bulge. The bulge luminosity is $L_{F160W}=(1.9 \pm 0.7)\times10^9~L_\odot$, with a S\`{e}rsic-to-disk ratio of 0.055. We input the age and metallicity estimates from single stellar population (SSP) fits to the SDSS spectrum, namely $T_{\rm SSP}=2.1$~Gyr and [Fe/H]$_{\rm SSP}=0.1$~dex, and a Kroupa initial mass function \citep{Kroupa}, into PEGASE \citep{pegase} and infer the mass-to-light ratio in the F160W filter to be 0.39. Assuming that the stellar population properties derived from the SDSS spectrum are representative of the infrared nuclear component, we determine the mass of the infrared nucleus to be $(7.3 \pm 2.7)\times10^8 M_\odot$. From the photometric decomposition, we find the S\`{e}rsic-disk ratio in the 3\arcsec\ diameter SDSS aperture to be 1.86. 

\subsubsection{Stellar Mass}
\label{sec:Mstar}
\citet{Chen17} report the total galaxy stellar mass of IC 750 to be $\log(M_*/M_\odot) = 9.1 \pm 0.3$, based on spectral energy distribution (SED) fitting of photometric measurements spanning from UV to mid-IR as well as a cross-check using SDSS photometry. The fit takes into account a possible contribution from the AGN. A fit for the mass was also done using only extinction corrected SDSS photometry and is reported to be consistent with the SED fits. However, the S$^4$G survey report a stellar mass of $\log(M_*/M_\odot) = 10.6$ based on {\it Spitzer} 3.6 $\mu$m photometry \citep{Sheth10}. Part of the difference is that \citet{Chen17} used a distance of 8.8 Mpc, deduced from their reported absolute and apparent r magnitudes, and S$^4$G used a distance of 23.518 Mpc. When the values are recalculated using 14.1 Mpc, the values become $\log(M_*/M_\odot) = 9.5$ for the \citet{Chen17} measurement and $\log(M_*/M_\odot) = 10.2$ for the S$^4$G measurement. When we use our measurements of the bulge mass and bulge-to-disk ratio, assuming that the mass-to-light ratios are the same for the bulge and disk, we derive a stellar mass of $\log(M_*/M_\odot) = 10.1 \pm 0.2$, consistent with the S$^4$G measurement. The \citet{Chen17} measurement could be low because IC 750 is highly reddened and the dust strongly affects the mid-IR and optical photometry used in the analysis, leading to an underestimation of the stellar contribution and/or overestimation of the AGN contribution. The {\it HST} and {\it Spitzer} measurements used in our analysis are less biased by dust. This suggests that IC 750 is a low mass galaxy instead of a dwarf galaxy.

\subsection{The Multi-wavelength Picture}

\begin{figure}[htb]
\begin{center}
\includegraphics[width=4.0in]{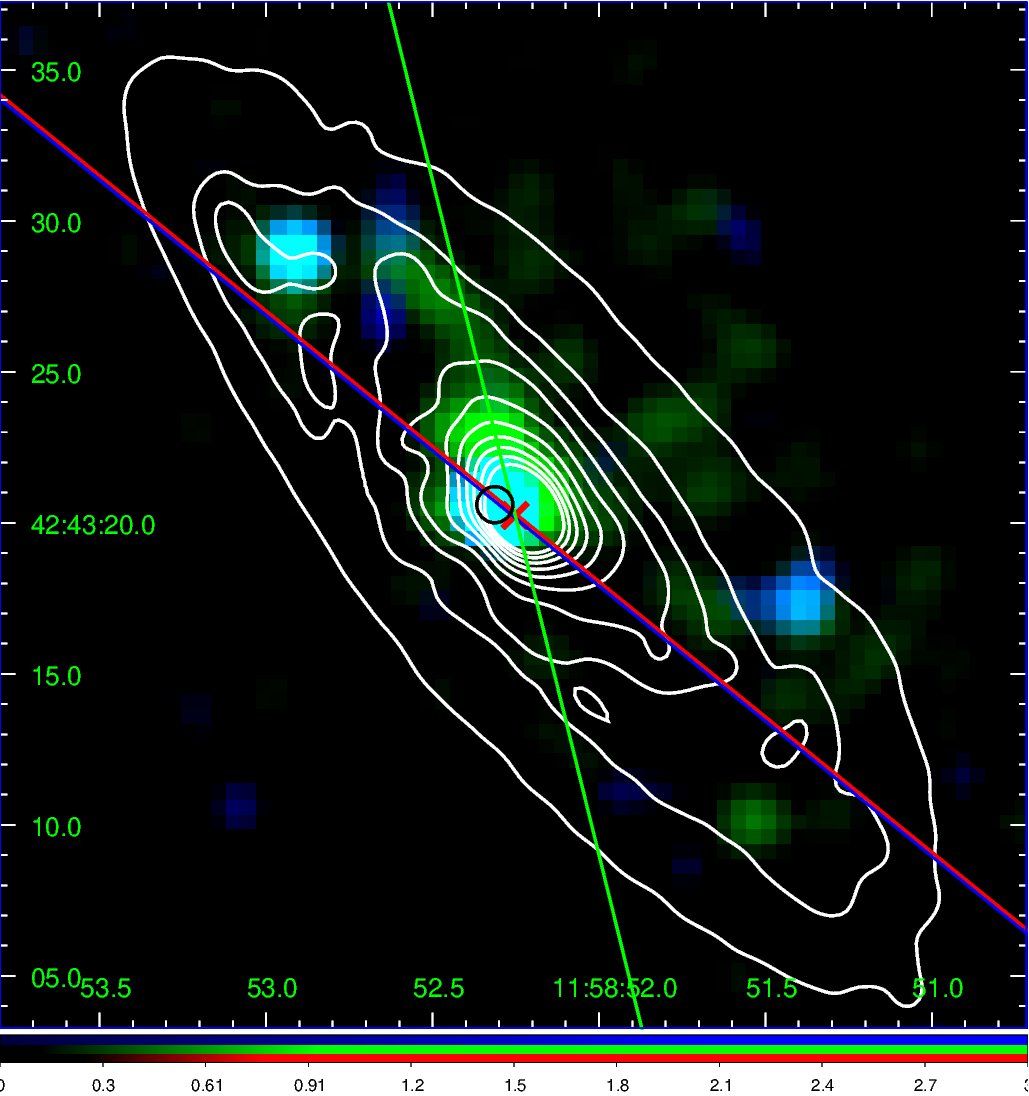}
\end{center}
\caption{\footnotesize \em{The Multi-wavelength Image of IC 750. The {\it Chandra} X-ray image (blue for 2-10 keV, green for 0.3-2.0 keV) is overlaid with infrared contours from {\it HST} NICMOS F160W data. The red, green, and blue lines show the orientations of the different sections of the maser disk. The black circle is the centroid position of the nuclear X-ray emission, with the radius corresponding to the positional error, and the red cross is the position of the maser emission (magnified by a factor of $\sim400$).}}
\label{fig:multiwavelength}
\end{figure}

The black hole, AGN, and galaxy properties measured for IC 750 from the multi-wavelength data are summarized in Table~\ref{tab:prop}. The composite image from the radio, X-ray, and infrared data is shown in Figure~\ref{fig:multiwavelength}, in order to see the relative locations of the emission from different wavelengths. The infrared contours form {\it HST} are overlaid on the {\it Chandra} image. The X-ray centroid position is marked with a black circle with a radius corresponding to the positional error. The position of maser disk is marked with a red cross with a size $\sim400$ times the extent of the disk to be visible in the image. The centers of the nuclear sources in the X-ray and infrared data are consistent with the position of the maser emission, indicating that the black hole is at the center of IC 750. Lines showing the extension of the redshifted, systemic, and blue-shifted parts of the maser disk are shown for relative orientation to the galaxy.

\begin{table}[htb]
   \caption{Summary of IC 750 Parameters}
    \begin{center}
    \begin{tabular}{lc}
        \hline\hline
        Parameter & Value  \\
        \hline
        $M_{\rm BH}$ & $(4.1 - 14) \times10^4~M_\odot^{~~a}$; $7.2 \times 10^4~M_\odot^{~~b}$ \\
        $M_{\rm BH}$ Upper Limit & $1.4\times10^5~M_\odot$ \\
        \hline
        $\sigma_*$ & $110.7^{+12.1}_{-13.4}$ km s$^{-1}$ \\
        \hline
        Balmer decrement, H$\alpha$/H$\beta$ & 10.3 \\
        $L_{{\rm [OIII]},cor}$ & $7.4 \times 10^{38}$ erg s$^{-1}$ \\
        $L_{2-10 keV,obs}$ & $3.4 \times 10^{38}$ erg s$^{-1}$\\
        \hline
        $M_{\rm Bulge}$ & $(7.3 \pm 2.7) \times 10^8~M_\odot$  \\
        $L_{{\rm Bulge},F160W}$ & $(1.9 \pm 0.7) \times 10^9~L_\odot$  \\
        $M/L_{F160W}$ & 0.39  \\
        $n$ & $1.34 \pm 0.11$ \\
        S\'{e}rsic $R_{eff}$ & 2$\farcs{30} \pm 0\farcs{39}$ \\
        S\'{e}rsic $R_{eff}$ & $0.157 \pm 0.027$ kpc \\
        $L_{\rm Bulge} / L_{\rm disk}$ & 0.055  \\
        $L_{\rm Bulge} / L_{\rm total}$ & 0.052  \\
        $L_{\rm Bulge} / L_{\rm disk}~(R<1.\!\!^{\prime \prime}5)$ & 1.86  \\
        \hline
        $\log(M_* / M_\odot$) & $10.1 \pm 0.2$  \\
         & 10.2$^c$ \\
         & $9.5 \pm 0.3^d$ \\
        \hline
    \end{tabular}
    \end{center}
    \begin{tablenotes}
    \item $^a$ Full range of the distribution of fitted masses
    \item $^b$ Mode of the distribution of fitted masses
    \item A distance of 14.1 Mpc is used for all relevant measurements.
    \item $^c$ Value from \citet{Sheth10}, recalculated for a distance of 14.1 Mpc.
    \item $^d$ Value from \citet{Chen17}, recalculated for a distance of 14.1 Mpc.
    \end{tablenotes}
    \label{tab:prop}
\end{table}

\section{Discussion}
\label{sec:discussion}
The analysis of the maser emission in IC 750 settles some open questions and challenges some theoretical predictions. The presence of the maser disk confirms IC 750 as an AGN, of which there was some room for doubt because the X-ray emission is low enough to be produced by a ULX \citep{Chen17}. Furthermore, a mass between $4.1 \times 10^4~M_\odot$ and $1.4 \times 10^5~M_\odot$ in a region $\sim$0.2 pc in diameter gives definitive proof that the source is a BH. 

The mass of the IMBH in IC 750 suggests that the seed mass was of order $10^4~M_\odot$ or smaller for the BH to remain $\lesssim10^5~M_\odot$ at $z=0$. It also provides a stringent challenge to the theoretical prediction that nuclear black holes in local galaxies must have masses $M_{\rm BH} \gtrsim 3 \times 10^5~M_\odot$ \citep{Alexander14}. While there are other reported IMBH candidates with masses less than $10^5~M_\odot$, the uncertainties and/or discrepancies in different measurements for the galaxies range from factors of a few to more than one order of magnitude \citep[e.g.,][]{Baldassare15, Chilingarian18, Nguyen19, Woo19}.

We find that the position and velocity of the black hole derived from the position and dynamics of the maser disk are consistent with the optical nucleus of the galaxy. The maser position is also consistent with the centers of the X-ray and infrared nuclei. However, recent simulations \citep[e.g.,][]{Bellovary19,Pfister19} suggest that BHs in $\gtrsim$ 50\% of dwarf galaxies, especially those which start from seeds of order $10^4~M_\odot$ should be significantly displaced from the center of the galaxy. Observationally, \citet{Reines20} report BHs $\gtrsim$ 2\arcsec\ from the optical center in seven out of 13 dwarf galaxies with active BHs identified by the detection of compact radio sources, although the authors point out that some fraction of the off-nuclear active BHs could be background radio galaxies at high redshift. Additional maser (and non-maser) observations are needed to test the theoretical predictions. Since the GBT beam is $\sim$ 34\arcsec\ in diameter, maser emission can be discovered from off-nuclear AGNs even when the observations are pointed at the optical center. The velocity of the maser emission will confirm association with the target galaxy and follow-up interferometry will determine whether the emission is offset from the nucleus.

\subsection{Black Hole-Galaxy Scaling Relations}

The mass of the nuclear BH has been observed to correlate with host galaxy properties, namely stellar velocity dispersion, $\sigma_*$, the bulge mass $M_{\rm Bulge}$, and stellar mass $M_*$. These results are interpreted as an indication of coevolution between the central BHs and their host galaxies, and are believed to be due to a combination of mergers, accretion, and feedback. While different intercepts and/or slopes for the scaling relations have been reported for different types of galaxies, e.g. quiescent vs. active, elliptical vs. spiral, classical vs. pseudo-bulges \citep[e.g.,][]{KormendyHo, Reines15, Greene16, Lasker16, McConnellMa, Greene19}, all studies find a single line over the full mass range. However, simulations indicate that where low mass galaxies and BHs fall on the $M_{\rm BH}-\sigma_*$ relation at $z=0$ can differentiate between different BH seed formation processes \citep{Volonteri10}. The scarcity of robust dynamical BH mass measurements at the low end has been a challenge to determining whether the relations hold for low mass galaxies and BHs.

We compare IC 750 to other galaxies and the fitted correlation lines from the literature. To be conservative, we use the BH mass upper limit rather than the measured mass and give errors obtained by adding in quadrature, the errors from the fitted lines with the errors in our measurements. Due to the availability of $\sigma_*$, $M_{\rm Bulge}$, and $M_*$ measurements, the galaxy samples shown in Figures~\ref{fig:mbh_sigma} to~\ref{fig:mbh_mstar} do not overlap completely. The details of the measurements gathered from literature are included as Online Data accompanying this paper. An illustrative excerpt is given in Table~\ref{tab:galparms}. 

There are many papers which report empirical correlations between the BH mass and galaxy properties, and the slopes and intercepts of the fitted lines vary depending on the sample used in a given paper. Comparing to every published BH-galaxy scaling relation is beyond the scope of this work. Our BH mass measurement for IC 750 is a dynamical measurement of a BH with very low mass. Therefore, we compare IC 750 to the $M_{\rm BH}-\sigma_*$ and $M_{\rm BH}-M_*$ relations from \citet{Greene19}, fit to galaxies with dynamically measured masses including BHs at the low mass end. For the $M_{\rm BH}-M_{\rm Bulge}$ relation, there is no reference that has dynamically measured BH masses at the low end. Therefore we compare to \citet{KormendyHo} which fit the relation based on dynamically measured masses, and to \citet{Schutte19} which adds reverberation mapped and single epoch BH masses to the fit at the low mass end. Table~\ref{tab:relations} contains comparisons to scaling relations taken from additional references.

\subsubsection{$M_{\rm BH}-\sigma_*$}

We first compare the $M_{\rm BH}$ and $\sigma_*$ of IC 750 to other galaxies and fitted $M_{\rm BH}-\sigma_*$ relations. The correlation between the mass of the nuclear BH and the stellar velocity dispersion ($\sigma_*$), a proxy for bulge mass, has been established over several orders of magnitude in black hole mass \citep[e.g.,][]{Ferrarese00, Gebhardt00, Tremaine02, Gultekin09, KormendyHo, Greene19} and it is the tightest relation of the three. The velocity dispersion is a proxy for the bulge mass but has fewer measurement uncertainties as it does not depend on the distance to the galaxy, choice of mass-to-light ratio, or bulge-disk decomposition, all of which affect the measurements of bulge mass.

\begin{figure}[htb]
\begin{center}
\includegraphics[width=4.0in]{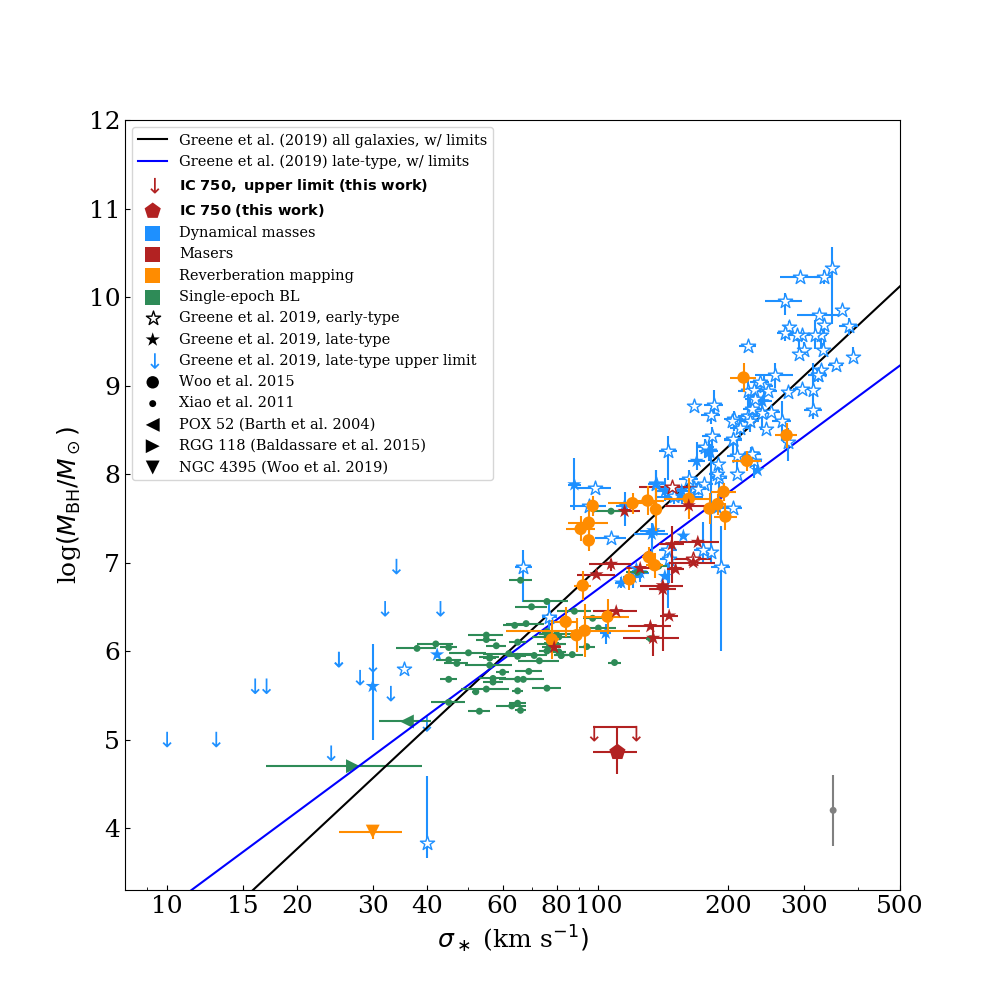}
\end{center}
\caption{\footnotesize \em{$M_{\rm{BH}}-\sigma_*$ Relation. Both lines show scaling relations fit only for dynamically measured masses from \citet{Greene19}, black for the full sample and blue for only late type galaxies. The BH mass for IC 750 is shown as the red pentagon (error bars represent the full range of fitted masses), and the mass upper limit is shown as the solid red line with downward arrows. The other galaxies are taken from the literature as indicated by the symbols and shown for comparison. The colors indicate the mass measurement method. The grey vertical line is the representative error of 0.4 dex for the single epoch spectroscopic masses from \citet{Xiao11}, obtained from adding in quadrature the 0.14 dex typical statistical error reported by the authors and (0.38$\pm$0.05) dex systematic error quoted by the authors based on \citet{McGill08}.}}
\label{fig:mbh_sigma}
\end{figure}

\citet{Greene19} presents the most recent review of the $M_{\rm BH}-\sigma_*$ relation for galaxies with dynamically measured BH masses. The sample includes the galaxies from \citet{KormendyHo} as well as more recent dynamical mass measurements. The fits are done with the full sample as well as to early type (elliptical and S0), late type (spiral) galaxies, separately. In addition, the fits are performed including and excluding the systems with BH mass upper limits at the low mass end. \citet{Greene19} state that the fits of late type galaxies without the systems with BH mass upper limits are biased towards the most massive galaxies for a given galaxy property. We therefore compare IC 750, a spiral galaxy, to the relation using late type galaxies with upper limits for some low mass BHs. For our mass upper limit, $1.3 \times 10^5~M_\odot$, and $\sigma_* = 110.7^{+12.1}_{-13.4}$ km s$^{-1}$, IC 750 is offset by $(-1.7 \pm 0.3)$ dex relative to the fitted relation, as shown in Figure~\ref{fig:mbh_sigma}. The error is obtained by adding in quadrature the errors in the fitted parameters of the relation and the errors in our measurements. This offset is roughly three times the reported intrinsic scatter in the relation of (0.58 $\pm$ 0.09) dex. When compared to the fit to the full sample of dynamical BH masses, IC 750 is offset by $-2.0^{+0.3}_{-0.2}$ dex relative to the relation, which has an intrinsic scatter of $(0.55\pm0.04)$ dex. Comparing to the fits for late type and all galaxies without the mass upper limits, IC 750 is more discrepant than comparing to the corresponding fits with upper limits, as listed in Table~\ref{tab:relations}. 

As noted in Section~\ref{sec:sigma}, the stellar velocity dispersion measurement could be contaminated by rotational broadening since IC 750 is highly inclined and the measurement is from a 3\arcsec\ diameter fiber rather than a spatially resolved long slit. We have applied average errors to account for the inclination and spatial resolution based on studies by \citet{Bellovary14} and \citet{Bennert15}. In addition, a toy model using the measured rotational properties of the disk in IC 750, described in detail in Appendix~\ref{ap:sigma}, shows consistency with the applied average errors. For IC 750 to fall on the $M_{\rm BH}-\sigma_*$ relation for late type galaxies using the fits with the upper limits from \citet{Greene19}, the stellar velocity dispersion would have to be 37.0 km s$^{-1}$. In order to be consistent within one standard deviation of the slope, intercept, and intrinsic scatter of the same relation, the stellar velocity dispersion would have to be 68.9 km s$^{-1}$. The former is low by roughly five times the error and the latter is low by roughly three times the error relative to our measurement, $110.7^{+12.1}_{-13.4}$ km s$^{-1}$.

More massive BHs ($M_{\rm BH} \sim 10^6-10^8~M_\odot$) in other maser systems have been seen to fall below the $M_{\rm BH}-\sigma_*$ relation, specifically $(-0.6 \pm 0.1)$ dex \citep{Greene10, Greene16} on average, though the stellar velocity dispersion measurements for the  maser galaxies in \citet{Greene16} are from spectra with different spatial resolutions, including SDSS, and some of the measurements may be contaminated by rotational broadening. The BH in IC 750 is under-massive even relative to the sample of galaxies with maser mass measurements and falls $(-1.8 \pm 0.4)$ dex relative to the \citet{Greene16} relation using only late type galaxies, which has an intrinsic scatter of (0.49 $\pm$ 0.07) dex.

\subsubsection{$M_{\rm BH}-M_{\rm Bulge}$ and $M_{\rm BH}-M_*$}
The bulge and stellar mass measurements for galaxies have higher uncertainties since they depend on the details of the bulge-galaxy decomposition, assumed mass-to-light ratios, and distance to the galaxy. The observed correlations show bigger scatter than the $M_{\rm BH}-\sigma_*$ relation. However, both the $M_{\rm BH}-M_{\rm Bulge}$ relation \citep[e.g.,][]{Haring04, KormendyHo, Lasker16, McConnellMa, Schutte19} and $M_{\rm BH}-M_*$ relation \citep[e.g.,][]{Greene19, Lasker16, Reines15} have been established in the literature.

\begin{figure}[hbt]
\begin{center}
\includegraphics[width=4.0in]{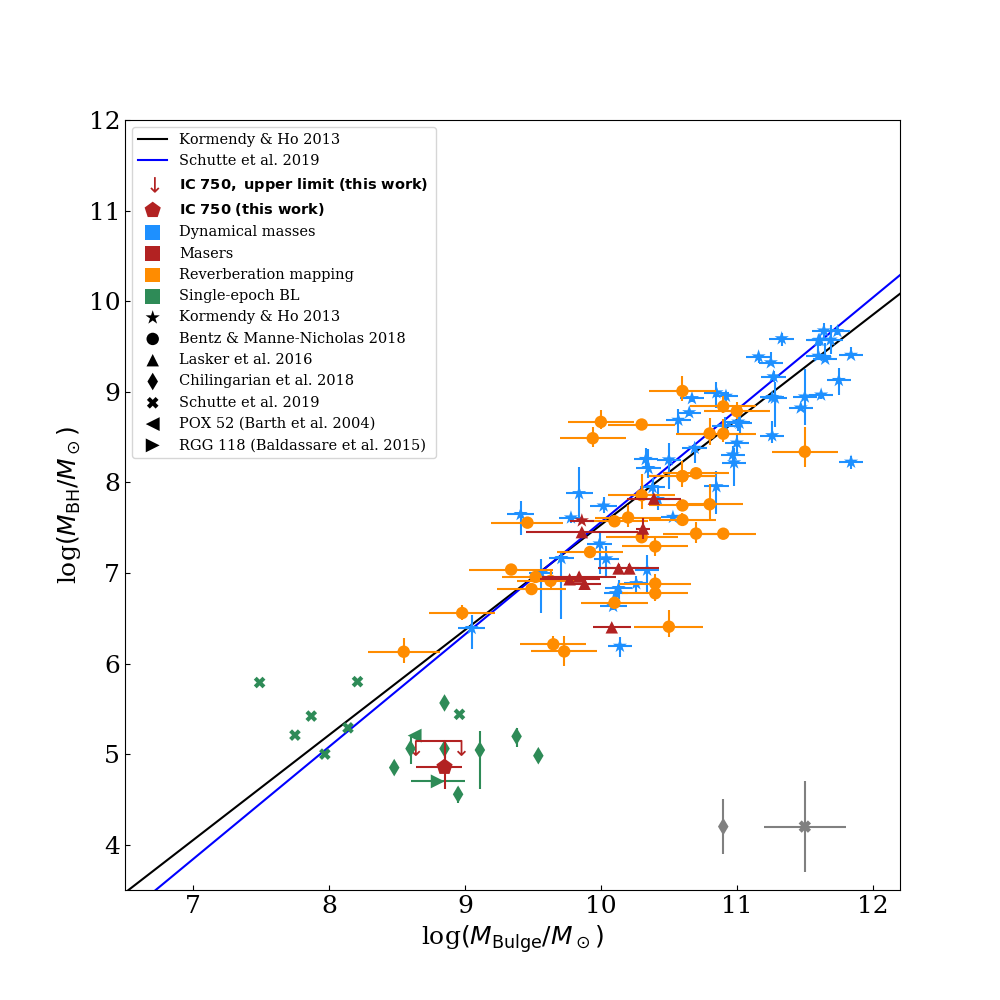}
\end{center}
\caption{\footnotesize \em{$M_{\rm{BH}}-M_{\rm{Bulge}}$ Relation. The black line shows the relation from \citet{KormendyHo}, which used only dynamically measured BH masses, and the blue line shows the relation from \citet{Schutte19}, which used low-mass BHs with masses from single epoch optical spectroscopy. The BH mass for IC 750 is shown as the red pentagon (the error bars represent the full range of fitted masses), and the mass upper limit is shown as the solid red line with downward arrows. The other galaxies are taken from the literature as indicated by the symbols. The colors indicate the mass measurement method. The grey cross plotted over grey ``x'' is the representative error for single epoch spectroscopic masses from \citet{Schutte19}, which do not have reported statistical errors, but for which the authors quote a systematic error of $\sim$0.3 dex in $M_{\rm{Bulge}}$ and $\sim$0.5 dex in $M_{\rm{BH}}$. The grey cross plotted over the grey diamond is the representative error for single epoch spectroscopic masses from \citet{Chilingarian18}, who quote a systematic error budget of 0.3 dex in $M_{\rm{BH}}$ in addition to the statistical errors already plotted here. The galaxies labeled as \citet{Schutte19} are only the galaxies whose masses were obtained in that work. The fit to the line by the authors include, dynamically measured and reverberation mapped masses, as well as some BHs in individual dwarf galaxies; in short, the \citet{Schutte19} fit includes all the galaxies shown in the plot except for those from \citet{Chilingarian18} and IC 750. The \citet{Chilingarian18} galaxies are from a search for IMBHs and has a mass cut of $M_{\rm BH} < 2 \times 10^5~M_\odot$.}}
\label{fig:mbh_mbulge}
\end{figure}

Figure~\ref{fig:mbh_mbulge} shows where IC 750 lies relative to the observed $M_{\rm BH}-M_{\rm Bulge}$ relation. One of the lowest masses in these samples, it is offset by ($-1.1\pm0.3$) dex relative to the expectation from dynamically measured masses ($M_{\rm BH} \gtrsim 10^6~M_\odot$) \citep{KormendyHo}, which has an intrinsic scatter of 0.29 dex, and $-1.0^{+0.3}_{-0.4}$ dex relative to the fitted relation from \citet{Schutte19}, which has an intrinsic scatter of 0.69 dex. The \citet{Schutte19} relation is fit to the dynamically measured BH masses from \citet{KormendyHo} with a different mass-to-light ratio for computing the bulge masses, reverberation mapped BH masses from \citet{Bentz18}, and low mass galaxies from their own fits and five others from literature including RGG 118 \citep{Baldassare15} and POX 52 \citep{Barth04}. \citet{Lasker16} measured the bulge mass of higher mass maser galaxies and found that, on average, masers lie $(-0.8\pm0.2)$ dex from the line fit to their sample of late type galaxies. IC 750 is $(-1.4\pm0.3)$ dex from the \citet{Lasker16} line, which has an intrinsic scatter of $(0.52\pm0.06)$ dex. 

IC 750 is roughly one order of magnitude below the fitted $M_{\rm BH}-M_{\rm Bulge}$ relations from \citet{KormendyHo} and \citet{Schutte19}, which have intrinsic scatters of 0.29 dex and 0.69 dex, respectively. This is smaller than the roughly two orders of magnitude offset from the $M_{\rm BH}-\sigma_*$ relation \citep{Greene19}, which has an intrinsic scatter of $(0.58 \pm 0.09)$ dex. In addition, there are several galaxies that fall around IC 750 in the $M_{\rm BH}$ vs. $M_{\rm Bulge}$ plot. Most are galaxies from \citet{Chilingarian18}, a search for IMBHs with a mass cut of $M_{\rm BH} < 2 \times 10^5~M_\odot$, and is, therefore, missing galaxies with higher BH masses. The others are RGG 118, which has been noted to fall below the line \citep{Baldassare15}, and POX 52. Most of the new measurements from \citet{Schutte19} fall above the fitted relation. All the galaxies that fall around IC 750 have total errors on the BH mass of at least 0.3 dex.

\begin{figure}[hbt]
\begin{center}
\includegraphics[width=4.0in]{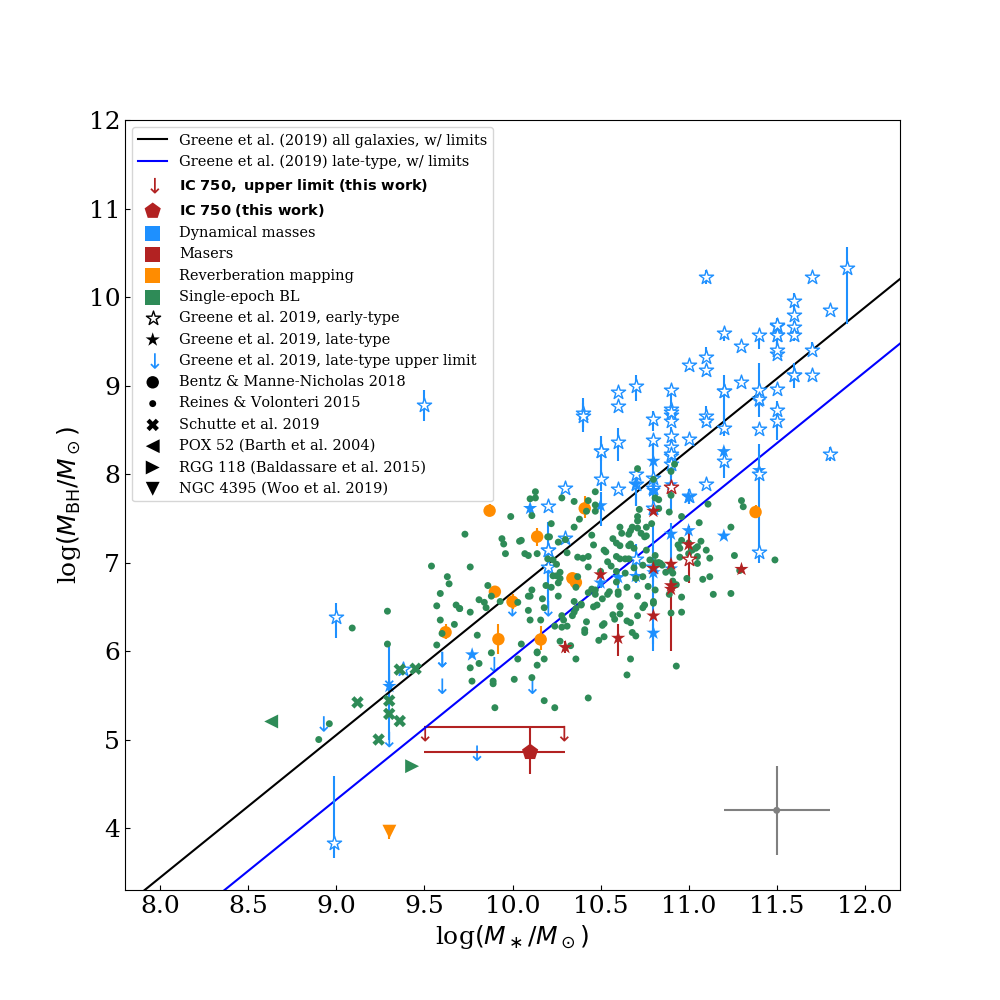}
\end{center}
\caption{\footnotesize \em{$M_{\rm{BH}}-M_*$ Relation. Both lines show scaling relations fit only for dynamically measured masses from \citet{Greene19}, black for the full sample and blue for only late type galaxies. The BH mass for IC 750 is shown as the red pentagon (error bars represent the full range of fitted masses), and the mass upper limit is shown as the solid red line with downward arrows. The central velue of $M_*$ is from our mesaurements and the error bars show the range of the measurements from the literature as discussed in section 3.4.4. The other galaxies are taken from the literature as indicated by the symbols. The colors indicate the mass measurement method. The grey cross is the representative error for the single epoch spectroscopic measurements from \citet{Schutte19} and \citet{Reines15}, both of which give a systematic error of 0.3 dex in $M_*$ and 0.5 dex in $M_{\rm{BH}}$.}}
\label{fig:mbh_mstar}
\end{figure}

Figure~\ref{fig:mbh_mstar} shows where IC 750 lies relative to the observed $M_{\rm BH}-M_*$ relation. The discrepancies in the measurements of the stellar mass for IC 750 are discussed in Section~\ref{sec:Mstar}. We use our $M_*$ measurement as the central value and use error bars to show the range of the other measurements. When comparing to the late type galaxies with dynamical BH mass measurements, including upper limits, from \citet{Greene19}, IC 750 is offset by ($-0.9\pm0.4$) dex. The relation has a reported intrinsic scatter of ($0.65\pm0.09$) dex. If the relation for late type galaxies excluding upper limits, which has an intrinsic scatter of ($0.60\pm0.08$) dex, is used, the offset increases to ($-1.4\pm0.3$) dex. When using fits to the full sample, the offset increases further as listed in Table~\ref{tab:relations}, since early type galaxies have a higher normalization than late type galaxies.

\begin{table}[htb]
\caption{Galaxies in the $M_{\rm BH}-\sigma_*$, $M_{\rm BH}-M_{\rm Bulge}$, and $M_{\rm BH}-M_*$ Relations, taken from literature}
\begin{center}
\begin{tabular}{lr@{}lccr@{}lcr@{}lccc}
        \hline
        (1) & \multicolumn{2}{c}{(2)} & (3) & (4) & \multicolumn{2}{c}{(5)} & (6) & \multicolumn{2}{c}{(7)} & (8) & (9) & (10) \\
        Name & \multicolumn{2}{c}{$M_{\rm BH}$} & Ref & Flag & \multicolumn{2}{c}{$\sigma_*$} & Ref & \multicolumn{2}{c}{$\log(M_{\rm Bulge}/M_\odot)$} & Ref & $\log(M_*/M_\odot)$ & Ref \\
         & \multicolumn{2}{c}{$(10^7~M_\odot)$} &  &  & \multicolumn{2}{c}{(km s$^{-1}$)} & & & & & &  \\
        \hline
		Milky Way & 0.43~ & $\pm0.036$ & b & D & 105~ & $\pm20$ & b & 10.09~ & $\pm0.10$ & b & $\cdots$ & $\cdots$ \\
		M32 & 0.245~ & $^{+0.101}_{-0.102}$ & a & D & 77~ & $\pm3$ & a & 9.05~ & $\pm0.10$ & b & 9.0 & a \\
		M31 & 14.3~ & $^{+9.1}_{-3.1}$ & a & D & 169~ & $\pm8$ & a & 10.35~ & $\pm0.09$ & b & 10.8 & a \\
		NGC 524 & 86.7~ & $^{+9.4}_{-4.6}$ & a & D & 247~ & $\pm12$ & a & 11.26~ & $\pm0.09$ & a & 11.2 & a \\
		NGC 821 & 16.5~ & $^{+7.4}_{-7.3}$ & a & D & 209~ & $\pm10$ & a & 10.98~ & $\pm0.09$ & a & 10.9 & a \\
		NGC 1023 & 4.13~ & $^{+0.43}_{-0.42}$ & a & D & 205~ & $\pm10$ & a & 10.53~ & $\pm0.09$ & a & 10.8 & a \\
		NGC 1068 & 0.838~ & $^{+0.045}_{-0.043}$ & a & M & 151~ & $\pm7$ & a & $\cdots$~ & $\cdots$ & $\cdots$ & 11.3 & a \\
		NGC 1194 & 7.1~ & $\pm0.3$ & a & M & 148~ & $\pm24$ & a & 10.39~ & $\pm0.20$ & l & 10.9 & a \\
		NGC 1300 & 7.55~ & $^{+7.2}_{-3.66}$ & a & D & 88~ & $\pm3$ & a & 9.84~ & $\pm0.10$ & a & 10.9 & a \\
		NGC 1277 & 1700.0~ & $^{+343.0}_{-286.0}$ & a & D & 333~ & $\pm17$ & $\cdots$ & $\cdots$~ & $\cdots$ & $\cdots$ & 11.1 & a \\
   \hline
\end{tabular}
\end{center}
\begin{tablenotes}
\item (1) Galaxy name; (2) mass of the black hole, with statistical errors; (3) reference for black hole mass; (4) flag indicating the method used to determine mass measurement, with M being H$_2$O maser dynamics, D being all other dynamical measurements, R being reverberation mapping, and B being single-epoch spectra of BLAGN; (5) stellar velocity dispersion; (6) reference for stellar velocity dispersion; (7) bulge or spheroid stellar mass; (8) reference for bulge or spheroid stellar mass; (9) log total stellar mass of the galaxy; (10) reference for log total stellar mass. References in columns (3), (6), (8), and (10) refer to (a) \citet{Greene19}, (b) \citet{KormendyHo}, (c) \citet{Bentz18}, (d) \citet{Woo15}, (e) \citet{Xiao11}, (f) \citet{Chilingarian18}, (g) \citet{Schutte19}, (h) \citet{Reines15}, (i) \citet{Barth04}, (j) \citet{Baldassare15}, (k) \citet{Woo19}, and (l) \citet{Lasker16}.
\end{tablenotes}
\label{tab:galparms}
\end{table}

\subsubsection{Implications}

The upper limit BH mass of IC 750 falls roughly two orders of magnitude below the $M_{\rm BH}-\sigma_*$ relation based on late type galaxies with dynamically measured BH masses, including upper limits at the low mass end, which has a scatter of $(0.58\pm0.09)$ dex \citep{Greene19}. IC 750 falls roughly one order of magnitude below $M_{\rm BH}-M_{\rm Bulge}$ relations: one which fits to only systems with dynamically measured masses with an intrinsic scatter of 0.29 \citep{KormendyHo} and one which fits to the dynamical sample plus additional masses from reverberation mapping and single epoch spectroscopy which has an intrinsic scatter of 0.69 dex \citep{Schutte19}. It also falls roughly one order of magnitude below $M_{\rm BH}-M_*$ relation fit to late type galaxies with dynamically measured BH masses, which has an intrinsic scatter of $(0.65\pm0.09)$ dex. There are other low mass galaxies with similar offsets relative to the $M_{\rm BH}-M_{\rm Bulge}$ and $M_{\rm BH}-M_*$ relations as IC 750. However, there are also low mass galaxies equally far from and above the lines.

There are two possible explanations why IC 750 lies well below the BH-galaxy scaling relations. The first is that there is larger scatter around the scaling relations at low masses. The second is that the scaling relations are different for low mass galaxies. Distinguishing between these two possibilities can shed light on the BH seeds and/or the growth of BHs over time. In simulations, models with heavier BH seeds predict a larger scatter around the $M_{\rm BH}-\sigma_*$ relation at $z=0$ while those with lighter Population III star seeds predict that low mass black holes would be preferentially below the relations that those with higher masses follow. 

Alternately, where the low mass BHs lie on the scaling relations can be an indication of their growth history. Many different simulations with a wide array of black hole accretion and feedback models find that black holes growth is severely limited at early times, when galaxies are small, and turns on only when the bulge or galaxy reaches a certain size \citep[e.g.,][]{BHsonFIRE, Bower17, Dekel19}. These simulations shows  ``L-shape'' tracks where the BH mass remains fairly constant relative to the bulge and stellar masses of the galaxies at low bulge and stellar masses, then grow quickly once the bulge or stellar mass reaches a certain threshold and end up on the scaling relations at larger bulge and stellar masses. These simulations predict that BHs in low mass galaxies will lie below a linear $M_{\rm BH}$ to $M_{\rm Bulge}$ or $M_*$ relation. The critical mass at which the rapid growth occurs is model dependent, e.g. $M_{\rm Bulge} \sim 10^{10}~M_\odot$ in the FIRE simulations \citep{BHsonFIRE} or $M_* \sim 10^{10.5}-10^{11}~M_\odot$ in the EAGLE \citep{Bower17} and New Horizons \citep{Dekel19} simulations. The bulge and stellar masses of IC 750 are well below the critical masses where black hole growth becomes efficient in these simulations. Tantalizingly, \citet{Lasker16} remarks that the stellar masses of maser host galaxies fall in a narrow range, $\log(M_*/M_\odot)$ = $10.5-11.1$, while the BH masses have a wide range, $\log(M_{\rm BH}/M_\odot)$ = $6.4-7.8$. More maser mass measurements in low mass galaxies, with a range of bulge and stellar masses, are needed to trace the true shape of black hole growth with bulge and galaxy masses and differentiate between models of BH seeds.

\begin{table}[htb]
\begin{center}
\caption{Offsets of the upper limit mass $M_{\rm BH}=1.4\times10^5~M_\odot$ from the BH-galaxy scaling relations taken from literature.}
\begin{tabular}{lccccccr}
\hline
Parameter & $\alpha$ & $\beta$ & $X_0$ & $\epsilon$ & Ref. & $\log(M_{\mathrm{BH}}/M_\odot)_{\mathrm{pred}}$ & $\Delta\log (M_{\mathrm{BH}}/M_\odot)$ \\
\hline
$\sigma*$ & $7.44\pm0.12$ & $3.61\pm0.50$ & 160 km s$^{-1}$ & $0.58\pm0.09$ dex & 1a & $6.9\pm0.3$ & $-1.7\pm0.3$ dex \\
$\sigma*$ & $7.40\pm0.10$ & $2.54\pm0.50$ & 160 km s$^{-1}$ & $0.50\pm0.07$ dex & 1b & $7.0\pm0.2$ & $-1.9\pm0.2$ dex \\
$\sigma*$ & $7.87\pm0.06$ & $4.55\pm0.23$ & 160 km s$^{-1}$ & $0.55\pm0.04$ dex & 1c & $7.1^{+0.2}_{-0.3}$ & $-2.0^{+0.3}_{-0.2}$ dex \\
$\sigma*$ & $7.88\pm0.05$ & $4.34\pm0.24$ & 160 km s$^{-1}$ & $0.53\pm0.04$ dex & 1d & $7.2^{+0.2}_{-0.3}$ & $-2.0^{+0.3}_{-0.2}$ dex \\
$\sigma*$ & $8.49\pm0.049$ & $4.377\pm0.290$ & 200 km s$^{-1}$ & 0.29 dex & 2 & $7.4^{+0.2}_{-0.3}$ & $-2.2^{+0.3}_{-0.2}$ dex \\
$\sigma*$ & $7.75\pm0.08$ & $3.48\pm0.21$ & 200 km s$^{-1}$ & $0.46\pm0.03$ dex & 3 & $6.9\pm0.2$ & $-1.7\pm0.2$ dex \\
$\sigma*$ & $8.33\pm0.09$ & $5.3\pm0.3$ & 200 km s$^{-1}$ & $0.46\pm0.04$ dex & 4a & $7.0\pm0.3$ & $-1.8\pm0.3$ dex \\
$\sigma*$ & $7.80\pm0.20$ & $3.4\pm0.7$ & 200 km s$^{-1}$ & $0.49\pm0.07$ dex & 4b & $6.9\pm0.4$ & $-1.8\pm0.4$ dex \\
\hline
$M_{\rm Bulge}$ & $8.69\pm0.05$ & $1.16\pm0.08$ & $10^{11}~M_\odot$ & 0.29 dex & 2 & $6.2\pm0.3$ & $-1.1\pm0.3$ dex \\
$M_{\rm Bulge}$ & $8.12\pm0.08$ & $0.88\pm0.10$ & $10^{10.66}~M_\odot$ & $0.52\pm0.06$ dex & 5 & $6.5\pm0.3$ & $-1.4\pm0.3$ dex \\
$M_{\rm Bulge}$ & $8.80\pm0.085$ & $1.24\pm0.081$ & $10^{11}~M_\odot$ & 0.69 dex & 6 & $6.1^{+0.3}_{-0.4}$ & $-1.0^{+0.4}_{-0.3}$ dex \\
\hline
$M_*$ & $6.70\pm0.13$ & $1.61\pm0.24$ & $3\times10^{10}~M_\odot$ & $0.65\pm0.09$ dex & 1e & $6.1\pm0.4$ & $-0.9\pm0.4$ dex \\
$M_*$ & $6.94\pm0.13$ & $0.98\pm0.27$ & $3\times10^{10}~M_\odot$ & $0.60\pm0.08$ dex & 1f & $6.6\pm0.3$ & $-1.4\pm0.3$ dex \\
$M_*$ & $7.43\pm0.09$ & $1.61\pm0.12$ & $3\times10^{10}~M_\odot$ & $0.81\pm0.06$ dex & 1g & $6.8\pm0.3$ & $-1.7\pm0.3$ dex \\
$M_*$ & $7.56\pm0.09$ & $1.39\pm0.13$ & $3\times10^{10}~M_\odot$ & $0.79\pm0.05$ dex & 1h & $7.0\pm0.3$ & $-1.9\pm0.3$ dex \\
$M_*$ & $8.95\pm0.09$ & $1.40\pm0.21$ & $10^{11}~M_\odot$ & 0.47 dex & 7a & $7.7\pm0.4$ & $-2.5\pm0.4$ dex \\
$M_*$ & $7.45\pm0.08$ & $1.05\pm0.11$ & $10^{11}~M_\odot$ & 0.24 dex & 7b & $6.5\pm0.3$ & $-1.4\pm0.3$ dex \\
\hline
\end{tabular}
\end{center}
\begin{tablenotes}
\item Reference numbers are as follows: 1 -- \citet{Greene19} (a) the $M_{\mathrm{BH}}-\sigma_*$ relation for late type galaxies including upper limit BH mass measurements, (b) excluding upper limits, (c), for all galaxies with dynamically measured BH masses including upper limits, (d) excluding upper limits, (e) the $M_{\mathrm{BH}}-M_*$ relation for late type galaxies including upper limits, (f) excluding upper limits, (g) for all galaxies with dynamically measured BH masses including upper limits, and (h) excluding upper limits; 2 --\citet{KormendyHo}; 3 -- \citet{Xiao11}, including only confirmed broad line AGNs; 4 -- \citet{Greene16}, (a) for early + late types, (b) for only late types; 5 -- \citet{Lasker16}; 6 -- \citet{Schutte19}; 7 -- \citet{Reines15}, (a) for ellipticals and classical bulges, (b) for AGN only.
\item Scaling relations are of the form $\log(M_{\mathrm{BH}}/M_\odot)=\alpha+\beta \log(x/X_0)$, with intrinsic scatter $\epsilon$. Error in the prediction is the result of the combination of uncertainties in the fit parameters for the relations and uncertainties in our measurements of $\sigma_* = 110 \pm 11.7~{\rm km}~{\rm s}^{-1}$, $M_{\rm Bulge} = (7.3 \pm 2.7) \times 10^8~M_\odot$, and $\log(M_*/M_\odot) = 10.1 \pm 0.2$ for IC 750.
\end{tablenotes}
\label{tab:relations}
\end{table}


\section{Conclusions}
\label{sec:conclusions}
We have constructed the first VLBI map of the circumnuclear water maser emission in IC 750, giving an unprecedented look at the subparsec-scale structure in the accreting nucleus of a low mass galaxy. The masers trace a nearly edge-on, warped disk $\sim$0.2 pc in diameter, coincident with the compact nuclear X-ray source at the base of a $\sim$kpc-scale extended emission. Systemic maser emission was not detected. Therefore, instead of knowing the BH position and velocity precisely, we can only constrain the BH position and velocity as being between the redshifted and blue-shifted emission. Furthermore, the maser emission does not fall along a single Keplerian curve, indicating additional contributions to the velocity of the maser spots, from a combination of the following possibilities: a wind, disk mass, fragmentation within the disk, a warp in the inclination angle, and disk eccentricities. An accurate modeling of these additional contributions require repeated mappings of the emission over many years. However, fitting Keplerian curves to the positions and velocities of different combinations of maser spots whose dynamics could be dominated by the gravity of the BH, and allowing for the uncertainties in the position and velocity of the BH, yield enclosed masses between $4.1 \times 10^4~M_\odot$ and $1.4 \times 10^5~M_\odot$, with the mode of the distribution at $7.2 \times 10^4~M_\odot$. All the maser emission falls below a Keplerian curve corresponding to $1.4 \times 10^5~M_\odot$, indicating that this value is a robust upper limit for the BH mass. This is currently the only IMBH and the lowest BH mass from an analysis of water maser emission, the method which yields the most accurate and precise BH masses beyond the local group and is fully independent of the $M-\sigma_*$ relation. 

In order to understand the relationship between the AGN and host galaxy, we have analyzed archival X-ray, optical, and infrared data. Fitting the optical spectrum, we obtain a stellar velocity dispersion, $\sigma_* = 110.7^{+12.1}_{-13.4}$ km s$^{-1}$. From near infrared photometry, we fit a bulge with a radius of $2\farcs{30} \pm 0\farcs{39}$, corresponding to $\sim160$ pc, and a mass of $(7.3 \pm 2.7) \times 10^8~M_\odot$. A S\`{e}rsic index $n=1.34 \pm 0.11$ indicates that it is not a classical bulge. We also derive a stellar mass of $M_* = 1.4 \times 10^{10}~M_\odot$.

The BH mass upper limit of IC 750, a spiral galaxy, falls roughly two orders of magnitude below the $M_{\rm BH}-\sigma_*$ relation derived from late type galaxies with dynamically measured BH masses which has an intrinsic scatter of $(0.58\pm0.09)$ dex \citep{Greene19}. It falls roughly one order of magnitude below the $M_{\rm BH}-M_{\rm Bulge}$ relations fit to galaxies with dynamically measured BH masses \citep{KormendyHo} and fit to galaxies with masses measured dynamically, from reverberation mapping, and single epoch spectroscopy \citep{Schutte19}, which have intrinsic scatters of 0.29 dex and 0.69 dex, respectively. It also falls roughly one order of magnitude below the $M_{\rm BH}-M_*$ relation derived from late type galaxies with dynamically measured BH masses which has an intrinsic scatter of $(0.65\pm0.09)$ dex \citep{Greene19}. These discrepancies could be due to larger scatter at the low mass end of the relations or because there is a different relation between the BH and one or more of the host galaxy parameters for low mass systems. Many simulations, with a wide variety of assumptions of black hole accretion and feedback, predict that black hole growth is largely suppressed in galaxies with bulge and stellar masses below $M_{\rm Bulge} \sim 10^{10}~M_\odot$ and $M_* \sim10^{10.5}-10^{11}~M_\odot$, well above the bulge and stellar masses of IC 750. More maser mass measurements of massive BHs in dwarf and low mass galaxies with a range of bulge and stellar masses are necessary to anchor the low mass end of the black hole-galaxy scaling relations, trace the true shape of black hole growth with bulge and galaxy masses, and constrain the range of seed BH masses.

\acknowledgements We thank the referee for his/her comments. We thank Jillian Bellovary, Daniel Angl{\'e}s-Alc{\'a}zar, Marta Volonteri, Hugo Pfister, Dmitry Makarov, and Anil Seth for helpful discussions. 

The research reported in this publication was supported by Mohammed Bin Rashid Space Centre (MBRSC), Dubai, UAE, under Grant ID number 201701.SS.NYUAD. I.K. acknowledges support by the Russian Science Foundation grant 17-72-20119. 

The National Radio Astronomy Observatory is a facility of the National Science Foundation operated under cooperative agreement by Associated Universities, Inc. This research has made use of data obtained from the Chandra Data Archive and the Chandra Source Catalog, and software provided by the Chandra X-ray Center (CXC) in the application packages CIAO, ChIPS, and Sherpa. Based on observations obtained with XMM-Newton, an ESA science mission with instruments and contributions directly funded by ESA Member States and NASA. Based on observations made with the NASA/ESA Hubble Space Telescope, obtained from the data archive at the Space Telescope Science Institute. STScI is operated by the Association of Universities for Research in Astronomy, Inc. under NASA contract NAS 5-26555. This research has made use of the NASA/ IPAC Infrared Science Archive, which is operated by the Jet Propulsion Laboratory, California Institute of Technology, under contract with the National Aeronautics and Space Administration. 

Funding for the Sloan Digital Sky Survey IV has been provided by the Alfred P. Sloan Foundation, the U.S. Department of Energy Office of Science, and the Participating Institutions. SDSS-IV acknowledges
support and resources from the Center for High-Performance Computing at
the University of Utah. The SDSS web site is www.sdss.org.

SDSS-IV is managed by the Astrophysical Research Consortium for the 
Participating Institutions of the SDSS Collaboration including the 
Brazilian Participation Group, the Carnegie Institution for Science, 
Carnegie Mellon University, the Chilean Participation Group, the French Participation Group, Harvard-Smithsonian Center for Astrophysics, 
Instituto de Astrof\'isica de Canarias, The Johns Hopkins University, Kavli Institute for the Physics and Mathematics of the Universe (IPMU) / 
University of Tokyo, the Korean Participation Group, Lawrence Berkeley National Laboratory, 
Leibniz Institut f\"ur Astrophysik Potsdam (AIP),  
Max-Planck-Institut f\"ur Astronomie (MPIA Heidelberg), 
Max-Planck-Institut f\"ur Astrophysik (MPA Garching), 
Max-Planck-Institut f\"ur Extraterrestrische Physik (MPE), 
National Astronomical Observatories of China, New Mexico State University, 
New York University, University of Notre Dame, 
Observat\'ario Nacional / MCTI, The Ohio State University, 
Pennsylvania State University, Shanghai Astronomical Observatory, 
United Kingdom Participation Group,
Universidad Nacional Aut\'onoma de M\'exico, University of Arizona, 
University of Colorado Boulder, University of Oxford, University of Portsmouth, 
University of Utah, University of Virginia, University of Washington, University of Wisconsin, 
Vanderbilt University, and Yale University.

\software{CASA (McMullin et al. 2007), SAS (v17.0; Gabriel et al. 2004), FTOOLS (Blackburn 1995), CIAO (v4.8.2; Fruscione et al. 2006), spec (v12.10.1; Arnaud 1996), lmfit (Newville et al. 2016), PEGASE (Le Borgne et al. 2004),  matplotlib (v3.1.1; Hunter 2007)}

\appendix
In the following appendices we give details of the series of maser spots excluded from and included in the BH mass fits. In addition to the variations in mass fitting described in Section~\ref{sec:BHmass}, we examined additional fit requirements. The results are consistent with those given in Section~\ref{sec:BHmass}. We also present a toy model to assess the effects of inclination and aperture size for the measurement of stellar velocity dispersion.

\section{Position-Velocity Structures of Series of Maser Spots}
\label{ap:regions}
\begin{figure}[htb]
\begin{center}
\includegraphics[width=3.5in]{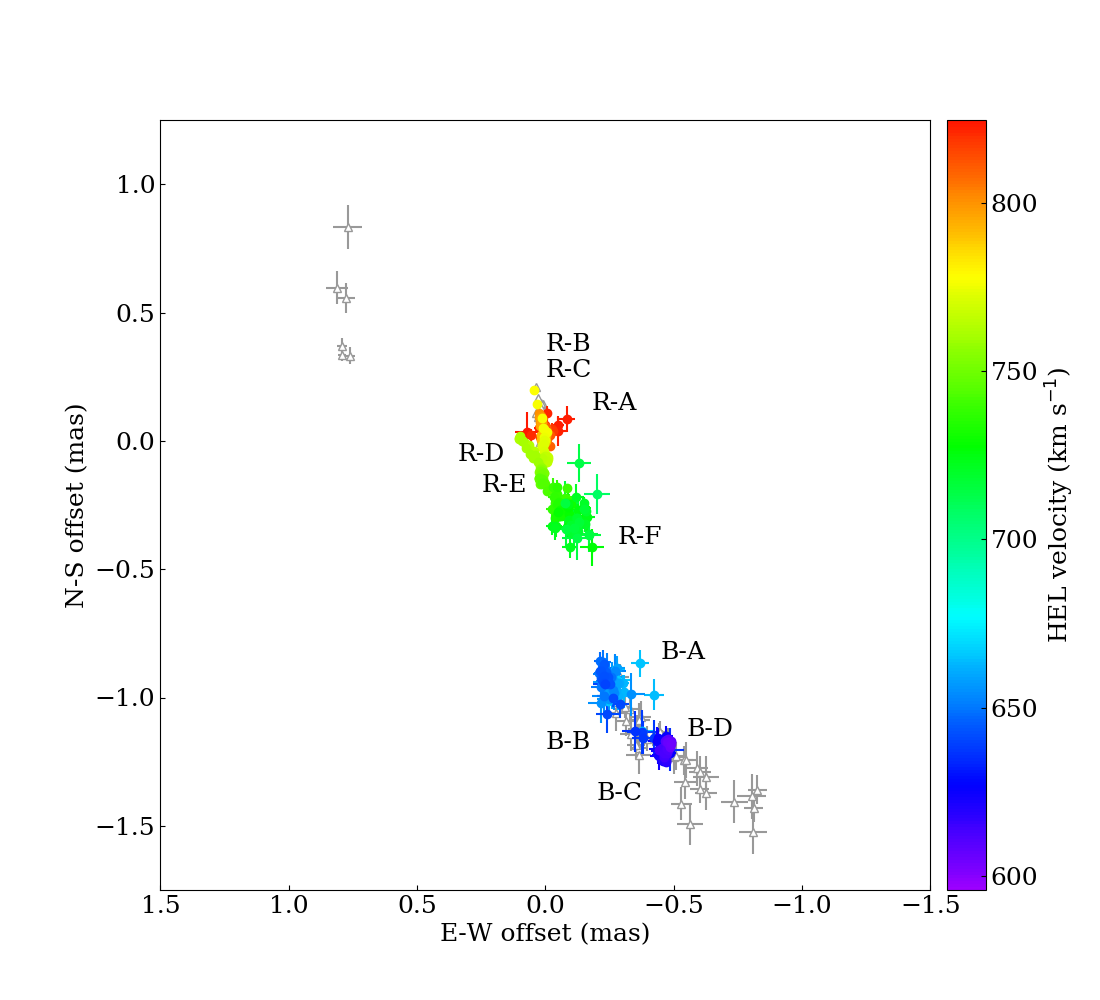}
\includegraphics[width=3.5in]{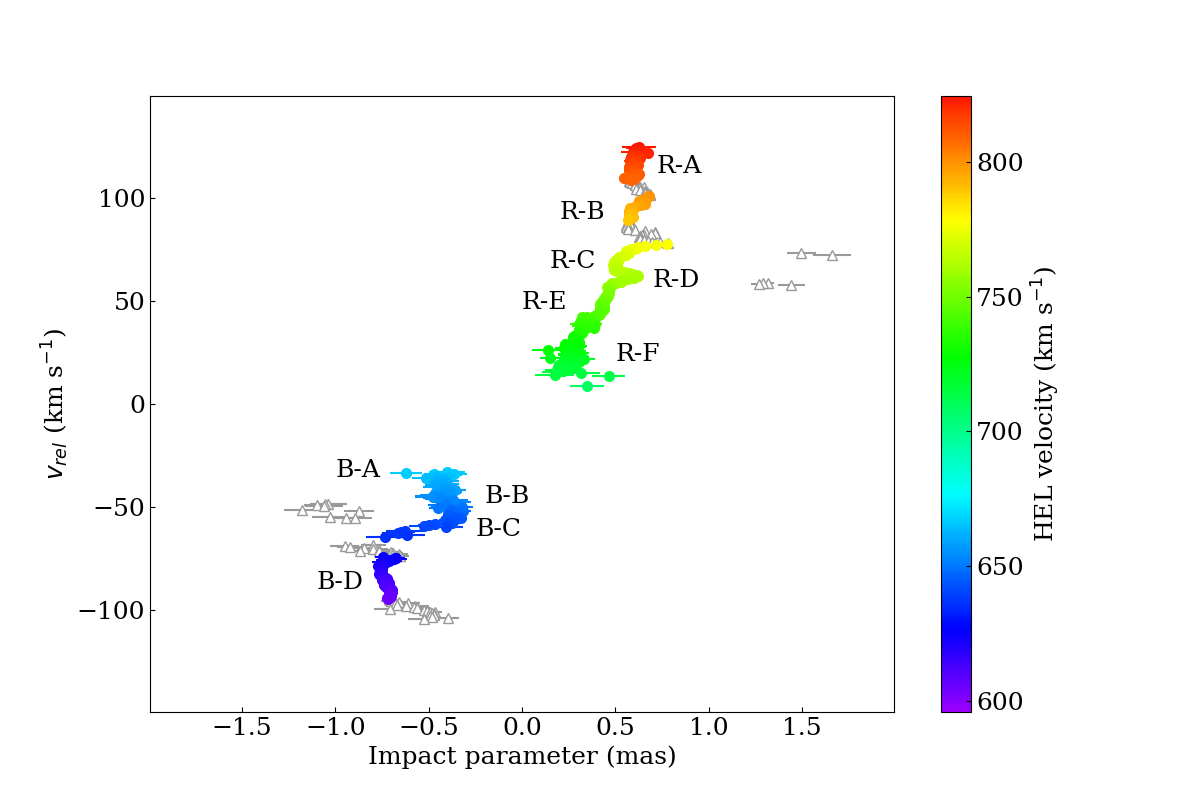}
\end{center}
\caption{\footnotesize \em{Map (left) and PV diagram (right) for the series of maser spots which are not consistent with being in the high velocity regions of the disk. In both plots, the filled circles indicate maser spots not used in the fits for BH mass and the grey triangles indicate maser spots used in the fits. The colors of the filled circles indicate the line-of-sight velocities of the maser spots excluded from the fits. For the PV diagram, the positions are relative to the center of the grid of putative BH positions and the velocities are relative to 701 km s$^{-1}$, the systemic velocity of the galaxy.}}
\vspace{-0.3cm}
\label{fig:nonkep}
\end{figure}

\begin{figure}[htb]
\begin{center}
\includegraphics[width=\textwidth]{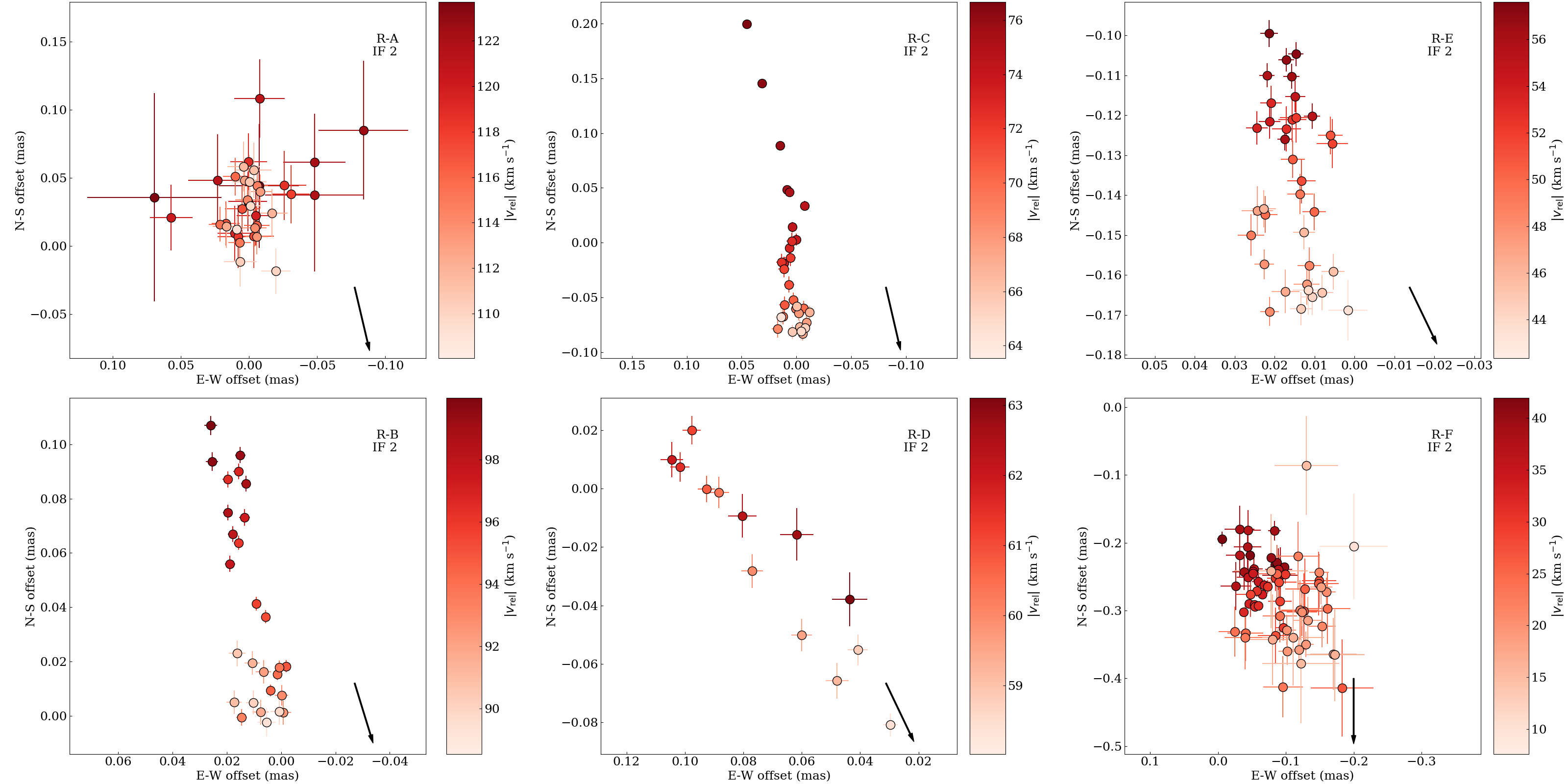}
\end{center}
\caption{\footnotesize \em{Maps of each series of redshifted maser spots not used in the fits for BH mass. The labels correspond to those in Figure~\ref{fig:kep}. The colors of the spots indicate their $|v_{rel}|$, the absolute value of the difference between the velocity of the maser spot and 701 km s$^{-1}$, the systemic velocity of the galaxy. The scale of each map is set by the velocity range of the maser spots in the map. The arrow points towards the center of the grid of the putative positions of the BH. In each series, the relative velocities of the maser spots increases, i.e. the markers get darker, with increasing distance from the dynamical center or show no correlation with distance.}}
\vspace{-0.3cm}
\label{fig:cutout_red_nonkep}
\end{figure}

\begin{figure}[htb]
\begin{center}
\includegraphics[width=\textwidth]{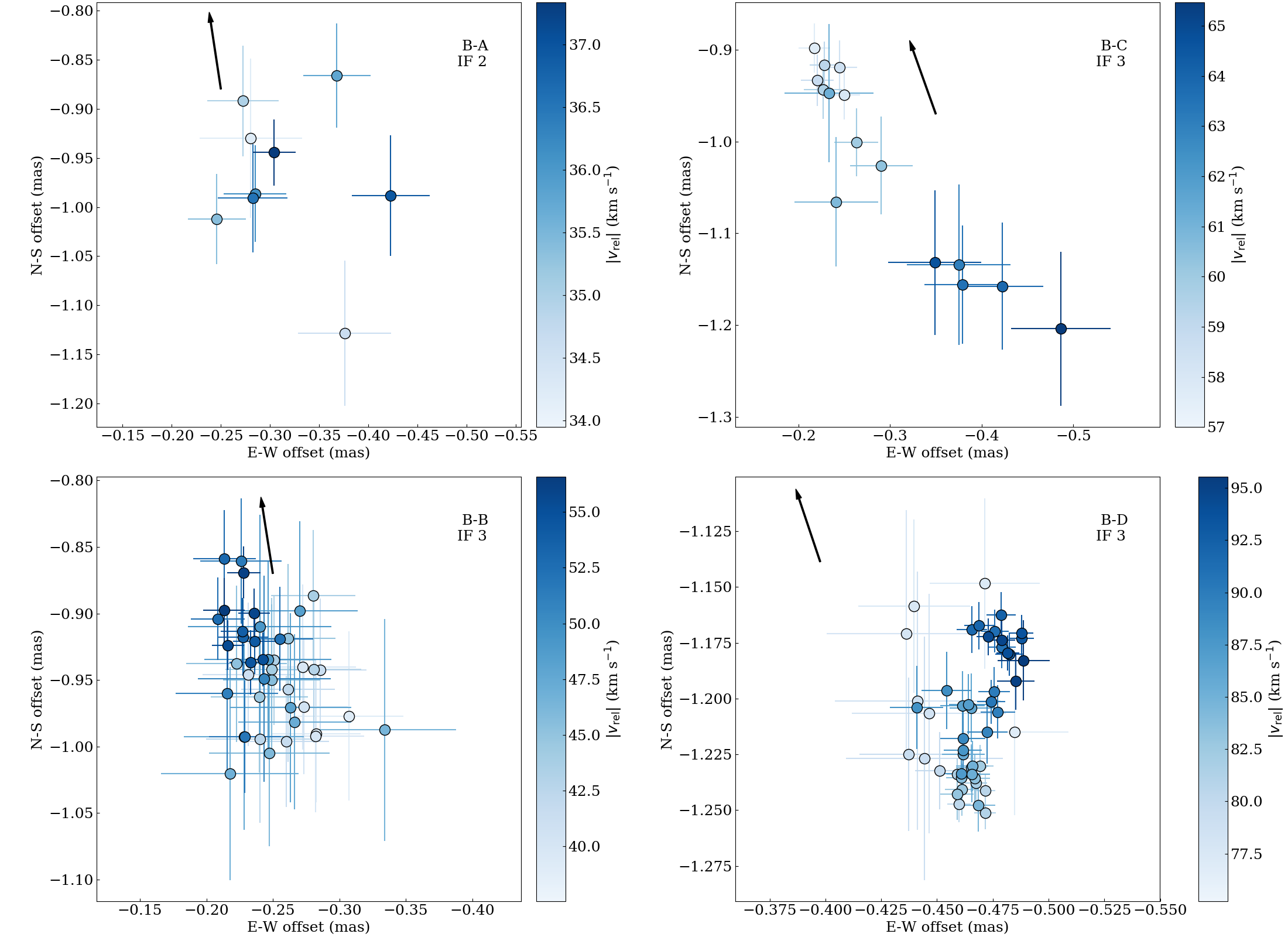}
\end{center}
\caption{\footnotesize \em{Maps of each series of blue-shifted maser spots used in the fits. The labels correspond to those in Figure~\ref{fig:kep}. The colors of the spots indicate their $|v_{rel}|$, the absolute value of the difference between the velocity of the maser spot and 701 km s$^{-1}$, the systemic velocity of the galaxy. The scale of each map is set by the velocity range of the maser spots in the map. The arrow points towards the center of the grid of the putative positions of the BH. In each series, the relative velocities of the maser spots increase, i.e. the markers get darker, with increasing distance from the dynamical center or shows no trend with distance.}}
\vspace{-0.3cm}
\label{fig:cutout_blue_nonkep}
\end{figure}

As discussed in Section~\ref{sec:massfits}, we analyze the position-velocity structure of series of maser spots by examining the maser spots in consecutive frequency/velocity channels. The high velocity maser spots in a disk should have decreasing relative velocities with increasing distance from the dynamical center. Series of maser spots which have increasing relative velocity with increasing distance or those which show no relating between relative velocity and distance are physically inconsistent with being in the high velocity regions of the disk and are excluded from the fits. Figure~\ref{fig:nonkep} shows the map and PV diagram of the series of maser spots excluded from the fits as filled circles, with colors indicating their line-of-sight velocities. Each of the excluded series of maser spots are labeled in the map and diagram. Figures~\ref{fig:cutout_red_nonkep} and~\ref{fig:cutout_blue_nonkep} show each individual series in a map. In these figures, the maser spots are colored according to their $|v_{rel}|$, the absolute value of the difference between the velocity of the maser spot and 701 km s$^{-1}$, the systemic velocity of the galaxy. The dynamic range of the colorbar in each map is set by the range of the $|v_{rel}|$ of the maser spots in the map to make the velocity structure clearer. An arrow shown on each map points towards the center of the grid of possible BH positions. 

\begin{figure}[htb]
\begin{center}
\includegraphics[width=3.5in]{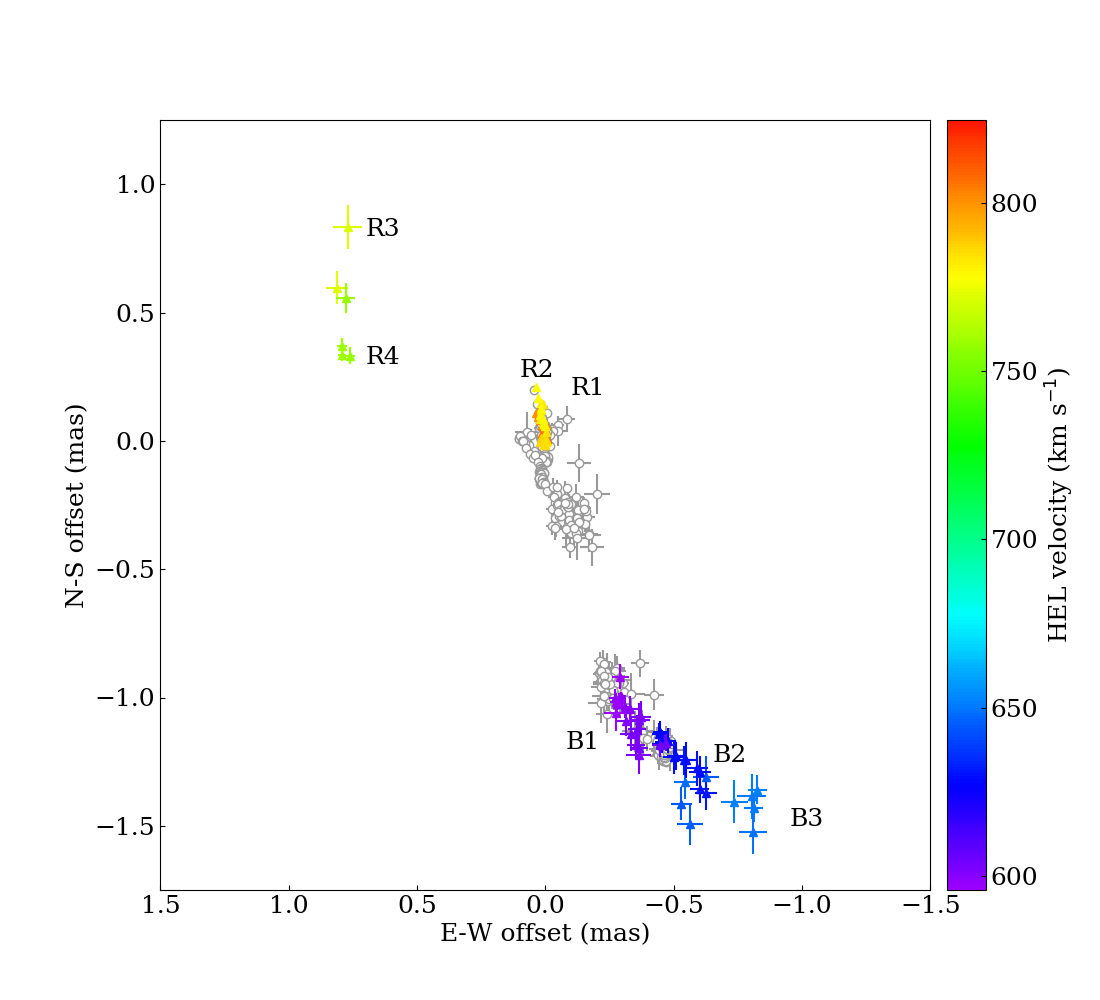}
\includegraphics[width=3.5in]{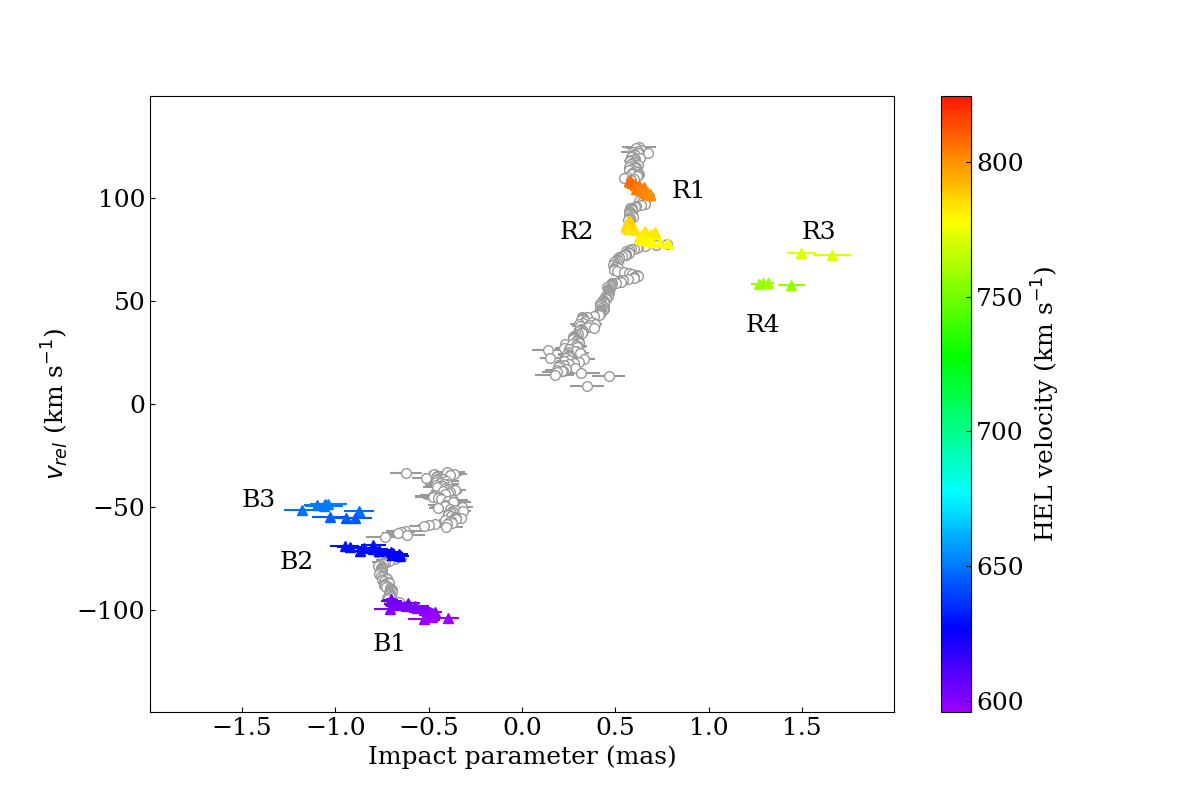}
\end{center}
\caption{\footnotesize \em{Map (left) and PV diagram (right) for the series of maser spots which are consistent with being in the high velocity regions of the disk. In both plots, the grey circles indicate maser spots not used in the fits for BH mass and the filled triangles indicate maser spots used in the fits. The colors of the filled triangles indicate the line-of-sight velocities of the maser spots. For the PV diagram, the positions are relative to the center of the grid of putative BH positions and the velocities are relative to 701 km s$^{-1}$, the systemic velocity of the galaxy.}}
\vspace{-0.3cm}
\label{fig:kep}
\end{figure}

\begin{figure}[htb]
\begin{center}
\includegraphics[width=5.5in]{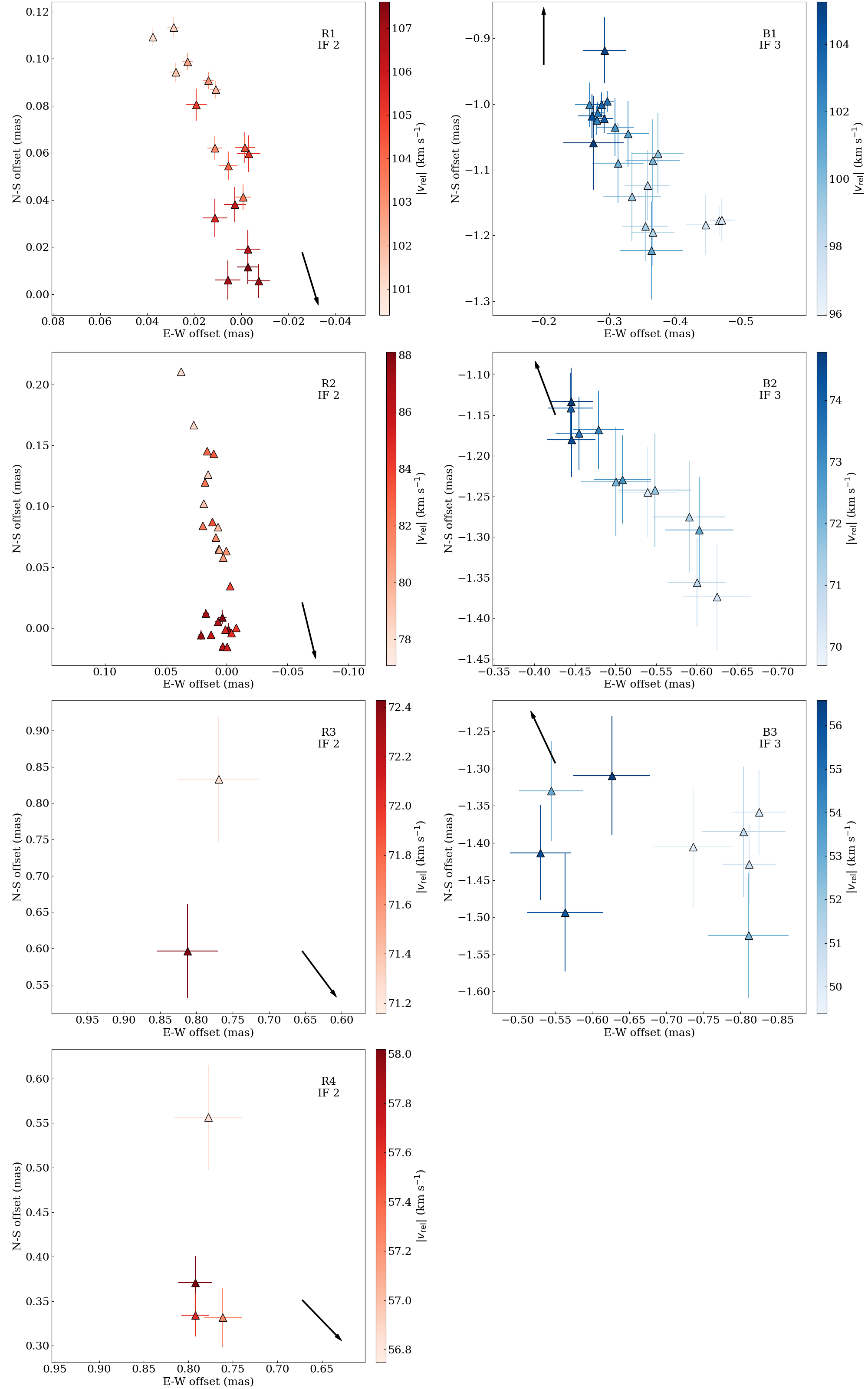}
\end{center}
\caption{\footnotesize \em{Maps of each series of redshifted (left) and blue-shifted (right) maser spots used in the fits for BH mass. The labels correspond to those in Figure~\ref{fig:kep}. The colors of the spots indicate their $|v_{rel}|$, the absolute value of the difference between the velocity of the maser spot and 701 km s$^{-1}$, the systemic velocity of the galaxy. The scale of each map is set by the velocity range of the maser spots in the map. The arrow points towards the center of the grid of the putative positions of the BH. In each series, the relative velocities of the maser spots decreases, i.e. the makers get lighter, with increasing distance from the dynamical center.}}
\vspace{-0.3cm}
\label{fig:cutout_kep}
\end{figure}

The map and PV of the series of maser spots which do have decreasing velocity with distance and are included in the fits for BH mass are shown in Figure~\ref{fig:kep} as filled triangles and labeled. The color of each included maser spot in Figure~\ref{fig:kep} indicates its velocity. The map of each individual series is shown in Figure~\ref{fig:cutout_kep}. In the individual maps, the maser spots are colored according to their $|v_{rel}|$, the absolute value of the difference between the velocity of the maser spot and 701 km s$^{-1}$, the systemic velocity of the galaxy. The dynamic range of the colorbar in each map is set by the range of the $|v_{rel}|$ of the maser spots in the map. An arrow shows on each map points towards the center of the grid of possible BH positions. 

A few spots within the series might not follow the trend of increasing/decreasing velocity with distance. However, the series of spots are excluded or included as a whole, based on the overall trend, to avoid biases in choice of maser spots. Some series of spots, e.g. B3, are well separated in position and/or velocity but many others are not. For example, there is no clear separation in velocity between R2, a series used in the BH mass fits, and R-B, a series not used in the BH mass fits. The boundaries were decided before mass fitting so as to not bias the resultant masses. After the fits, we varied the boundary by one to three points and found that it causes a difference of only a few percent in $M_{red}$ or $M_{blue}$ and $\lesssim$1\% difference in the combined BH mass from the fits.

\section{Disk Mass}
\label{ap:diskmass}
If we know the true position and velocity of the BH and the mass of the maser disk is negligible relative to the mass of the BH, the mass derived from the redshifted maser emission, $M_{red}$, and from the blue-shifted maser emission, $M_{blue}$, should be equal, i.e. $M_{red}/M_{blue} = 1$. However, in our mass fits, we vary the putative position and velocity of the BH, since we do not have systemic maser emission to pinpoint the BH, assuming that the BH is centered with respect to the disk which orbits it. 

If the assumed velocity is correct but the position is shifted towards the redshifted emission, $M_{red}$ will decrease and $M_{blue}$ will increase, proportional to the position; the opposite will happen if the position is shifted towards the blue-shifted emission. Similarly, if the assumed position is correct but the velocity is shifted redward, $M_{red}$ will decrease and $M_{blue}$ will increase, proportional to the square of the velocity. Therefore, an error in position can be offset by an error in velocity (i.e. shifting the position towards the redshifted emission and the velocity towards the blue-shifted emission or vice-versa) to give $M_{red}/M_{blue}=1$, depending on the relative shifts in position and velocity. The gap in position in between the redshifted and blue-shifted maser emission is $\sim$10\% of the size of the disk and we vary the relative velocity by $\sim$4\%. Ignoring the slight asymmetry of the redshifted and blue-shifted emission around the systemic velocity of the galaxy, assuming that $M_{red,center}/M_{blue,center}=1$ from a position and velocity at the center of the allowed positions and velocities, and shifting the position maximally towards the redshifted emission and the velocity maximally towards the blue shifted emission, we get $M_{red,new}/M_{red,center} = 0.9*1.04^2 = 0.97$, $M_{blue,new}/M_{blue,center} = 1.1*0.96^2 = 1.01$, and $M_{red,new}/M_{blue,new}=0.96$. Shifting in the opposite direction gives $M_{red,new}/M_{blue,new}=1.04$. Therefore, we allow the BH mass fits to have $0.95 \leq M_{red}/M_{blue} \leq 1.05$.

In addition, the disk in IC 750 could be massive relative to the BH. For maser systems with geometrically thin disks, reported disk masses which are negligible relative to the BH mass for NGC 4945 \citep{Herrnstein99}, $\sim$40\% of the BH mass for Circinus \citep{Greenhill03}, and comparable to the BH mass for NGC 3393 \citep{Kondratko08} and NGC 1068 \citep{Hure02, Lodato03}. The reported disk mass for NGC 3079 is $\sim$3.5 times the mass of the BH but it does not apply to IC 750 since the disk in NGC 3079 is thick and more disrupted. In a standard thin disk model \citep{SS73}, the density decreases as $r^{-15/8}$. The area of the disk increases as $r^2$. Assuming a uniform thickness for the maser disk, the mass enclosed scales with radius as $r^{1.25}$. The redshifted maser emission extends out to a radius of $\sim$1.7 mas while the blue-shifted emission extends to a radius of $\sim$1.2 mas, roughly a factor of 1.4. Therefore, the redshifted emission in IC 750 will enclose, relative to the blue-shifted emission, the same mass if the ratio of disk to BH mass is negligible, $\sim$10\% higher mass if the ratio of disk mass to BH mass is $\sim$40\%, and 25\% higher mass if the disk mass and BH mass are roughly equal. 

\begin{figure}[htb]
\begin{center}
\includegraphics[width=3.5in]{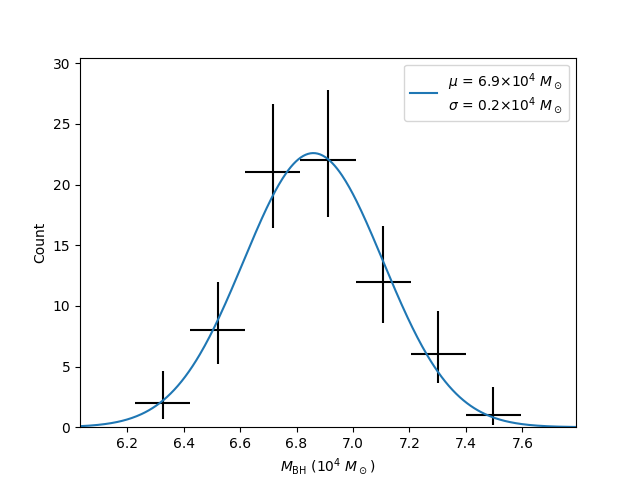}
\includegraphics[width=3.5in]{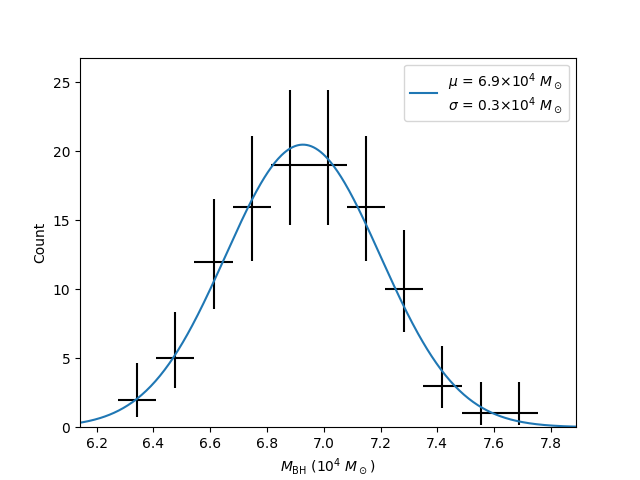}
\end{center}
\caption{\footnotesize \em{Left: The distribution of masses from the fits from the fits where the BH velocity and position are varied and the ratio of the masses from the redshifted and blue-shifted maser emission are required to be $0.95 < M_{red}/M_{blue} < 1.05$. Right: With the requirement $0.95 < M_{red}/M_{blue} < 1.10$}}
\vspace{-0.3cm}
\label{fig:masshists_5_10}
\end{figure}

In our main analysis, we allowed the disk in the maser radii range to be as massive as the BH, the maximum reported for a thin disk system, i.e. $0.95 < M_{red}/M_{blue} < 1.25$. Here, we check the results for different assumptions of disk mass. We repeat the exercise of allowing the position and velocity of the BH to vary and require $0.95 \leq M_{red}/M_{blue} \leq 1.05$ assuming negligible disk mass and $0.95 \leq M_{red}/M_{blue} \leq 1.10$, assuming the disk mass is $\sim$40\% of the BH mass. The distributions of the masses from the acceptable fits are shown in Figure~\ref{fig:masshists_5_10}. The means of the mass distributions are consistent with the results from the main analysis and the widths of the Gaussian distributions are narrower, as expected.

\section{No $M_{red}-M_{blue}$ Agreement}
\label{sec:masscut}
\begin{figure}[htb]
\begin{center}
\includegraphics[width=3.5in]{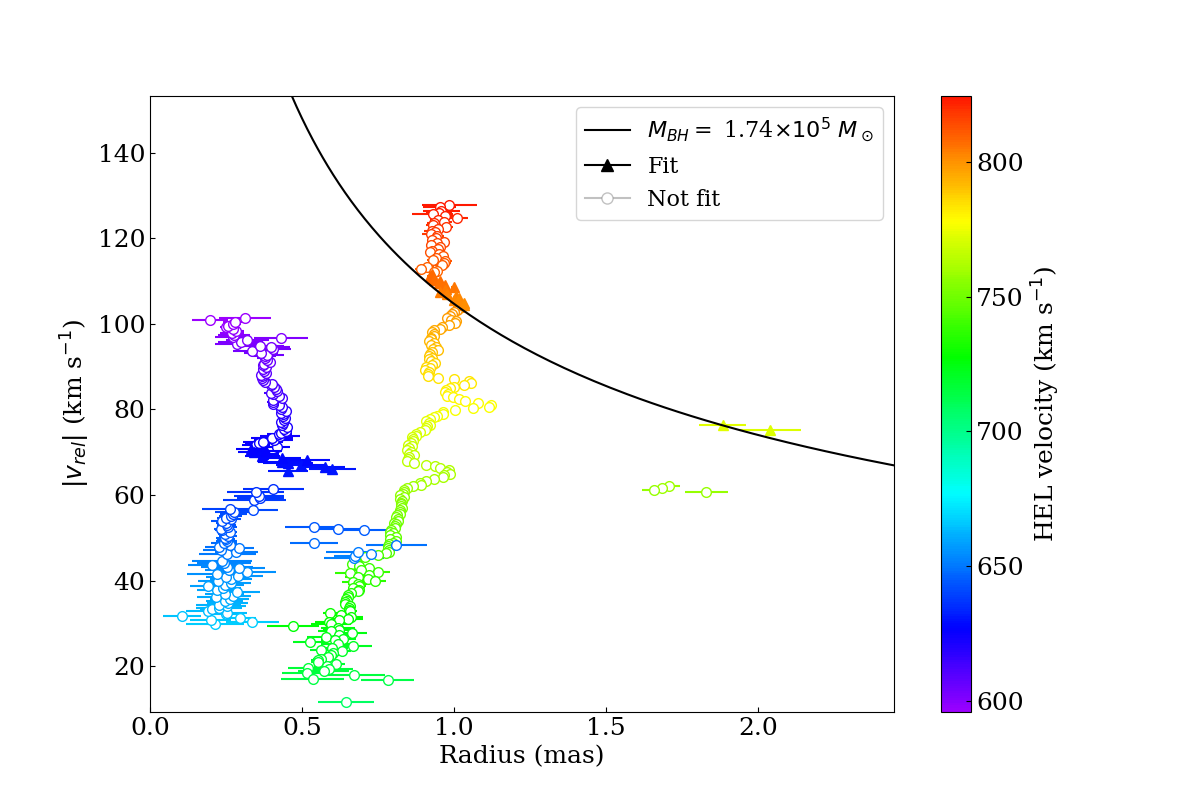}
\includegraphics[width=3.5in]{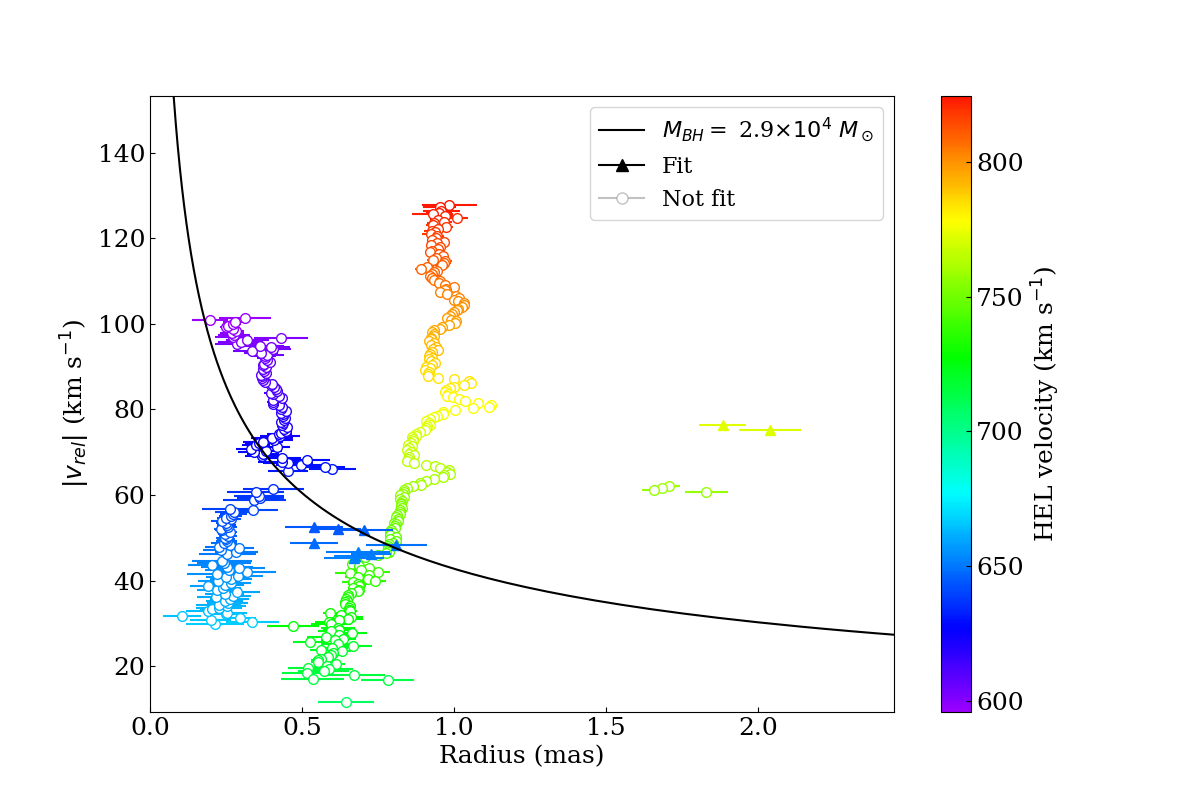}
\end{center}
\caption{\footnotesize \em{PV diagrams for the Keplerian fits when varying the sets of maser spots used in the fits at the same time as varying the BH position and velocity. In both plots, the open circles indicate maser spots not used in the fit and the solid triangles indicate maser spots used in the fit. The colors indicate the line-of-sight velocities of the maser spots. The radius is the distance from the putative BH position to the maser spot and the relative velocity, $v_{rel}$, is the difference between the putative BH velocity and the maser line-of-sight velocity. Left: The PV diagram where the fit yields the maximum mass. Right: The PV diagram where the fit yields the minimum mass.}}
\vspace{-0.3cm}
\label{fig:varyallanymax}
\end{figure}

In our main analysis, the highest fitted mass were obtained by varying which series of maser spots were used at the same time as varying the BH position and velocity, requiring $0.95 < M_{red}/M_{blue} < 1.25$. We examine what happens when no agreement is required between the masses from the masers on the two sides of the disk. The highest mass obtained is $1.74 \times 10^5~M_\odot$. The PV diagram is shown in the left panel of Figure~\ref{fig:varyallanymax}. The fitted curve completely misses the blue-shifted maser emission and $M_{red}$ is 5.8 times more massive than $M_{blue}$. Furthermore, the systemic part of the blue-shifted emission is highly shifted in (absolute) position with respect to the systemic part of the redshifted emission (when masses are required to agree the systemic emission from the two sides are aligned). All these facts indicate that this fit is unphysical. Even so, this maximum mass is less than 50\% higher than the physically well motivated upper limit discussed in Section~\ref{sec:upperlimit}. Similarly, the lowest mass from the fits, $2.9 \times 10^4~M_\odot$ results from a fit that completely misses the redshifted emission with $M_{red}$ 7.3 times larger than $M_{blue}$, and a PV diagram where the redshifted and blue-shifted systemic emission are offset from each other. We, therefore, conclude that a requirement for agreement between the masses from the two sides is necessary and we adopt the one based on the maximum reported ratio between the masses of the disk and the BH, for a geometrically thin maser disk.

\section{Widening the Possible Velocity Range for the BH}
\begin{figure}[htb]
\begin{center}
\includegraphics[width=2.25in]{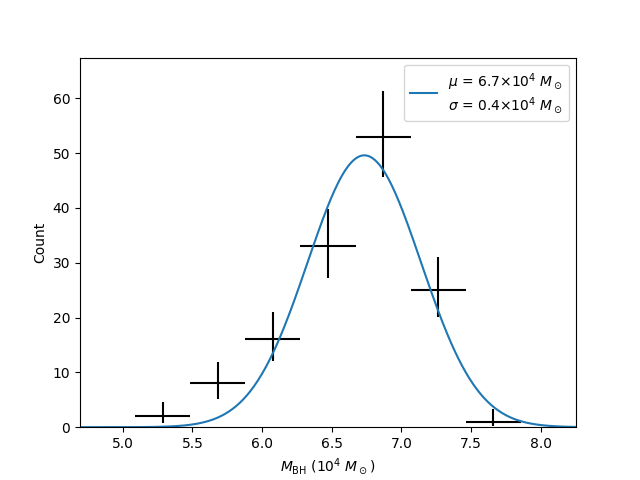}
\includegraphics[width=2.25in]{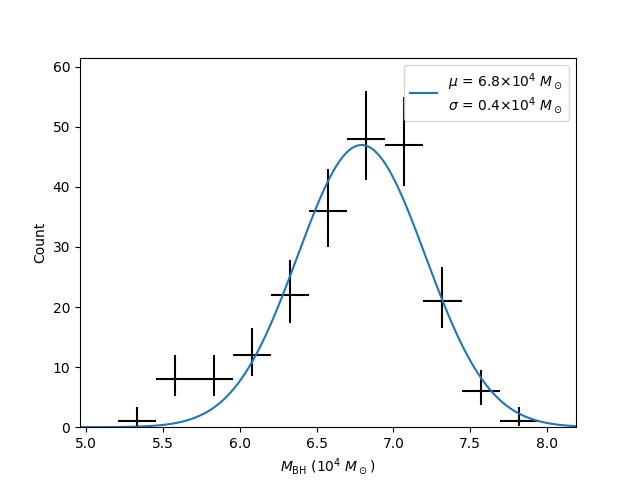}
\includegraphics[width=2.25in]{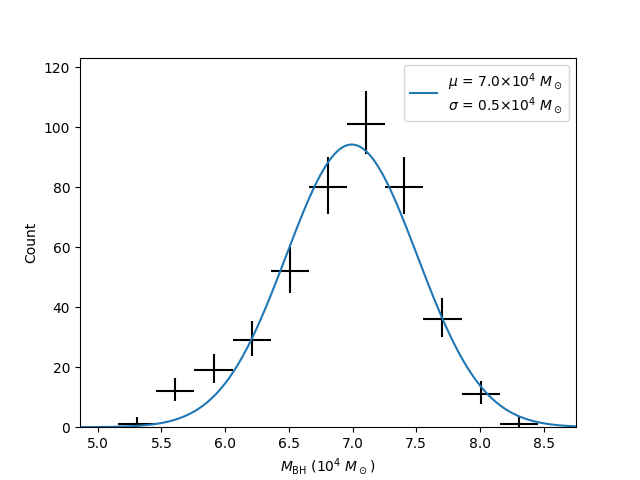}
\end{center}
\caption{\footnotesize \em{The distribution of masses from the fits where the BH position is varied of the same grid but the varied velocity range is increased to 667 km s$^{-1}$ $<~v~<$ 708 km s$^{-1}$. Left: Accepted fits with the mass agreement requirement of $0.95 < M_{red}/M_{blue} < 1.05$. Middle: Accepted fits with the mass agreement requirement of $0.95 < M_{red}/M_{blue} < 1.10$. Right: Accepted fits with the mass agreement requirement of $0.95 < M_{red}/M_{blue} < 1.25$.}}
\vspace{-0.3cm}
\label{fig:masshists_extendv}
\end{figure}

We have also tried increasing the possible velocity range for the BH. In this study, we keep the same grid for the possible BH location as in Section~\ref{sec:BHmass} but allow the velocity to vary between the lowest velocity redshifted emission and highest velocity blue-shifted emission, namely 667 km s$^{-1}$ $<~v~<$ 708 km s$^{-1}$. We again require $0.95 < M_{red}/M_{blue} < 1.05$, $0.95 < M_{red}/M_{blue} < 1.10$, $0.95 < M_{red}/M_{blue} < 1.25$. The velocity range of the viable fits is 687 km s$^{-1}$ to 708 km s$^{-1}$ for the up to 5\% and 10\% mass differences and 685 km s$^{-1}$ to 708 km s$^{-1}$ for the up to 25\% mass difference. The histograms for the masses from these fits are shown in Figure~\ref{fig:masshists_extendv}. The means of the Gaussian fits and the peaks are consistent with the means from our standard fits. However, the mass distributions are skewed towards lower masses.

\section{Additional maser spots in the Fit}
\begin{figure}[htb]
\begin{center}
\includegraphics[width=3.5in]{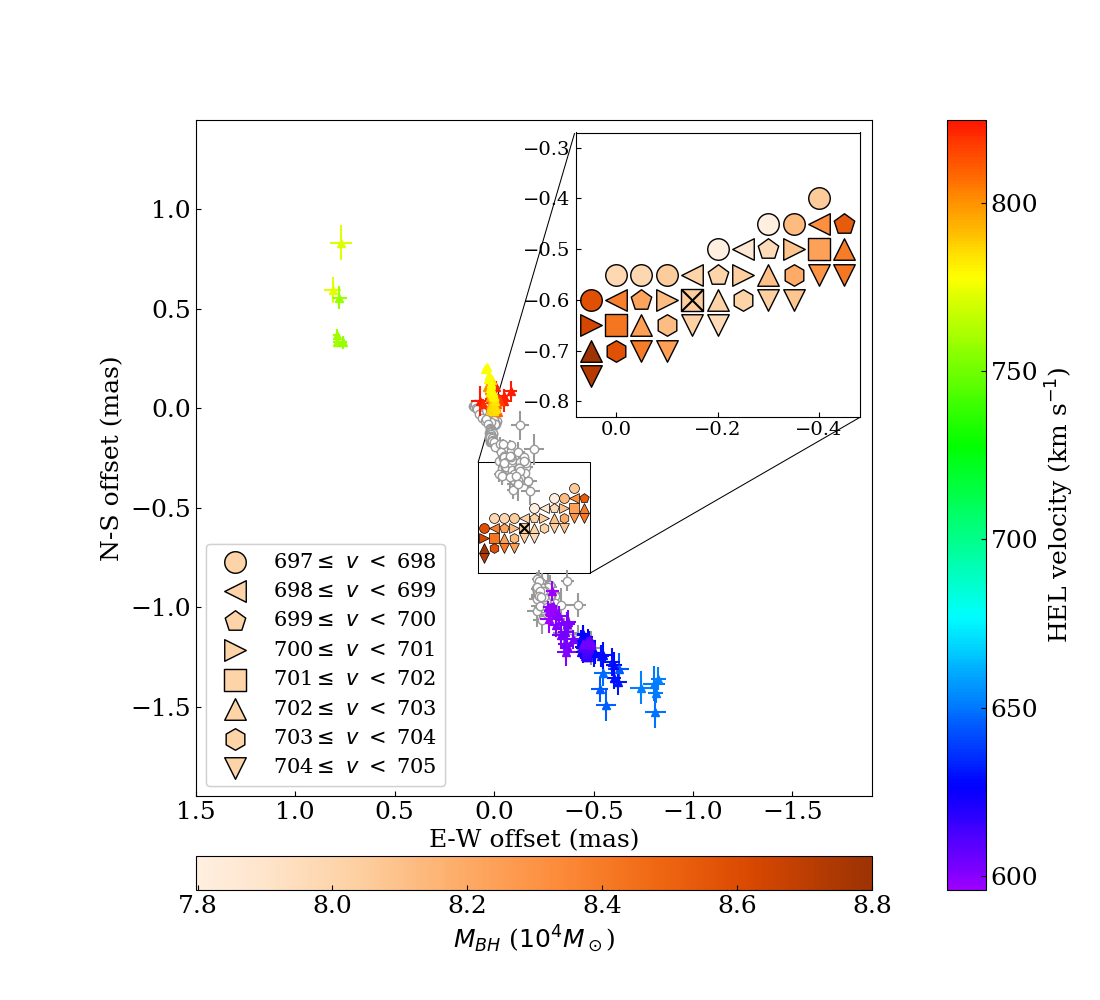}
\includegraphics[width=3.5in]{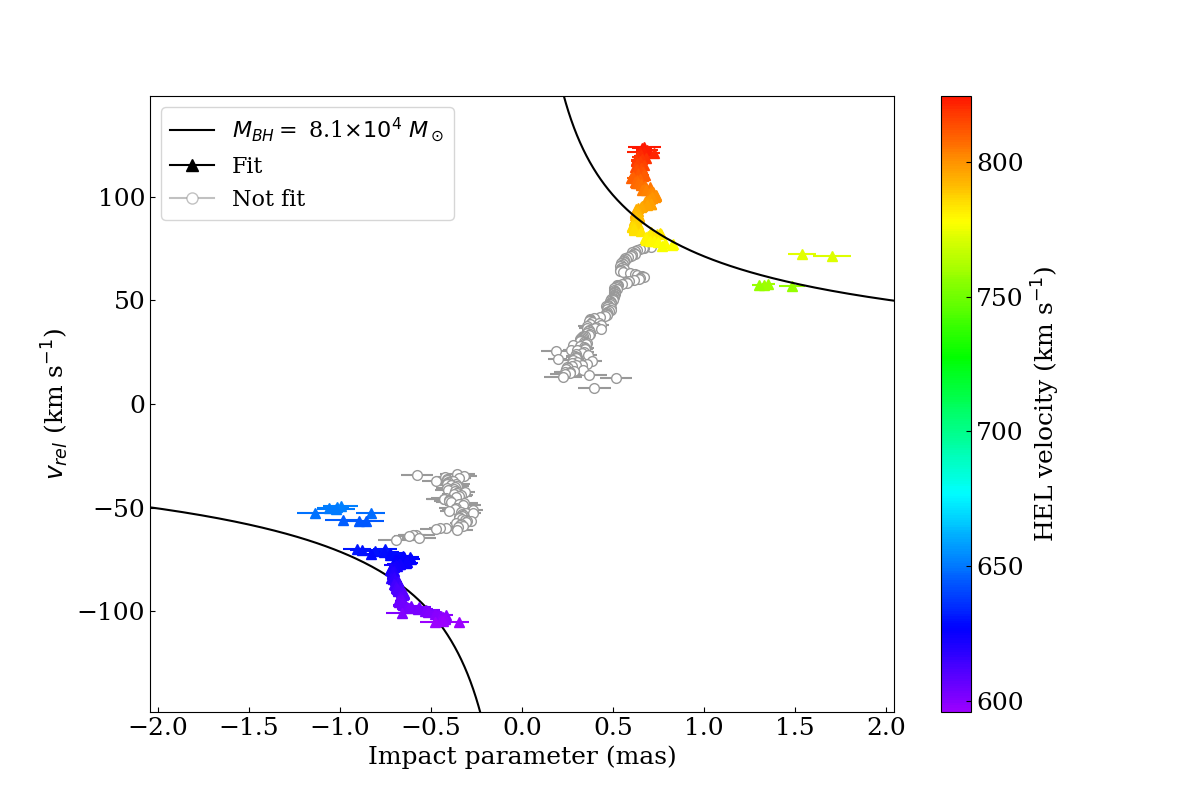}
\end{center}
\caption{\footnotesize \em{Map and PV diagram for the Keplerian fits. In both plots, the open circles indicate maser spots not used in the fit and the solid triangles indicate maser spots used in the fit. The colors of the solid triangles indicate the line-of-sight velocities of the maser spots. Left: The map of the viable positions and velocities for the central black hole. The 0.5 mas $\times$ 0.5 mas box outlined in black shows the region over which we vary the putative position of the BH, in steps of 0.05 mas. We also vary the putative velocities between 697 km s$^{-1}$ and 705 km s$^{-1}$, in steps of 1 km s$^{-1}$. A fit is considered to be acceptable if $0.95 < M_{red}/M_{blue} < 1.05$ and the BH position and velocity are considered viable. The symbols indicate the viable velocity at the given location. Each position has one to three viable velocities. For the positions with more than one viable velocity, we average the velocities and the resulting masses. The orange color bar indicates the averaged viable mass at each position. Right: The PV diagram for the BH location at an East-West offset of -0.15 mas and a North-South offset of -0.6 mas, marked with a black cross on the map, and a velocity of 701 km s$^{-1}$. The impact parameter is the great-circle angular distance from (-0.15, -0.6) and the relative velocity, $v_{rel}$, is the difference between 701 km s$^{-1}$ and the maser line-of-sight velocity. The line is the Keplerian fit of the maser spots. This position and velocity gives a fitted mass of $(8.1 \pm 0.1) \times 10^4~M_\odot$.}}
\vspace{-0.3cm}
\label{fig:bestfitmass_allhv}
\end{figure}

\begin{figure}[htb]
\begin{center}
\includegraphics[width=2.25in]{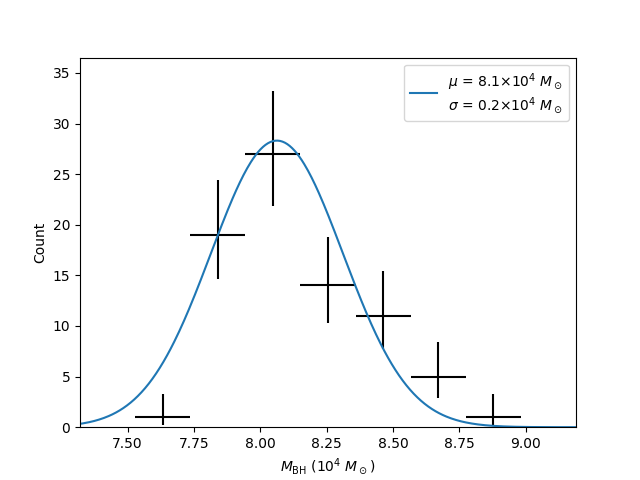}
\includegraphics[width=2.25in]{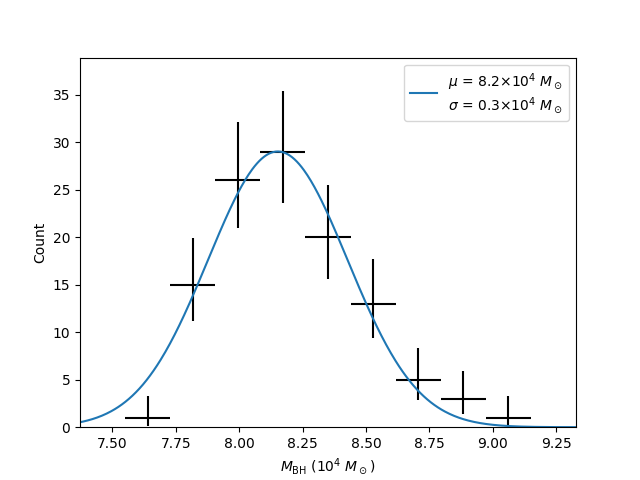}
\includegraphics[width=2.25in]{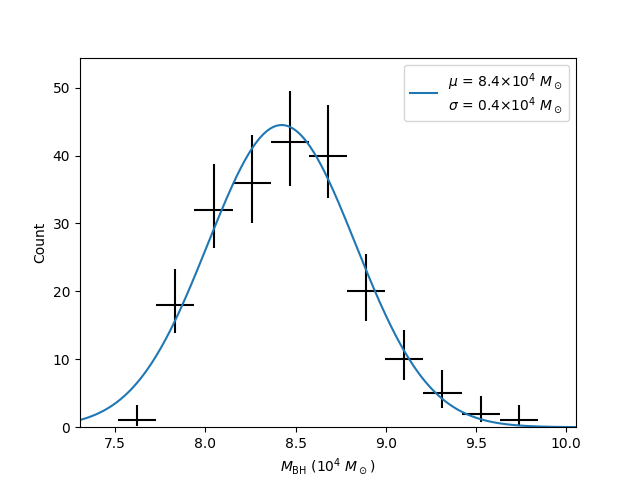}
\end{center}
\caption{\footnotesize \em{Distribution of the masses from the fits with additional maser spots. Left: Accepted fits with the mass agreement requirement of $0.95 < M_{red}/M_{blue} < 1.05$. Middle: Accepted fits with the mass agreement requirement of $0.95 < M_{red}/M_{blue} < 1.10$. Right: Accepted fits with the mass agreement requirement of $0.95 < M_{red}/M_{blue} < 1.25$.}}
\vspace{-0.3cm}
\label{fig:masshists_allhighv}
\end{figure}

We also tested the fits by adding all the blue-shifted maser spots with relative velocities greater than B2 and all the redshifted maser spots with relative velocities greater than R3. The left plot in Figure~\ref{fig:bestfitmass_allhv} shows the map of the viable BH positions and velocities with the corresponding masses requiring $0.95 < M_{red}/M_{blue} < 1.05$. The right plot shows the PV diagram for (-0.15 mas, -0.55 mas; 700 km s$^{-1}$), with the position marked on the map with a black cross, which yields a mass of $8.1 \times 10^4~M_\odot$. The mean of the distribution masses from these fits, shown in the left panel of Figure~\ref{fig:masshists_allhighv}, is higher than when only the maser spots consistent with being in a disk are kept but is within 1.5 $\sigma$ from our measured mass in Section~\ref{sec:BHmass}. Note that the mass is artificially high due to the highest velocity redshifted maser spots which are inconsistent with being in a Keplerian disk since there is no correlation between their positions and velocities, and may be maser emission in a wind. Figure~\ref{fig:masshists_allhighv} shows the mass distributions for mass agreement requirements of $0.95 < M_{red}/M_{blue} < 1.05$, $0.95 < M_{red}/M_{blue} < 1.10$, and $0.95 < M_{red}/M_{blue} < 1.25$. Unlike the mass distributions for the physically well-motivated choice of high velocity emission, these distributions are skewed towards higher masses, but the fitted widths are comparable to those in Section~\ref{sec:BHmass} and Appendix~\ref{sec:masscut}, indicating that the error due to the uncertainty in BH position and velocity is not highly sensitive to the choice of maser spots used in the fits. In addition, the masses from these fits never exceed $9.6 \times 10^4~M_\odot$, indicating that the upper limit in Section~\ref{sec:BHmass} is robust.

\section{Stellar Velocity Dispersion Measurement Toy Model}
\label{ap:sigma}
\begin{figure}[htb]
\begin{center}
\includegraphics[width=\textwidth]{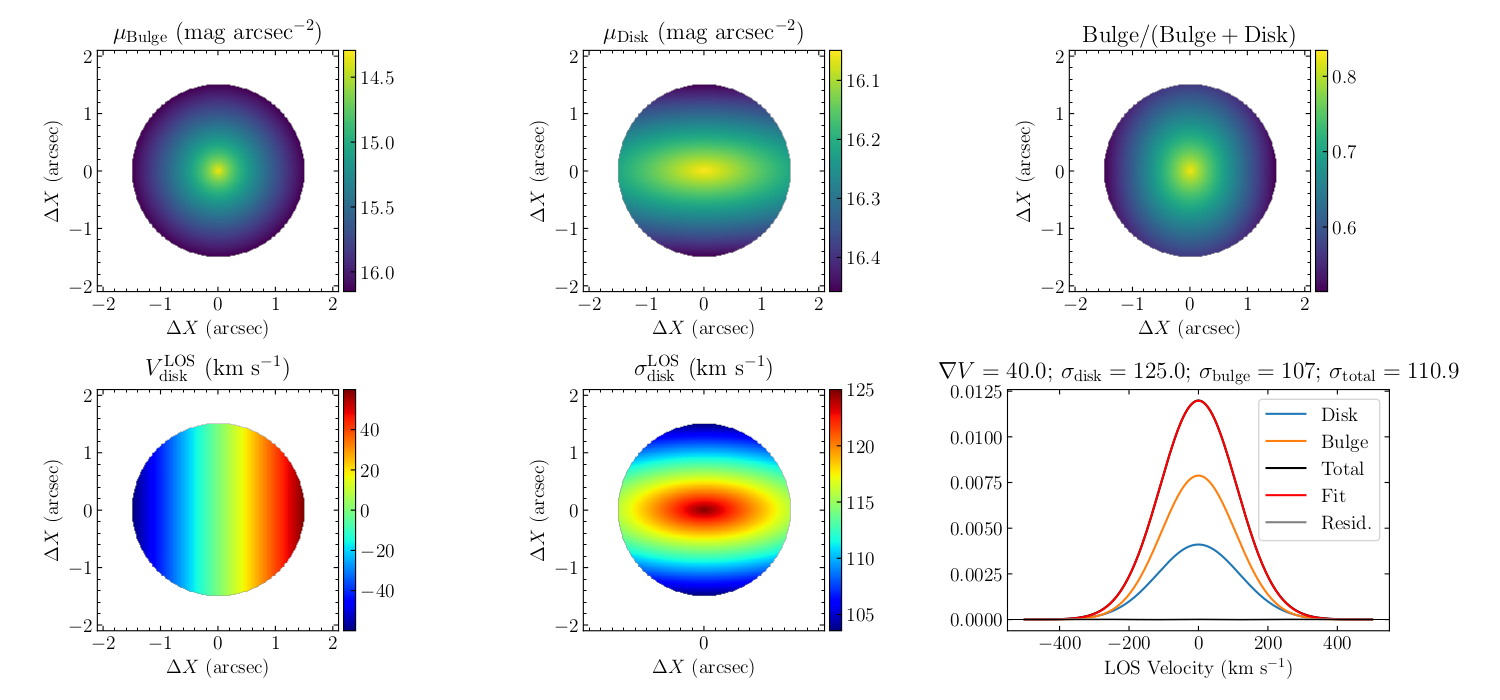}
\end{center}
\caption{\footnotesize \em{Illustration of the toy model for stellar velocity dispersion. Top left: Light density of the bulge. Top middle: Light density of the disk. Top right: Light density of bulge/(bulge+disk). Bottom left: Line-of-sight velocity profile of the disk. Bottom middle: Line-of-sight stellar velocity dispersion of the disk. Bottom right: Gaussian line of sight velocity distributions for the bulge, the disk, the total, with the fit and residuals. The bulge has a Gaussian LOSVD by definition. The disk and total LOSVDs have complex shapes in general. However, we fit the total LOSVD with a Gaussian to mimic the procedure for determining stellar velocity dispersion from a spectrum. For the particular combination shown here, the disk and total LOSVDs appear Gaussian.}}
\vspace{-0.3cm}
\label{fig:sigma_model}
\end{figure}

In order to assess the effects of disk inclination angle and SDSS fiber spatial resolution on the determination of the bulge stellar velocity dispersion, we construct a toy model using the measured properties of IC 750. We model the galaxy as a combination of a S\`{e}rsic bulge and an exponential disk, using the parameters derived from the {\it HST} and {\it Spitzer} images, as described in Section~\ref{sec:IRnuc}. In particular, in order to account for projection effects, we model the disk as having an inclination angle of 66.4 degrees, the value from our isophotal decomposition, in agreement with previously published values \citep{Verheijen01, Mao10}.  We assume that the bulge is not rotating and that it has a constant velocity dispersion. For the disk, we assume a linearly increasing rotation curve which is parametrized by the velocity gradient. A linear velocity gradient is valid within the 3\arcsec\ diameter, which corresponds to the central $\sim$200 pc, based on published measurements of the disk rotation curves, in both stars \citep{Heraudeau99} and gas \citep{Catinella05}. Assuming an isothermal stellar disk having a constant thickness, we model the velocity dispersion of the disk with an exponential fall-off, $\sigma = \sigma_0 \cdot e^{-r/2h}$, where $\sigma_0$ is the disk velocity dispersion at the center, $r$ is the radius and $h$ is the scale length of the exponential disk. We then construct the two dimensional rotational velocity and velocity dispersion for the bulge and disk, assuming Gaussian line-of-sight velocity distribution (LOSVD) for both components at each point in space. Based on this model, we can sum up the LOSVD within the SDSS aperture. While the LOSVD of the bulge is a Gaussian, the disk and total LOSVDs will have complex shapes in general. However, we fit the total LOSVD with a Gaussian to mimic the procedure for determining stellar velocity dispersion from a spectrum. Figure~\ref{fig:sigma_model} illustrates this model.

\begin{figure}[htb]
\begin{center}
\includegraphics[width=4.5in]{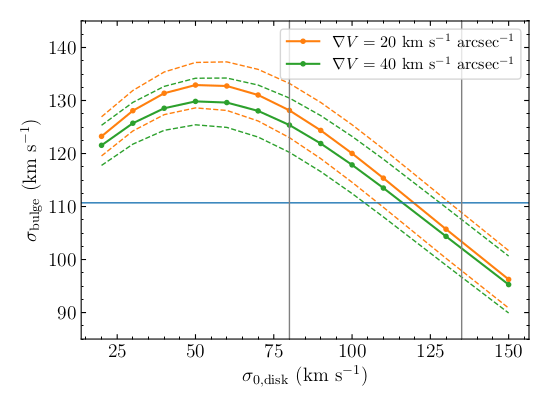}
\end{center}
\vspace{-0.5cm}
\caption{\footnotesize \em{The combinations for bulge stellar velocity dispersion and central disk stellar velocity dispersion which yield an observed stellar velocity dispersion of 110.7 km s$^{-1}$ for the reported stellar velocity gradient of 20 km s$^{-1}$ arcsec$^{-1}$ \citep[solid orange line]{Heraudeau99} and gas stellar velocity gradient of 40 km s$^{-1}$ arcsec$^{-1}$ \citep[solid green line]{Catinella05}. The dashed orange and green lines indicate the same for observed stellar velocity dispersions of $(110.7+3.6)$ =  114.3 km s$^{-1}$ and $(110.7-3.6)$ = 107.1 km s$^{-1}$, since the statistical error on the observed value of $\sigma_*$ is 3.6 km s$^{-1}$. The vertical lines indicate the upper limit for the central disk stellar velocity dispersion 135 km s$^{-1}$, as described above, and the spectral resolution 80 km s$^{-1}$, which we take as the lower limit since there are pixels outside the bulge dominated region which have roughly this value.}}
\vspace{-0.1cm}
\label{fig:sigma_summary}
\end{figure}

In addition to the parameters from our isophotal decomposition of the {\it HST} and {\it Spitzer} images, we use the disk stellar kinematics and the ionized gas kinematics reported in literature. The reported stellar velocity gradient is 20 km s$^{-1}$ arcsec$^{-1}$ \citep{Heraudeau99} and the gas velocity gradient is 40 km s$^{-1}$ arcsec$^{-1}$ \citep{Catinella05}. Even with spatially resolved spectroscopy, the central stellar velocity dispersion of the disk is not known because the central pixels contain contributions from the disk and the bulge. We, therefore, extrapolate from the disk stellar velocity dispersion at large radii which do not contain a bulge contribution. Based on the radial stellar velocity dispersion reported in \citet{Heraudeau99}, the stellar velocity dispersion is $\sim$70 km s$^{-1}$ at $r\sim$13\arcsec. Beyond this radius, the stellar velocity dispersion fall well below the spectral resolution of the \citet{Heraudeau99} spectrum, 80 km s$^{-1}$. Applying the exponential model, we find that the upper limit for the central disk stellar velocity dispersion is $\sim$135 km s$^{-1}$. Putting this data together with the model, we can find the combinations of bulge stellar velocity dispersion, central disk stellar velocity dispersion, and disk velocity gradient, which would yield an observed stellar velocity dispersion of 110.7 km s$^{-1}$ when the LOSVD is integrated within the 3\arcsec\ diameter SDSS aperture and fit with a single Gaussian. Figure~\ref{fig:sigma_summary} shows the combinations for bulge stellar velocity dispersion and central disk stellar velocity dispersion which yield an observed stellar velocity dispersion of 110.7 km s$^{-1}$ for the reported stellar velocity gradient of 20 km s$^{-1}$ arcsec$^{-1}$ \citep{Heraudeau99} (solid orange line) and gas stellar velocity gradient of 40 km s$^{-1}$ arcsec$^{-1}$ \citep{Catinella05} (solid green line). The dashed orange and green lines indicate the same for observed stellar velocity dispersions of $(110.7+3.6)$ =  114.3 km s$^{-1}$ and $(110.7-3.6)$ = 107.1 km s$^{-1}$, since the statistical error on the observed value of $\sigma_*$ is 3.6 km s$^{-1}$. The vertical lines indicate the upper limit for the central disk stellar velocity dispersion 135 km s$^{-1}$, as described above, and the spectral resolution 80 km s$^{-1}$, which we take as the lower limit since there are pixels outside the bulge dominated region which have roughly this value. In this model, for the upper limit of the central disk velocity dispersion, using the higher velocity gradient, we find that a bulge velocity dispersion of $\gtrsim$95 km s$^{-1}$ yield an observed value one (statistical) standard deviation lower than the central value. This indicates that the velocity dispersion we report in Section~\ref{sec:sigma}, $\sigma_*=110.7^{+12.1}_{-13.4}$ km s$^{-1}$ is not a large overestimate of the bulge stellar velocity dispersion and not a large underestimate of the error, since $(110.7 - 13.4)$ = 97.3 km s$^{-1}$.

\end{document}